%% file: FMintSDE_arXiv_Mar2026.tex
\documentclass[fleqn,11pt]{article}

\usepackage{graphicx}
\usepackage{float}
\usepackage{standalone}
\usepackage{mathrsfs}
\usepackage{wrapfig}
\usepackage{float}
\usepackage{stmaryrd}
\usepackage{rotating}
\usepackage{amsmath}
\usepackage{subfigure}
\usepackage{pgf,tikz}
\usetikzlibrary{arrows}
\usetikzlibrary{babel}
\usepackage{wrapfig}
\usetikzlibrary{calc,patterns,angles,quotes}
\usepackage{vmargin}
\usepackage{longtable}
\usepackage{subfigure}
   \usepackage{pdflscape}

\usepackage{subcaption}
\usepackage{algpseudocode}
\usepackage{array}
\usepackage{multirow}
\usepackage{multicol}
\usepackage{longtable}
\usepackage{booktabs}

\usepackage{titlesec}

\usepackage{pdfpages}
\usepackage{pdflscape}
\usepackage{xfrac}
\usepackage{authblk}
\usepackage[utf8]{inputenc}
\usepackage{enumitem} 
\usepackage{natbib}
\usepackage{extarrows}
\usepackage[ruled,linesnumbered]{algorithm2e}
\usepackage{multirow}
\usepackage{pdflscape}
\usepackage[toc,page]{appendix}

\usepackage{amssymb}
\usepackage{amsmath}
\usepackage{amsthm}
\usepackage{mathabx}
\usepackage{empheq}
\usepackage{mathtools}

\usepackage{amsopn}

\usepackage{array}
\newcolumntype{P}[1]{>{\centering\arraybackslash}p{#1}}

\def\dotminus{\mathbin{\ooalign{\hss\raise1ex\hbox{.}\hss\cr
  \mathsurround=0pt$-$}}}

\usepackage{arydshln}

\begin{document}

\title{FMint-SDE: A Multimodal Foundation Model for Accelerating Numerical Simulation of SDEs\\ via Error Correction}
\author[1]{Jiaxin Yuan\thanks{jyuan98@umd.edu}}
\author[1]{Haizhao Yang\thanks{hzyang@umd.edu}}
\author[1]{Maria Cameron\thanks{mariakc@umd.edu}}
{\scriptsize
\affil[1]{Department of Mathematics, University of Maryland, College Park, MD 20742, USA}
}

\maketitle

\abstract{
Fast and accurate simulation of dynamical systems is a fundamental challenge across scientific and engineering domains. Traditional numerical integrators often face a trade-off between accuracy and computational efficiency, while existing neural network-based approaches typically require training a separate model for each case.
To overcome these limitations, we introduce a novel multi-modal foundation model for large-scale simulations of differential equations: FMint-SDE (Foundation Model based on Initialization for stochastic differential equations). 
Based on a decoder-only transformer with in-context learning, FMint-SDE leverages numerical and textual modalities to learn a universal error-correction scheme. It is trained using prompted sequences of coarse solutions generated by conventional solvers, enabling broad generalization across diverse systems. We evaluate our models on a suite of challenging SDE benchmarks spanning applications in molecular dynamics, mechanical systems, finance, and biology. Experimental results show that our approach achieves a superior accuracy-efficiency tradeoff
compared to classical solvers, underscoring the potential of FMint-SDE as a general-purpose simulation tool for dynamical systems.
}


\tableofcontents

\section{Introduction}
Stochastic differential equations (SDEs) are widely used to describe the evolution of physical states over time while accounting for randomness. They serve as a fundamental framework for modeling systems subject to uncertainty or noise across diverse fields, including conformational changes in biomolecules and clusters of interacting particles \cite{MDsimulations_book2017,BolhuisTPS1998}, protein folding~\cite{protein_folding2002}, noise-driven transitions in nonlinear oscillator systems \cite{lorenz_Rand1978, Cycles_Tori_Dankowicz,CILENTI2024104582}, genetic switches~\cite{GeneticToggle2014}, financial market dynamics \cite{finance}, predator–prey interactions \cite{predator_prey_LiLiuWei}, and animal movements~\cite{McBride2023}.

Typically, simulation plays a crucial role in studying these systems, helping to understand how parameter variations or interventions affect their behavior. Large-scale simulations based on a given set of initial conditions are indispensable across diverse domains in many real-world applications. A key challenge in studying such systems effectively through simulation lies in balancing the trade-off between the accuracy and efficiency of traditional numerical solvers \cite{GENG2024112589,Zhongzhan2023,geng2025parallelintimesolutionallencahnequations,song2024fmintbridginghumandesigned}. 
In this work, we focus our attention on systems governed by Ito SDEs
\begin{equation}
\label{eqn:sde}
dX(t) = b(X_t, t)dt + \sigma(X_t, t)dW_t,\quad X_0 =x\in\mathbb{R}^d,
\end{equation}
where $b(x, t)$ denotes the drift term, $\sigma(x, t)$ the diffusion matrix, and $W_t$ a Wiener process modeling the stochastic noise. 
Given an initial condition $X_0$, one usually uses a numerical integrator, e.g., {Euler-Maruyama} or Milstein method \cite{Desmond_SDEs,KloedenPlaten1992}. 
For systems described by the SDE~\eqref{eqn:sde}, the time interval $[0, T]$ is discretized into $n$ equal subintervals, each with length $\Delta t = \frac{T}{n}$. Let $w = [w_1, ..., w_n]$ be a vector of $n$ independent Gaussian random variables with mean 0 and covariance matrix $(\Delta t)I_{d\times d}$. The Euler-Maruyama method computes the numerical solution using the following step function:
\begin{equation}
\label{eq:euler-maruyama}
X_{j+1} = X_j + b(X_j)\Delta t + \sigma(X_j) w_{j+1}, \quad X_0 = x.
\end{equation}
Milstein's method is essentially restricted to one-dimensional SDEs~\cite{Desmond_SDEs}. Its time steps are defined as
\begin{equation}
\label{eq:milstein}
X_{j+1} = X_j + b(X_j)\Delta t + \sigma(X_j) w_{j+1} + \frac{1}{2}\sigma(X_j)\sigma'(X_j)[w_{j+1}^2 - \Delta t], \quad X_0 = x.
\end{equation}

A larger time step $\Delta t$ accelerates the simulation but increases numerical error, and if it is too large, it causes instability issues. Conversely, a smaller time step reduces error at the expense of computational speed. This inherent trade-off between accuracy and efficiency makes large-scale simulations based on traditional numerical schemes computationally expensive.

Several efforts have been devoted to developing simulation schemes that accelerate the solution of SDEs. For instance, \cite{osti_10423299} and \cite{error_correct_EM_2015} present direct improvements over the {Euler-Maruyama} method, enabling faster simulations with higher accuracy. Liu et al. \cite{liu2021sevenleagueschemedeeplearning} proposed an instance-based deep learning approach for large time-step Monte Carlo simulations of stochastic differential equations. 
However, these methods primarily focus on designing new integrators by refining conventional schemes and still suffer from the inherent trade-off between accuracy and efficiency in large-scale simulations.

A foundation model is a large-scale AI model pre-trained on broad and diverse data, capable of adapting to a wide range of downstream tasks with minimal additional training \cite{ICON,Liu2024PROSEFD,Sun2025PROSEPDE,Ye2025PDEformer2,Herde2024POSEIDON,Ye2024PDEformer,Herde2024POSEIDON,song2024fmintbridginghumandesigned}. Such models have demonstrated strong transfer learning capabilities across numerous domains, avoiding repeated training for problem-dependent deep learning models \cite{Zhongzhan2023}. Motivated by this, we aim to design a foundation model that addresses the challenge of large-scale simulation of SDEs by leveraging initialization from traditional numerical integrators.

This work introduces FMint-SDE, a multi-modal foundation model initialized from classical numerical integrators for stochastic differential equation systems. FMint-SDE is a pre-trained model designed to accelerate large-scale SDE simulations while maintaining high accuracy through error correction.

Inspired by the FMint framework \cite{song2024fmintbridginghumandesigned} for large-scale simulation of ODE systems, we extend the approach to SDEs and train a large model using an in-context learning strategy~\cite{radford2019language,ICON}. The model first observes numerical demos consisting of coarse trajectories simulated with large time steps, the corresponding noise realizations at each timestamp, and the associated error correction terms that recover the fine trajectories generated with smaller time steps.
Using these demos, the model learns to reconstruct fine-grained trajectories by predicting the error terms from the coarse trajectories and noise realizations. The architecture is based on a decoder-only transformer~\cite {vaswani2023attentionneed,yang2024ICON-LM} that processes tokenized numerical data, along with an optional textual prompt, and predicts correction terms in a next-token prediction manner.

Pre-trained on four families of stochastic systems, we evaluate FMint-SDE on 12 distinct SDE families, including oscillators with time-periodic drift and state-dependent diffusion, and magnetic-core systems with colored noise. We demonstrate the effectiveness of FMint-SDE in terms of accuracy and practical applicability and further discuss its use cases and potential failure modes.

To the best of our knowledge, FMint-SDE is the first multi-modal foundation model for large-scale SDE simulation. It integrates the strengths of traditional numerical solvers with the learning capabilities of language models. Leveraging a decoder-only transformer architecture within an in-context learning framework, FMint-SDE achieves high efficiency and accuracy across a wide range of stochastic systems.

{The rest of the paper is organized as follows. Section \ref{sec:background} provides an overview of SDE solvers, their error metrics, and background on foundation models and in-context learning. The proposed FMint-SDE model is presented in Section \ref{sec:methodology}. Implementation details and diverse numerical test results on a large variety of SDEs are reported in Section \ref{sec:numerical}. Conclusions are drawn in Section \ref{sec:conclusion}. Supplementary Materials (SM) provide additional technical information on implementation and testing. We will refer to sections, figures, and tables from the Supplementary Materials by their numbers preceded by ``SM'', e.g., Section SM1, etc. }


\section{Background}
\label{sec:background}
\subsection{An overview of SDE simulators}
Commonly used conventional numerical integrators for SDEs include the first-order Euler-Maruyama~\eqref{eq:euler-maruyama} and the second-order Milstein~\eqref{eq:milstein} integrators.
In 1996, Roberts and Tweedie [13] proposed the Metropolis-Adjusted Langevin Algorithm (MALA) to address instability arising from the failure of the global Lipschitz condition in the overdamped Langevin SDE.
In 1988, Van Gunsteren and Berendsen \cite{Van_Gunsteren_leapfrog} proposed the leapfrog algorithm. This symplectic integration scheme updates positions and momenta in an interleaved fashion, with momenta evaluated at half-time steps relative to positions. 
Targeting systems with small noise -- stochastically perturbed nonlinear dynamical systems -- Breunung and Balachandran \cite{osti_10423299} proposed a computationally efficient integrator by leveraging deterministic numerical integration routines.

Some modern simulation schemes use neural networks to parameterize drift and diffusion.
In 2018, Yang et al. \cite{yang2018physicsinformedgenerativeadversarialnetworks} used physics-informed GANs (PI-GANs) with embedded governing SDEs. They employed Wasserstein GANs with gradient penalty (WGAN-GP) to improve stability when learning SDEs from scattered sensor snapshots.
Kidger et al.~\cite{kidger2021neuralsdesinfinitedimensionalgans} proposed learning SDEs as a Wasserstein GAN problem. They treated a neural SDE as the generator and a neural controlled differential equation as the discriminator, allowing arbitrary drift and diffusion in the infinite-data limit.
Furthermore, the backpropagation of neural SDE solvers can be made more accurate and faster, enabling reliable model training~\cite{kidger2021efficientaccurategradientsneural}.
Oh et al.~\cite{oh2025stableneuralstochasticdifferential} introduced three stable classes of neural SDEs robust to irregular, partially observed time series.
Ryder et al.~\cite{ryder2018blackboxvariationalinferencestochastic} proposed a fast, “black-box” variational inference method for SDEs that jointly infers model parameters and latent diffusion paths from discrete noisy observations using the Euler–Maruyama method.
Liu et al. \cite{liu2021sevenleagueschemedeeplearning} introduced a data-driven numerical scheme for solving SDEs with large time-step Monte Carlo simulations. This discretization scheme was designed using a polynomial chaos expansion method.

\subsection{The notion of strong and weak convergence of SDE solvers}
Let $Y(t)$ be the exact solution to SDE~\eqref{eqn:sde}, and $\{X_j\}$ be its numerical solution computed at times $\{t_j\}$.
A method is said to have a \emph{strong order of convergence} $\gamma$ if there exists a constant $C_1$ such that { for all small enough time steps $\Delta t$}
\begin{equation}
\label{eq:strong}
\mathbb{E} \left[\sup_{0 \leq t_j \leq T} \left|X_j - Y(t_j)\right| \right] \leq C_1{ \Delta t}^{\gamma}.  
\end{equation}
In other words, a method has a strong order of convergence $\gamma$, if the expected maximal path-wise error over a fixed time interval $[0,T]$ decays as $O({ \Delta t}^\gamma)$.

A method has a \emph{weak order of convergence} $\gamma$ if there exists a constant $C_2$ such that  { for all small enough time steps $\Delta t$}
\begin{equation}
\label{eq:weak}
\sup_{0 \leq t_j \leq T} \left|\mathbb{E}[X_j] - \mathbb{E}[Y(t_j)]\right| \leq C_2{ \Delta t}^{\gamma}.
\end{equation}
I.e., a method has a weak order of convergence $\gamma$ if the maximal error of the mean
over realizations of Brownian motions over a fixed time interval $[0,T]$ decays as $O({ \Delta t}^\gamma)$.

If the diffusion term is position-dependent, Euler-Maruyama has a strong order of $\sfrac{1}{2}$ and a weak order of $1$, while Milstein's method has both a strong and weak order of $1$ \cite{Desmond_SDEs}. The strong order of convergence of MALA  is $\sfrac{3}{4}$ \cite{Bou_Rabee_2009}.

\subsection{Foundation models and in-context learning for scientific computing}

\subsubsection{Foundation models}
The transformer architecture was first introduced by Vaswami et al. (2017) in their seminal paper \emph{``Attention is all you need''}~\cite{vaswani2023attentionneed} for the task of language translation.
The introduced attention mechanism, combined with positional encoding, enables large-scale training and has since been widely adopted to build large language models such as GPT-4 \cite{openai2024gpt4technicalreport} and Llama \cite{touvron2023llamaopenefficientfoundation}.
This architecture has also been applied to a variety of other domains, including recommendation systems \cite{sun2019bert4recsequentialrecommendationbidirectional,Liu_Yuan_csrec}, image classification and segmentation with Vision Transformer (ViT)-style backbones \cite{VIT, kirillov2023segmentanything, ramesh2021zeroshottexttoimagegeneration}, safety in medicine \cite{yang2024ensuringsafetytrustanalyzing}, and time-series forecasting \cite{zhou2021informerefficienttransformerlong, nie2023timeseriesworth64}.

Through extensive pre-training, foundation models can adapt to downstream tasks with little or no additional training.
Given these advances, the scientific computing community has also begun exploring the potential of leveraging such models.
Subramanian et al.~\cite{subramanian2024towards} studied how pre-trained and fine-tuned foundation models perform across PDE tasks and showed that the models remain effective even for out-of-distribution PDEs.
McCabe et al.~\cite{mccabe2023multiple} employed a transformer architecture to embed heterogeneous physical systems into a shared embedding space.
Shen et al.~\cite{shen2024ups}  proposed a unified PDE solver using the Fourier Neural Operator (FNO)~\cite{li2021fourier} in 1D and 2D.
On the other hand, Rahman et al.~\cite{rahman2024pretraining} introduced the Codomain Attention Neural Operator (CoDA-NO), which tokenizes functions with varying properties along the codomain.
This work also builds upon a decoder-only transformer architecture as the backbone. We show that, when pre-trained on various stochastic systems, our model exhibits strong transfer capabilities to downstream tasks with minimal fine-tuning.

\subsubsection{In-context learning for dynamical systems}
Yang et al. adapted the idea of in-context learning for operator learning in In-Context Operator Networks (ICON) \cite{ICON}. 
The model is trained to learn the operator from multiple demos, each consisting of a condition and a quantity of interest (QoI) and then to apply it to the query condition.
For instance, for a scalar ODE, $u' = \alpha u(t) + \beta c(t) + \gamma$, 
a condition consists of values of the discretized input function $c: [0, T] \rightarrow \mathbb{R}$ at given times, and the initial condition $u(0)$, while the QoI is the output function values $u: [0, T] \rightarrow \mathbb{R}$ at query times $\tau_0, \ldots, \tau_m$.

ICON~\cite{ICON} utilizes an encoder–decoder transformer architecture. The input to the encoder consists of all the demo pairs (conditions and QoIs) along with the condition for which the QoI is to be predicted:
\begin{equation} 
\left\{\{\text{Cond}_1, \text{QoI}_1\}, ..., \{\text{Cond}_n, \text{QoI}_n\}, \text{Cond}_{n+1}\right\}. \end{equation}
The input to the decoder is a query term, which is the vector $[\tau_0, \ldots, \tau_m]$ in the ODE example. 
Information from all demos and the query condition is processed by self-attention and fused through cross-attention with the query term for the QoI. 

Later, Yang et al.~proposed ICON-LM \cite{yang2024ICON-LM}, which applies the idea of in-context learning to operator learning.
It uses a decoder-only transformer architecture and uses existing pre-trained language models by incorporating textual prompts as input.
The model is trained in a language-like manner, predicting all QoI functions based on previous examples.
The data is prepared to consist of condition-QoI pairs, as in ICON, with an additional query concatenated at the end, e.g., {Cond, QoI, Query}. The model is then trained to predict the QoI value for each example at the query. 

The decoder uses the masked multi-head attention, where the mask prevents information flow within the decoder to preserve the autoregressive property.
In this case, the mask is designed to satisfy the following requirements: (1) predictions for each query are independent, and thus the model should be invariant to the order of the queries; and (2) for the current example, the query token attends to the tokens of the conditions and QoIs in previous examples, but not to those in the current example.

FMint \cite{song2024fmintbridginghumandesigned} expanded the use of in-context learning to numerical simulations of ordinary differential equations (ODEs).
It considered the case of simulating ODE systems with a large step size $k\Delta t$ for computational efficiency. Large time steps result in larger numerical errors. 
To eliminate this issue, FMint builds a corrective foundation model that approximates the error terms 
for various dynamical systems:
\begin{equation}
\hat{\mathbf{u}}_{k(n+1)}=\hat{\mathbf{u}}_{k n}+S\left(f, \hat{\mathbf{u}}_{k n}, k \Delta t\right)+\operatorname{FMint}\left(\hat{\mathbf{u}}_{k n}; \Theta\right), \quad \hat{\mathbf{u}}_0=\mathbf{c}_0, \quad n=0,1,\cdots,
\end{equation}
where $S(\cdot)$ represents the numerical integration scheme. In this work, we extend 
FMint to an error auto-corrector for SDEs.


\section{Proposed methodology}
\label{sec:methodology}
\subsection{The setup}
We will simulate stochastic differential equations (SDEs) of the form~\eqref{eqn:sde}. To explain our approach, we apply the Euler-Maruyama method~\eqref{eq:euler-maruyama} with step size $\Delta t$ to a scalar SDE. 
In addition, we simulate the same SDE with the same noise realization with a much larger time step $k\Delta t$, where $k \in \mathbb{N}$:
\begin{equation}
    X(t + k\Delta t) = X(t) + b(X(t),t) k\Delta t + \sigma(X(t),t) W_{k\Delta t},\quad W_{k\Delta t} \sim \mathcal{N}(0,k\Delta t).
\end{equation}
From the stochastic Taylor expansion\cite{JentzenKloeden2011_notes,KloedenPlaten1992}, the exact solution to SDE~\eqref{eqn:sde} is  
\begin{equation}
\label{eq:EMdesign}
    X_t = X_{t_0} + b(X_{t_0},t_0)\int_{t_0}^t ds + \sigma(X_{t_0},t_0)\int_{t_0}^t dW_s + R(t, t_0),
\end{equation}
where the error term $R(t,t_0)$ is
\begin{multline}
\label{eq:R}
    R(t,t_0) = \int_{t_0}^t \int_{t_0}^sL^0 b(u, X_u)du ds + \int_{t_0}^t \int_{t_0}^sL^1 b(u, X_u)dW_u ds \\
    + \int_{t_0}^t \int_{t_0}^s L^0 \sigma(u, X_u)du dW_s + \int_{t_0}^t \int_{t_0}^sL^1 \sigma(u, X_u)dW_u dW_s.
\end{multline}
The operators $L^0$ and $L^1$ are defined as 
\begin{equation}
    L^0 = \frac{\partial}{ \partial t} + b \frac{\partial}{ \partial x} + \frac{1}{2}\sigma^2 \frac{\partial^2}{ \partial x^2}, \quad L^1 = \sigma \frac{\partial}{ \partial x}.
\end{equation}
Equations \eqref{eq:EMdesign} and \eqref{eq:R} reveal the error term for Euler-Maruyama:
\begin{equation}
    \underbrace{X(t + k\Delta t) = X(t) + b(X(t),t) k\Delta t + \sigma(X(t),t) W_{k\Delta t}}_{\text{Euler-Maruyama}} + \underbrace{R(t+k\Delta t, t)}_{\text{ err}_n(k,\Delta t, X(t)) }.
\end{equation}
Equation~\eqref{eq:R} shows that the error term $\text{err}_n(k,\Delta t, X(t))$  is $O(k\Delta t)$ when $\sigma$ is position-dependent and $O((k\Delta t)^{3/2})$ otherwise. Hence, it is large if $k\Delta t$ is large.

We aim to design a model approximating the error terms for numerical solutions obtained with large time step $k \Delta t$. 
{ We will use the adjective ``coarse'' for these solutions.}
These coarse trajectories are used as an initialization to produce solutions with accuracy equivalent to those obtained with a small step size $\Delta t$ via error correction.
Our model with in-context learning  is of the form
\begin{equation}
\begin{aligned}
\hat{X}_{k(n+1)}&=\hat{X}_{k n}+\text{\bf Simul}\left(b, \sigma, \hat{X}_{k n}, k \Delta t\right)+\operatorname{\bf FMint-SDE}\left(\hat{X}_{k n}, \Delta W_{kn} ; \Theta\right),\\
\hat{X}_0&=c_0, \quad n=0,1, \ldots, {\rm round}(\sfrac{T}{k\Delta t}),
\end{aligned}
\end{equation}
where $\hat{X}_{kn}$ is the simulated solution at time step $n$ with step size $k\Delta t$,
$\Theta$ represents all the model parameters, $\text{\bf Simul}(\cdot)$ denotes the integration scheme, and $\Delta W_{kn}\sim\mathcal{N}(0,k\Delta t)$. This allows us to simulate with a large time step $k\Delta t$ while maintaining the same accuracy as when using the refined time step $\Delta t$.
Note that the scheme is independent of the numerical integration method and can be easily applied to higher-order simulators and to systems of SDEs.
For example, for Milstein’s method, FMint-SDE is designed to recover
\begin{equation}
\begin{aligned}
    \hat{X}_{k(n+1)} &= \hat{X}_{kn} + b(\hat{X}_{kn})\Delta t + \sigma(\hat{X}_{kn}) w_{k(n+1)}\\& + \frac{1}{2}\sigma(\hat{X}_{kn})\sigma'(\hat{X}_{kn})[w_{k(n+1)}^2 - \Delta t ] 
    + \operatorname{\bf FMint-SDE}\left(\hat{X}_{k n}, \Delta W_{kn}; \Theta\right).
    \end{aligned}
\end{equation}

\subsection{Data preparation}
\label{sec:demos}
We aim to train the model to predict the correction terms under the in-context learning framework. Therefore, we prepare data pairs consisting of coarse trajectories and their corresponding error terms.
These pairs serve as demos for the model, helping it learn the relationship between the coarse trajectories and the desired error correction.
{ To prepare the demos, we first simulate ``fine'' solutions with a small step size $\Delta t$, using the Euler–Maruyama method:}
\begin{equation*}
    X^{\rm fine}_{m+1} = {\rm EM}(X^{\rm fine}_m) = X^{\rm fine}_m + b(X^{\rm fine}_m, t_m) \Delta t + \sigma(X^{\rm fine}_m, t_m) \Delta W_m,\quad \Delta W_m \sim \mathcal{N}(0, \Delta t).
\end{equation*}
Then, we compute coarse Euler-Maruyama solutions { at time indices $m = nk$, $n = 0,1,\ldots,N$, and $k\in\mathbb{N}$ is fixed.} 
The increments of Brownian motions  obtained by summing those for the fine solutions: 
\begin{equation}
\begin{aligned}
    X^{\rm coarse}_{m+k} &= {\rm EM}(X^{\rm coarse}_m ) \\
    &= X^{\rm coarse}_m + b(X^{\rm coarse}_m, t_m) k\Delta t + \sigma(X^{\rm coarse}_m, t_m) \sum_{i = 0}^{k-1} \Delta W_{m + i}.
    \end{aligned}
\end{equation} 
The error term is computed as the difference between fine trajectories and coarse trajectories at each time stamp $nk \Delta t$, $n=0,1,2,\ldots, N$: 
\begin{equation}
    {\rm err}_n = X^{\rm fine}(nk\Delta t) - X^{\rm coarse}(nk\Delta t).
\end{equation}
For in-context learning, the data is constructed so that FMint-SDE learns to produce the correction term ${\rm err}_n$ from multiple demos of the same SDE with different noise realizations.
Hence, the model is first presented with a few sets of demos consisting of the time stamps {$t_n = nk\Delta t$}, a coarse trajectory, noise values, and error values relative to the corresponding fine trajectory:
\begin{equation}
    (t_n, X^{\rm coarse}_n, \Delta W_n, {\rm err}_n),\quad n = 0,1,...,N.
\end{equation}

For a $d$-dimensional SDE, each demo consists of the following arrays:
\begin{itemize}
    \item the time stamps $0 \ldots t_N$, 
    \item the realization of noise at each dimension for the given time stamp, $\Delta W^1_0$, $\ldots$, $\Delta W^1_N$, $\Delta W^2_0$, $\ldots$, $\Delta W^2_N$, etc.,
    \item the value of the coarse initialization for each dimension, $\hat{X}_1(0) \ldots \hat{X}_1(t_N)$, $\hat{X}_2(0) \ldots \hat{X}_2(t_N)$, etc., and
    \item the error correction term { for each dimension}, $\text{err}_{\hat{X}_1}(0) \ldots \text{err}_{\hat{X}_1}(t_N)$.
\end{itemize}
These demos are organized into arrays as follows (also see Section SM1):
\begin{center}
    \includegraphics[width = \textwidth]{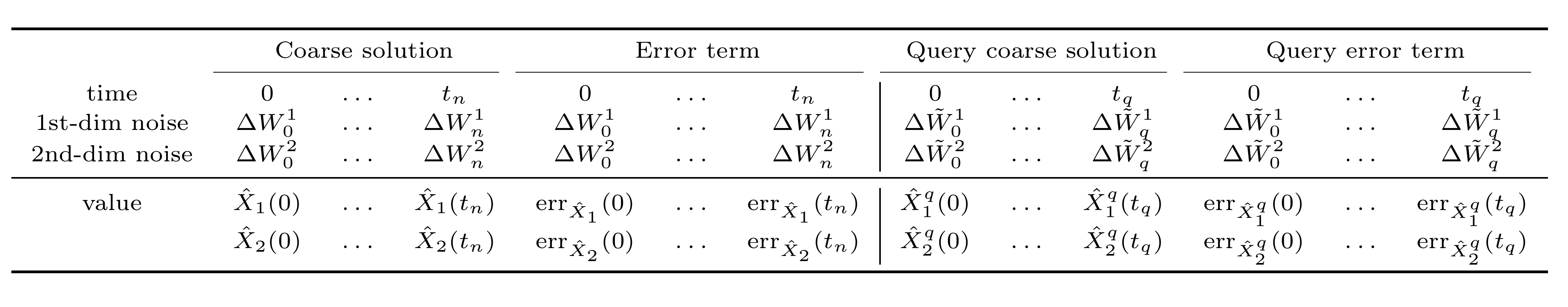}
\end{center}

From these demos, the model captures the relationship between the conditions $(t_n, X^{\rm coarse}_{n}, \Delta W_n)$ and the desired output error term ${\rm err}_n$.
Given $K$ demos prepared from trajectories of the same SDE with different initial conditions and noise realizations, the model is trained to recover ${\rm err}^{\rm new}_n$ from the query conditions:
\begin{equation}
    {\rm Cond.} = (t^{\rm new}_n, X^{\rm new, coarse}_{n}, \Delta W^{\rm new}_n), 
\end{equation}
generated from the same SDE with different initial conditions and noise realizations. Table SM1 in the Supplementary Material shows an example of an input data structure.

{ Our implementation of the model supports input dimensions up to $d = 3$. When $d < 3$, the unused dimensions are filled with zeros.}

\subsection{Tokenization, model architecture and modality fusion}
The model treats each column { of a demo array} 
as one input token, 
which is first processed by an embedding layer that maps all the data into an embedding vector space.
A learnable positional encoding is applied to the input embeddings to differentiate between different types of input data and different examples.
Specifically, the positional embeddings are the same for each data category, e.g., coarse solution or error term, within the same example.
For the optional textual data, we utilize a pre-trained language model. The textual data are first mapped into a separate embedding space. The embedding vectors of the numerical data are then concatenated with the text embeddings.

After concatenation, the resulting input sequence is fed into a decoder-only transformer model.
To ensure that the model does not “see” the error terms it is asked to predict, we use a masking design in the attention blocks similar to those used in FMint \cite{song2024fmintbridginghumandesigned} and ICON-LM \cite{yang2024ICON-LM}.
In particular, the mask is designed such that when predicting the error terms for the previous examples, the textual information, time stamp, noise values, coarse solution values, and error terms are not masked out; for the current example, the token attends to the time stamp, noise values, and coarse solution values but not to the error terms.

The output from the transformer blocks is passed through a multi-layer perceptron head to align the dimensions, producing the final corresponding error correction terms for the queried coarse solution.

\subsection{Training and inference}
We first pre-trained FMint-SDE on diverse simulated SDE data using an in-context learning scheme, with or without providing textual information. 
The training minimizes the mean squared difference (MSD)~\eqref{eqn:mse}, which is the difference between the predicted error corrections and the ground truth error
\begin{equation}
\text{MSD} = \frac{1}{N}\sum_{i = 1}^N\|\widehat{\text{err}}^i - \text{err}^i\|_2,
\label{eqn:mse}
\end{equation}
where $\widehat{\text{err}}$ is the model's predicted error term.

{ During training, we provide the model with $K$ demos.} To better utilize the data, the loss is computed for all { demos} from the second to the last within the same pass. { Using a masking mechanism,} the model learns dynamically to make predictions with a random number of demo pairs, up to $K$ demos.
When textual information is included, an additional loss is computed from the first { demo} prediction to the last. In this case, the model learns to perform zero-shot prediction on the first { demo}, relying solely on textual information, time stamps, and noise realizations.
The total loss for multimodal training combines both components, whereas the loss that depends only on numerical values excludes the zero-shot prediction term.
Furthermore, to improve robustness to missing values, data at several time stamps are masked, so the model learns to make predictions without access to all inputs. 

During inference, we provide the model with up to $K$ demos of the form described in Section \ref{sec:demos}.
FMint-SDE then makes an error prediction for the query coarse solution.

\subsection{Roll-out for long time horizon}
\label{sec:rollout}
Given computational resource constraints, the input trajectory length for FMint-SDE is predefined. As a result, FMint-SDE can only perform error correction over a limited number of time steps.
In practical applications, however, simulations often need to run for durations much longer than those predefined for FMint-SDE.
To address this issue, we develop a roll-out scheme enabling FMint-SDE to do error correction for arbitrarily long simulations.

Suppose the model processes $N$ steps at a time, and we aim to recover the error of coarse-grained data over $\alpha N$ steps, where $\alpha > 1$.
For the first $N$ steps, inference can be performed as usual.
However, from step $N$ to $2N-1$, error propagation in the coarse solution can cause the original correction scheme to produce inaccurate predictions.
Thus, we first shift the coarse solution to have the same value as the fine solution at time $t_N$:
\begin{equation}
    \rm X^{fine}_N = \rm \tilde{X}^{coarse}_{N}
\end{equation}
where $\tilde{X}^{\mathrm{coarse}}_{N}$ denotes the coarse solution at time step $N$ after alignment.
Before feeding the coarse solution into FMint-SDE, we shift all values in $\{X^{\mathrm{coarse}}\}_{N}^{2N-1}$ by $\mathrm{err}_N$. During inference, although the true error is unknown, we approximate it using the model’s predicted error, such that
\begin{equation}
    {\rm \hat{X}^{fine}_N} = {\rm \tilde{X}^{coarse}_{N}} = {\rm X^{coarse}_{N}} + \widehat{\rm err}_N.
\end{equation}
The predicted fine solution on $[t_N,t_{2N}]$, given the predicted error term from step $N$ to $2N$, can be written as
\begin{align}
    \{\rm \hat{X}^{fine}\}_{N}^{2N} & = \{\widehat{\rm err}\}_N^{2N} + \{\rm \tilde{X}^{coarse}\}_{N}^{2N} \\
    \notag & = \{\widehat{\rm err}\}_N^{2N} + \left(\{\rm X^{coarse}\}_{N}^{2N} + \rm err_N\right).
\end{align}
This scheme is applied iteratively until the model makes error corrections for all steps.

\subsection{Novelty relative to FMint for ODEs}
{ 
FMint-SDE is an AI-based tool for the autocorrection of numerical errors arising from the Euler-Maruyama SDE integrator. Its core methodology and training paradigm are transferred from FMint for ODEs~\cite{song2024fmintbridginghumandesigned}. However, FMint-SDE is not a trivial extension of FMint. 

First, a trajectory of an SDE is determined not only by the initial data but also by the noise realization, i.e., a Brownian path. This requires an additional component for each demo of the time-series type. Furthermore, small differences in Brownian paths may lead to drastic differences in the SDE trajectories. This is especially evident in metastable systems, where the drift field bears multiple attractors. Therefore, for each SDE and each initial condition, we must construct a set of demos corresponding to different noise realizations.

Second, the additional memory required to learn the dependence of the SDE trajectory on the noise realization limits the length of the time interval in each demo and, consequently, the inference time interval. To address this difficulty, we developed the roll-out scheme that allows us to use FMint-SDE on time intervals of arbitrary length, provided that they are split into portions of the demo's time interval size. 
}


\section{Numerical experiments}
\label{sec:numerical}
In this section, we conduct numerical experiments to evaluate FMint-SDE's effectiveness and performance under various conditions.
Specifically, we present results on in-distribution and out-of-distribution SDE systems. We examine use-case applications and systems that pose greater challenges for FMint-SDE and analyze the amount of additional training data required to achieve effective transferability.
We further assess its performance on SDEs that exhibit distinct behaviors under varying parameter settings.

\subsection{Basic setup}
\label{sec:exp_setup}
\subsubsection{Pre-training data preparation}
We created a pre-training dataset using the following four SDE systems:
\begin{enumerate}
    \item the geometric Brownian motion in 1D;
    \item the overdamped Langevin dynamics with Mueller's potential in 2D (\emph{OLD-Mueller}, Section SM3), a popular test problem in chemical physics literature \cite{EVE2010,yuan2023optimalcontrolsamplingtransition};
    \item a noisy nonlinear oscillator with a periodic modulation in the drift term and state-dependent noise in 2D (\emph{periodic nonlinear oscillator}) ~\cite{Cycles_Tori_Dankowicz}, \begin{equation}
    \label{eq:pno}
\begin{aligned}
    dx_1 &= \frac{2\pi}{\omega}\!\left( -\Omega x_2 + x_1 g(x_1,x_2,t)\right) dt 
            + \sigma \sqrt{\frac{2\pi}{\omega}}\, x_1 x_2\, dW_t, \\
    dx_2 &= \frac{2\pi}{\omega}\!\left( \Omega x_1 + x_2  g(x_1,x_2,t) \right) dt 
            + \sigma \sqrt{\frac{2\pi}{\omega}}\, x_2^2\, dW_t,\\
            &{\rm where}\quad g(x_1,x_2,t) = 1 + \sqrt{x_1^2 + x_2^2}\, (\cos 2\pi t - 1);
\end{aligned}
\end{equation}
    \item the 3D Lorenz'63 system~\cite{lorenz1963deterministic} with added additive noise (\emph{stochastic Lorenz})\cite{lorenz_Cameron_2019}.
\end{enumerate}
The definitions and settings for these systems, including coefficient ranges, time step sizes $\Delta t$, stride values $k$, and the range of initial conditions, are found in Section SM3. 

We first sampled $N_{\rm par} = 1000$ different parameters from the predefined parameter range for all pre-training SDE systems. For each set of parameters, we sampled $N_{\rm ic} = 10$ initial conditions. Given each combination of parameters and initial conditions, $N_{\rm noise} =40$ trajectories were generated with different noise realizations.
Hence, the pre-training dataset consisted of $N_{\rm par}\cdot N_{\rm ic}\cdot  N_{\rm noise} = 400,000 $ trajectories for each system. 
During training, each training sample consists of $K = 4$ demos and one query, randomly drawn from trajectories simulated with identical parameters.
This means that the model first observes demos constructed from four trajectories of the same SDE, each with different initial conditions and noise realizations, and then predicts the error term of the fifth trajectory.
The test set for each system consisted of trajectories with 25 different initial conditions and 40 noise realizations for each initial condition.

For the multi-modal feature, we constructed $N_{\rm text} = 10$ textual descriptions for each system. The descriptions include the systems' mathematical form in LaTeX, their applications to different fields, the meaning of the parameters, and the types of behavior for different ranges of their parameters.
The supplemental textual data were generated with the assistance of GPT-5 for well-known systems such as stochastic Lorenz and OLD-Mueller. 
For other systems, e.g., predator-prey~\cite{predator_prey_LiLiuWei} and fluxgate sensor~\cite{PhysRevE_fluxgate_sensor}, textual information was manually created. 
The details of the textual data generation and some examples are provided in Section SM5.

\subsubsection{Implementation details}
FMint-SDE is a six-layer decoder-only transformer with eight heads. The embedding vector dimension used for numerical input, e.g., coarse solutions, error corrections, and all conditions, is 256, and the hidden dimension of the feed-forward networks is set to 1024. The network uses the Glorot uniform initialization~\cite{pmlr-v9-glorot10a} and no dropout. 
The model involves approximately 15.8M parameters.
For the multi-modal feature, we used GPT-2 \cite{radford2019language} as the pre-trained language model for obtaining the embedding of the textual data.

We trained the model for 100 epochs, whereas each training epoch comprised 10,000 iterations.
We used the AdamW optimizer with a warmup-cosine-decay schedule, the peak learning rate of $10^{-4}$, $\beta_1 = 0.9$ and $\beta_2 = 0.999$, and the weight decay of  $10^{-4}$. 
We conducted training { and all simulations by Euler-Maruyama} on NVIDIA A6000 GPUs with 48 GB of memory {with JAX, a Python library designed for high-performance numerical computing and large-scale machine learning~\cite{jax2018github}}. 

\subsubsection{Evaluation metrics}
Inspired by the strong and weak orders of convergence for numerical integration schemes \eqref{eq:strong}, \eqref{eq:weak}, we designed the following two metrics to evaluate the performance of FMint-SDE. The Averaged Maximum Difference (AMD) and the Maximum Averaged Difference (MAD) are defined, respectively, as
\begin{equation}
\label{eq:AMD}
   \text{AMD} =  \frac{1}{N}\sum_{i = 1}^N \frac{1}{M}\sum_{\omega = 1}^M \max_j |\hat{X}_j^{\omega,i} - X_j^{\omega,i}|,
\end{equation}
\begin{equation}
\label{eq:MAD}
   \text{MAD} =  \max_j \frac{1}{N}\sum_{i = 1}^N  \left|\frac{1}{M}\sum_{\omega = 1}^M\hat{X}_j^{\omega,i} - \frac{1}{M}\sum_{\omega = 1}^M X_j^{\omega,i}\right|,
\end{equation}
where $\hat{X}_j^{\omega, i}$ is the predicted SDE solution for the $i$-th equation with $\omega$-th noise realization at $j$-th time stamp. 
It is obtained via $\hat{X}^{\omega, i}_j = \tilde{X}^{\omega, i}_j + \widehat{\text{err}}^{\omega, i}_j$, where $\tilde{X}^{\omega, i}_j$ represents the coarse solution and $\widehat{\text{err}}^{\omega, i}_j$ is the predicted output by FMint-SDE. 
The ground-truth $X_j^{\omega, i}$ is the fine solution at $j$-th time stamp.

In addition, we used the standard evaluation metrics: Mean Absolute Error (MAE) and Root Mean Squared Error (RMSE):
\begin{align}
    \text{MAE}& = \frac{1}{N}\sum_{i = 1}^N \frac{1}{M} \sum_{\omega = 1}^M \|\hat{X}^{\omega, i} - X^{\omega, i}\|_2, \\
    \text{RMSE} &= \sqrt{\frac{1}{N}\sum_{i = 1}^N \frac{1}{M} \sum_{\omega = 1}^M \|\hat{X}^{\omega, i} - X^{\omega, i}\|_2^2}.
    \end{align}

\subsubsection{Baselines}
{ The problem solved by FMint-SDE is novel; hence, no direct baselines exist in the literature. Thus, we compare FMint-SDE against two ad hoc learning-based baseline models\footnotemark[1]: (i) a single-SDE specialized model and (ii) a black-box surrogate model. \footnotetext[1]{We thank the reviewer for proposing these baselines.}
The single-SDE specialized model is a neural network that takes coarse trajectories and their corresponding noise realizations as inputs and predicts the error-correction term. 
The black-box surrogate model directly predicts the solution at each time step from the initial condition and the noise realization, without requiring the coarse solution as input. For a fair comparison, we use the same architecture and training settings for the black-box surrogate as for FMint-SDE, but replace coarse-trajectory inputs with initial conditions. 
Both baseline models are trained and tested on NVIDIA A6000 GPUs with 48 GB of memory. The configuration and training details of these baseline models are provided in Section SM2.}

\subsection{In-distribution SDEs}

\input{Tables/in-distribution-ft}

We evaluated FMint-SDE on the test split of the pre-training dataset using four different schemes: few-shot inference with and without textual prompts, and fine-tuning with and without textual prompts.
{ ``Few-shot'' means that the model is provided with a few demos as context, but no additional training is done. ``Fine-tune'' means that further training is done on the query SDE data.
}

The test split contains 1000 trajectories that differ from the pre-training dataset in parameters, initial conditions, and noise realizations.
Fine-tuned results are obtained by training the model for 1000 or 2000 epochs using a fine-tuning dataset of size 50.
{ The single-SDE specialized models are each trained for 500 epochs on 5000 trajectories simulated under the same parameter setting but different initial conditions.}
Table~\ref{tab:in_dist_error_all} compares the four FMint-SDE schemes, { the black-box surrogate model, the single-SDE specialized models} and the corresponding coarse solution errors. The the best results are highlighted in bold.
All error metrics are computed by averaging over 1000 trajectories, each evaluated with four demo pairs during inference.

Table \ref{tab:in_dist_error_all} shows that the best strategy of FMint-SDE improves the accuracy 
\begin{itemize}
    \item by nearly two orders of magnitude for geometric Brownian motion,
    \item by the factors from 4 to 8 for the OLD-Mueller, 
    \item by the factors from 2 to 4 for the periodic nonlinear oscillator,
    \item by the factors from 6 to 11 for the stochastic Lorenz. 
\end{itemize}
{Across all systems, the proposed FMint-SDE consistently achieves the lowest errors in all four metrics, substantially outperforming coarse simulations and baseline models.}

\begin{figure}[h]
    \centering
    \vspace{-1em}
    \includegraphics[width=0.45\linewidth]{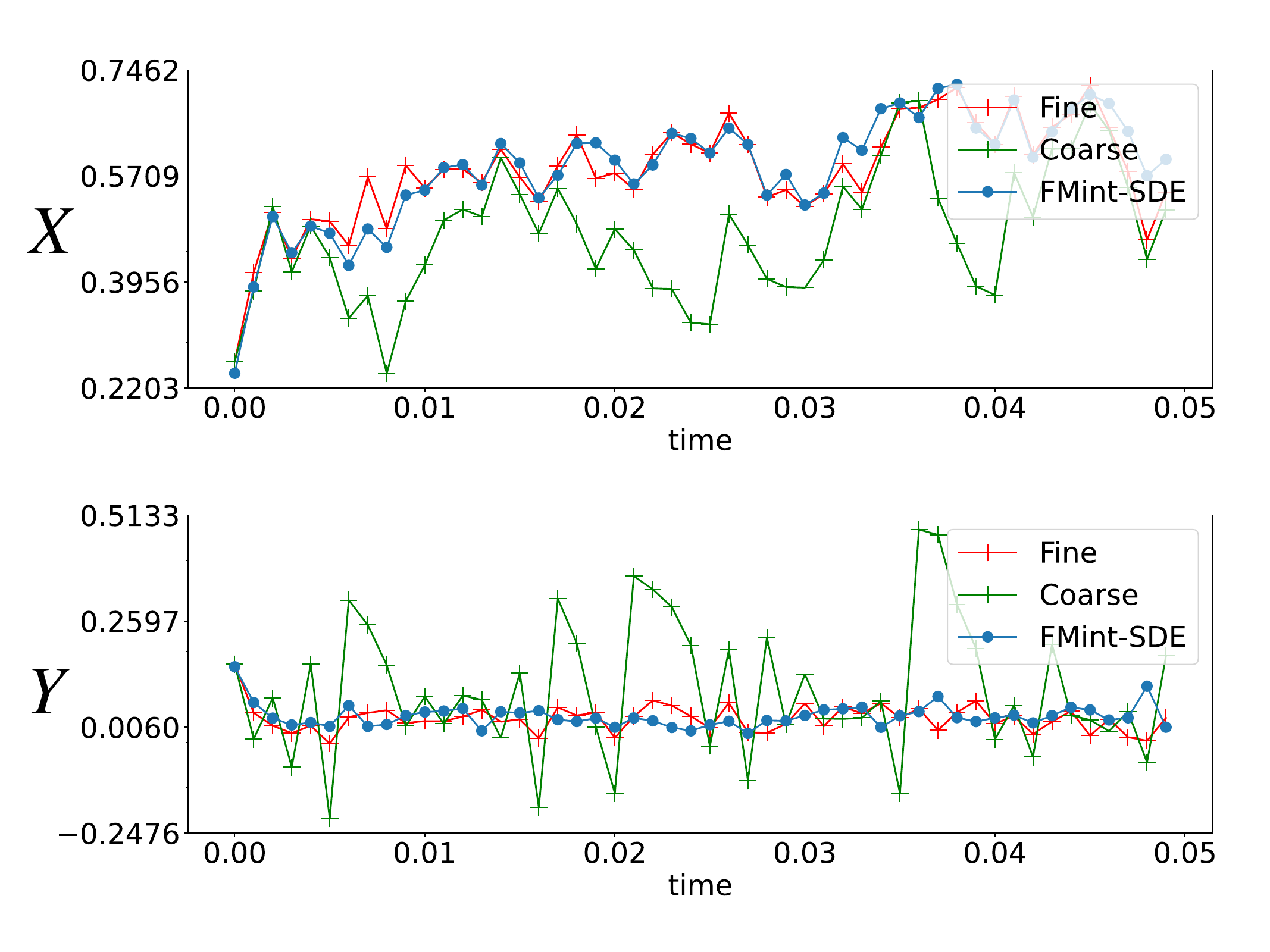}
    \includegraphics[width=0.53\linewidth]{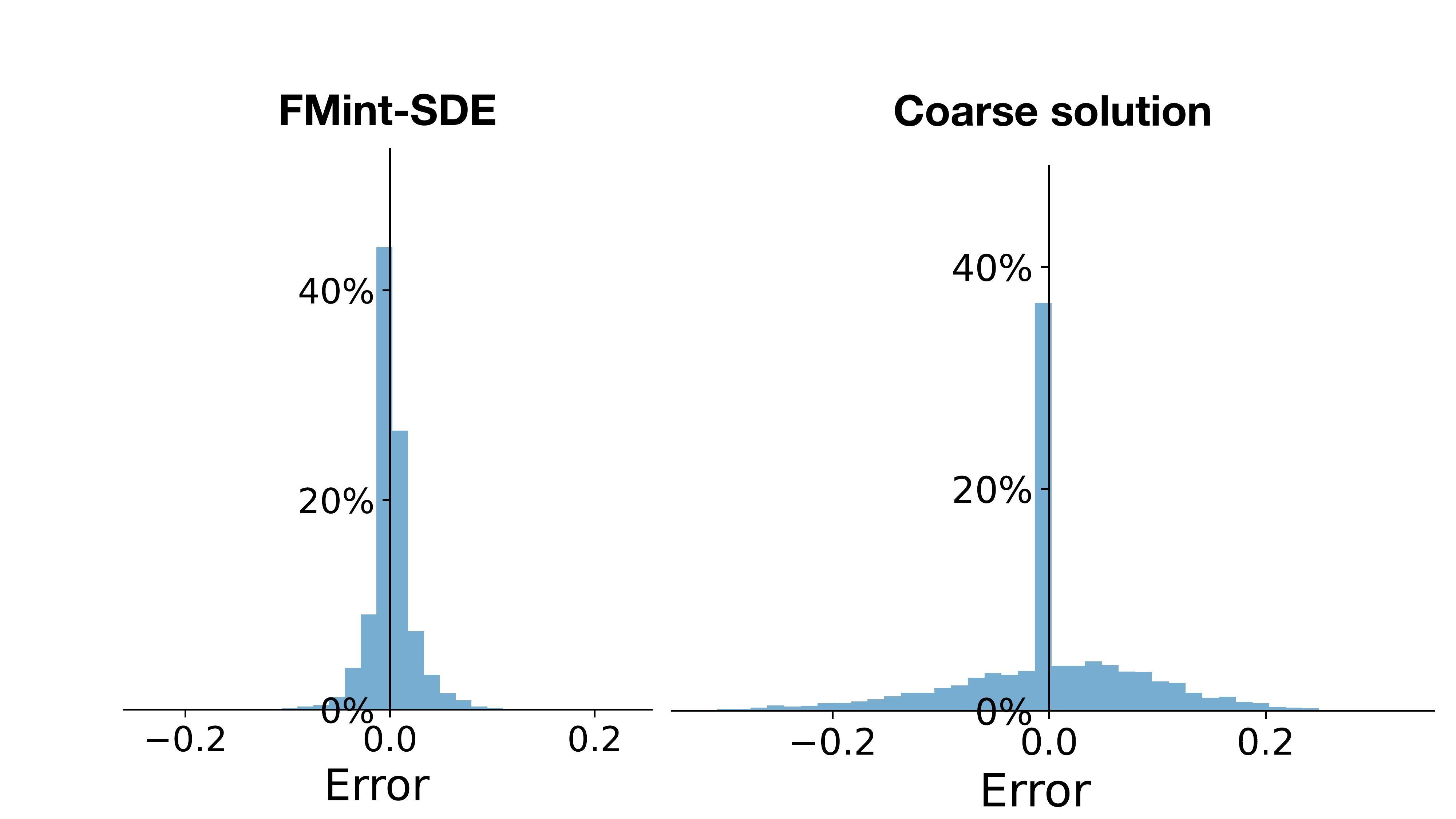}
    \caption{\footnotesize Left: A comparison of the coarse, fine, and FMint-SDE-corrected $X$- and $Y$-components of a trajectory of the OLD-Mueller.  Right: The error distributions with and without the correction by FMint-SDE. The fine and coarse time steps are $\Delta t = 10^{-5}$ and $100\Delta t = 10^{-3}$, respectively.}
    \label{fig:Mueller_visual}
\end{figure}

{ Table 1 shows that for the SDE with the simplest drift and diffusion coefficient, the geometric Brownian motion, the few-shot without a textual prompt gives the best result. Few-shot with a textual prompt yields the best results for SDEs with more complex drift, such as the OLD-Mueller and the stochastic Lorenz. Fine-tuning seems to lead to overfitting in these cases. By contrast, fine-tuning, along with a textual prompt, is beneficial for the periodic nonlinear oscillator~\eqref{eq:pno}. This SDE exhibits drastically different behaviors under different parameter settings--see Fig. \ref{fig:periodic_traj_main}. This hinders learning error patterns. }

\begin{figure}[h]
    \centering
    \vspace{-1em}\includegraphics[width=0.75\linewidth]{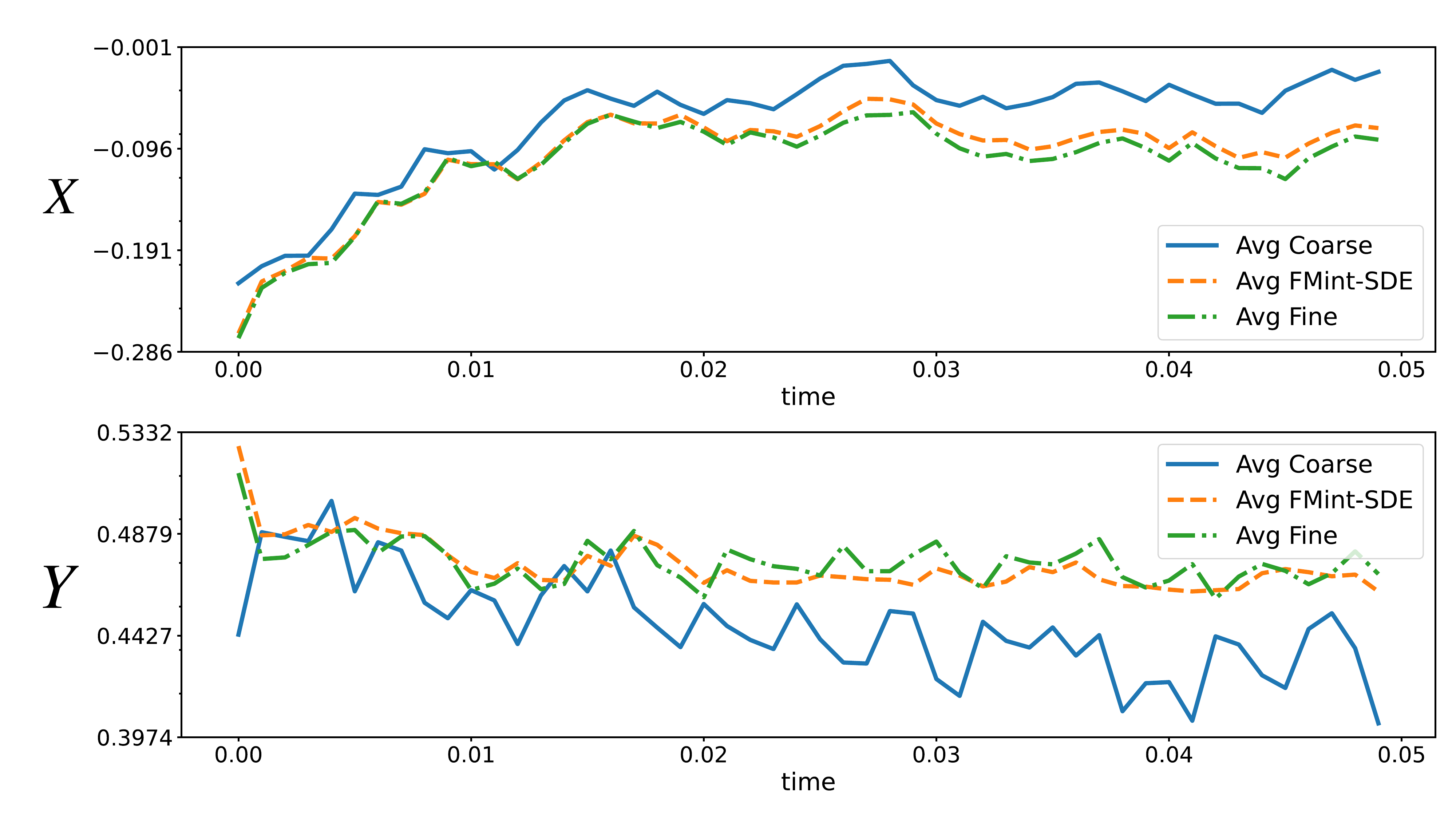}
    \caption{\footnotesize Averaged trajectories of coarse solution, fine solution and FMint-SDE over forty different noise realizations for the overdamped Langevin SDE with Mueller's potential (OLD-Mueller).}
    \label{fig:traj_avg}
\end{figure}

Fig. \ref{fig:Mueller_visual}(left) displays a sample trajectory of the OLD-Mueller. The FMint-SDE trajectory closely aligns with the fine one, whereas the coarse trajectory deviates significantly from it. Fig.~\ref{fig:Mueller_visual}(right) compares the FMint-SDE and coarse error distributions. The latter one is considerably wider. This observation is supported by Fig.~\ref{fig:traj_avg}, which shows averaged trajectories of the fine, coarse, and FMint-SDE-corrected solutions over forty noise realizations of the OLD-Mueller. 
The average trajectory of FMint-SDE aligns well with that of the fine solution, whereas the coarse solution substantially deviates from it.

{Figures SM12, SM15, and SM18 visualize FMint-SDE's error correction performance on the other three in-distribution SDE systems.}

\begin{figure}[h]
    \centering
    \includegraphics[width = \textwidth]{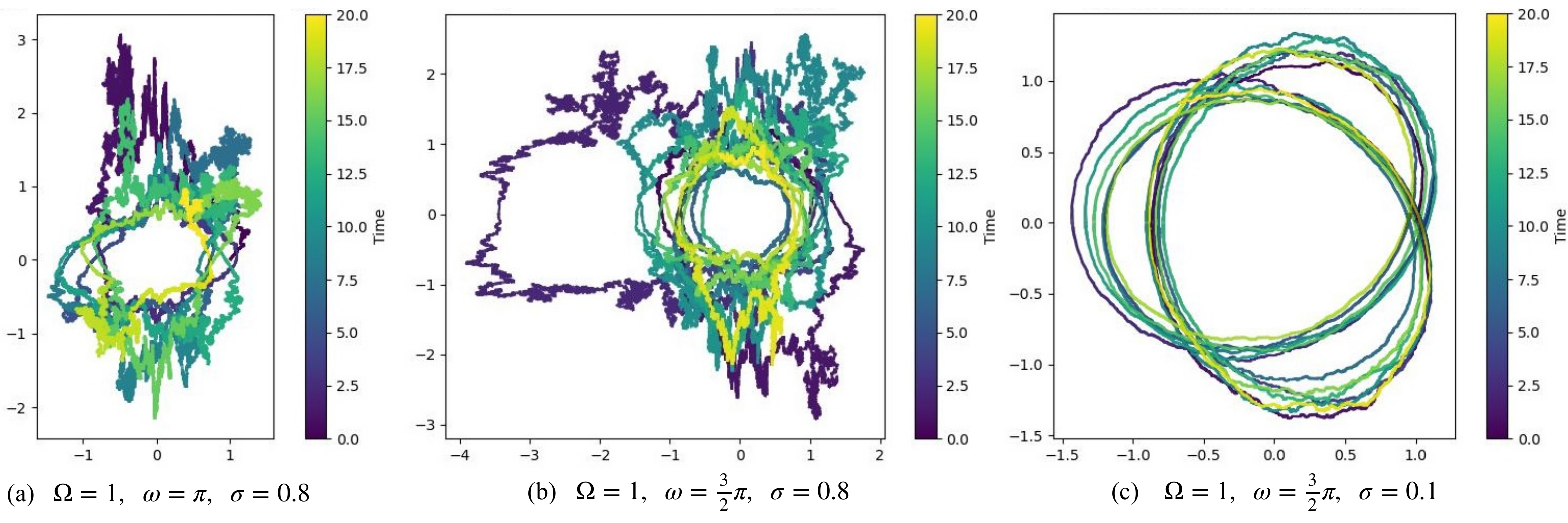}
    \caption{ \footnotesize Sample trajectories of the periodic nonlinear oscillator \eqref{eq:pno}. The time step is $\Delta t = 10^{-5}$. The time interval is $[0,20]$. The initial condition is $[1, 0]$.}
    \label{fig:periodic_traj_main}
\end{figure}

\begin{figure}[h]
    \centering    
    \includegraphics[width=0.4\linewidth]{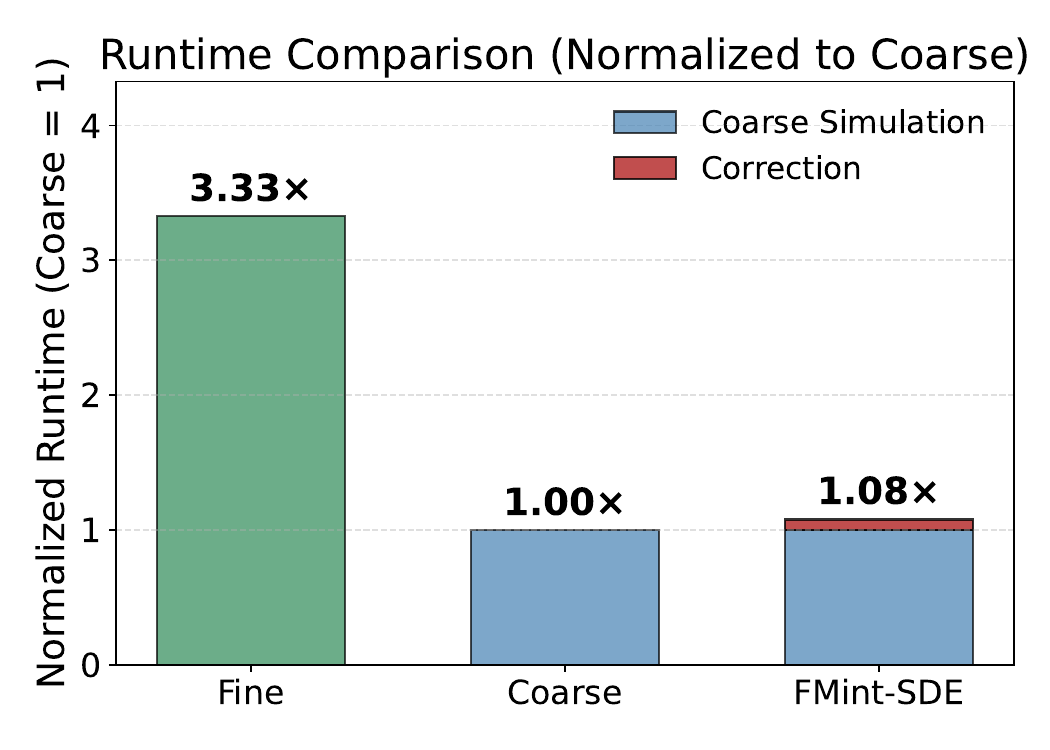}
    \caption{\footnotesize Normalized runtime for the stochastic Lorenz system.}
    \label{fig:runtime}
    \vspace{-1.5em}
\end{figure}

\subsection{Efficiency}

We further examined the runtime of FMint-SDE in comparison with the fine and coarse solution generation. 
The experiment was conducted on the stochastic Lorenz, { with 25 sets of initial conditions}
and 40 noise realizations for each initial condition. 
The fine trajectories are simulated with $\Delta t = 10^{-5}$ and coarse trajectories with $k\Delta t = 10^{-3}$.
Fig.~\ref{fig:runtime} presents the normalized runtime, where the runtime for obtaining the coarse solutions is taken as one unit. { The runtime of FMint-SDE includes running coarse simulation and error correction.}
FMint-SDE achieves the accuracy of the fine solution while maintaining nearly the same runtime as the coarse solution.

\subsection{Out-of-distribution SDEs}
Next, we test the performance of FMint-SDE on unseen SDEs: 
\begin{enumerate}
    \item {\bf 1D}: Ornstein-Uhlenbeck (OU) $dX_t = \theta (\mu - X_t) dt + \sigma dW_t$ and Inhomogeneous Ornstein-Uhlenbeck (IOU) \cite{time_inhomo_OU} $dX_t = [a\cos(\omega t) - \theta X_t] dt + \sigma dW_t$;
    \item {\bf 2D}: overdamped Langevin with the \emph{double-well} and \emph{coupled double-well} potentials, $ V(x) = (x_1^2 - 1)^2 + \alpha x_2^2$ and $ V(x) = (x_1^2 - 1)^2 + \tfrac{1}{2}x_2^2 +\alpha x_1x_2$, respectively; 
    \item {\bf 2D}: a periodically forced noisy Duffing oscillator (\emph{Duffing})~\cite{Zhang_2022,yuan2023optimalcontrolsamplingtransition}: 
    \begin{equation}
    \label{eq:Duffing}
    dX_t = V_tdt,\quad dV_t = -\delta V_t - \alpha X_t - \beta X_t^3 + \gamma \cos(\omega t)dt + \sqrt{\epsilon} dW_t;
    \end{equation}
    \item {\bf 2D}: a stochastically \emph{perturbed nonlinear oscillator} \cite{Cycles_Tori_Dankowicz}: 
    \begin{equation}
        \label{eq:Dank}
        \begin{aligned}
   dx_1 &= T\!\left(x_1 - x_2 - x_1\bigl(x_1^2 + x_2^2\bigr)\right) dt
         + \sigma \sqrt{T}\, (x_1 x_2)\, dW_t,\\          dx_2 &= T\!\left(x_1 + x_2 - x_2\bigl(x_1^2 + x_2^2\bigr)\right) dt
         + \sigma \sqrt{T}\, (x_2^2)\, dW_t;
         \end{aligned}
          \end{equation}
    \item {\bf 3D}: a predator-prey model with role reversal \cite{predator_prey_LiLiuWei} with geometric noise in all species (\emph{predator-prey}); the predator population is split into the juvenile and adult subpopulations:
    \begin{equation}
    \label{eq:predator-prey}
        \begin{aligned}
    dx & = [x(r - ax + sy_1 - by_2)]dt + \sigma_1 x dB_1(t)\\
    dy_1 & = [kxy_2 - y_1(gx + D + v_1)]dt + \sigma_2 y_1 dB_2(t)\\
    dy_2 & = (D y_1 - v_2 y_2) dt + \sigma_3 y_2 dB_3(t);
\end{aligned}
    \end{equation}
    \item {\bf 3D}: a \emph{fluxgate sensor}~\cite{PhysRevE_fluxgate_sensor} with colored noise in the nonlinear terms:
    \begin{equation}
    \label{eq:fluxgate}
\begin{cases}
    dx_1 & = -x_1 + \tanh[c(x_1 + \lambda x_2 + y_1)], \quad dy_1 = -\omega y_1 dt + \omega \sqrt{\epsilon} dW\\
    dx_2 & = -x_2 + \tanh[c(x_2 + \lambda x_3 + y_2)], \quad dy_2 = -\omega y_2 dt + \omega \sqrt{\epsilon} dW\\
    dx_3 & = -x_3 + \tanh[c(x_3 + \lambda x_1 + y_3)], \quad dy_3 = -\omega y_3 dt + \omega \sqrt{\epsilon} dW.\\
\end{cases}
\end{equation}
\end{enumerate}

\begin{figure}[h]
    \centering
    \includegraphics[width=0.85\linewidth]{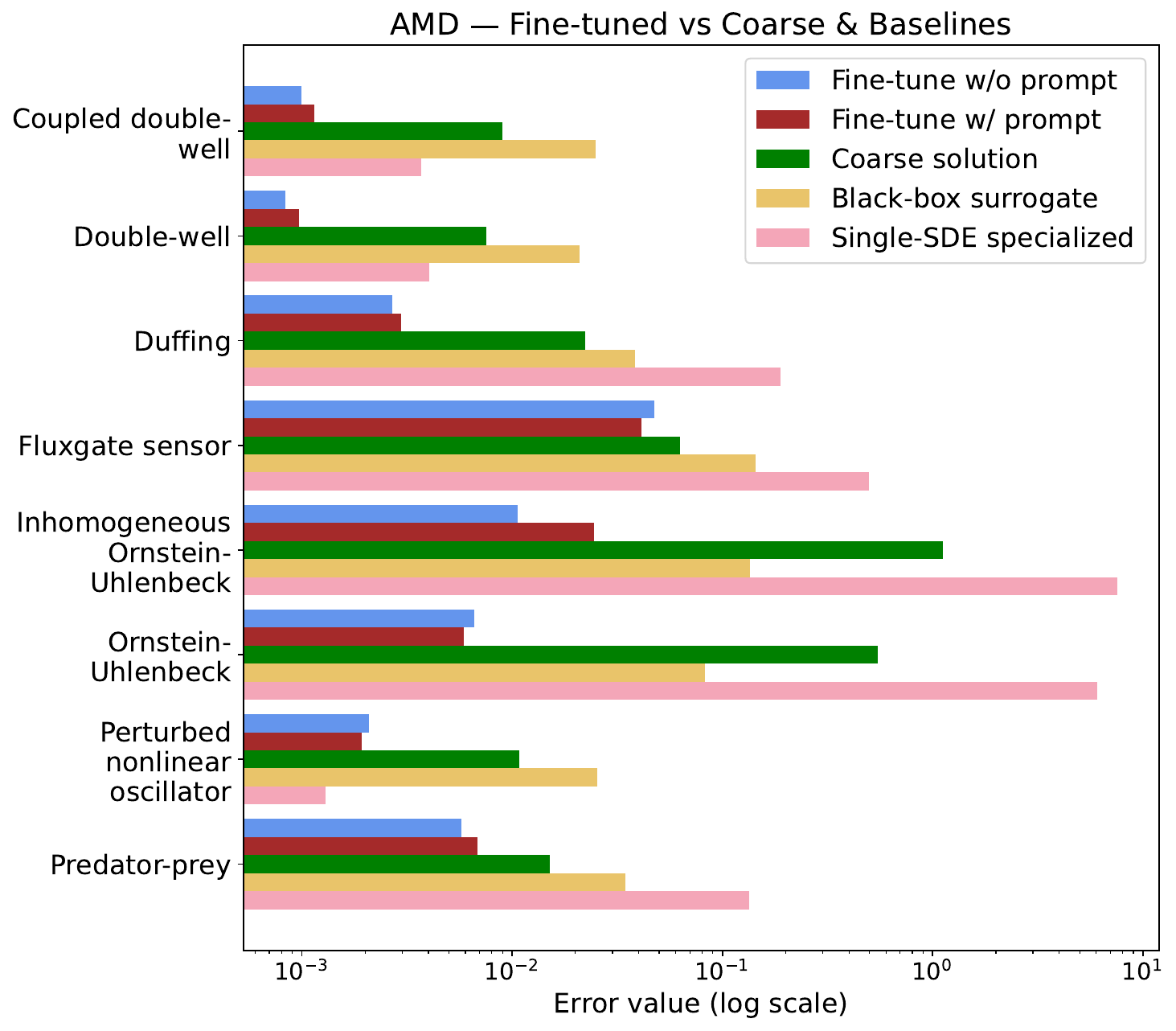}
    \caption{\footnotesize Performance of fine-tuned FMint-SDE on unseen SDE systems in log scale.}
    \label{fig:AMD-ft}
    \vspace{-1em}
\end{figure}

These test SDEs exhibit qualitatively different behaviors from the systems in the pre-training set and come from different applications. 
We fine-tuned the model for 1000 or 2000 iterations for each system on the new data of size $N$. 
We then evaluate the performance of FMint-SDE on 1000 testing trajectories.
Details on these test systems, settings, and fine-tuning epochs are provided in Section SM4.
The resulting AMD errors in the log scale are shown in Fig.~\ref{fig:AMD-ft}, and a complete report of the performance of FMint-SDE with four error metrics is given in Table SM2. 

{\bf FMint-SDE vs coarse solutions.} { Fig.~\ref{fig:AMD-ft} demonstrates the AMD error reduction by FMint-SDE without prompt for all test systems compared to the coarse solution.  FMint-SDE reduces the AMD error by one to two orders of magnitude for all 1D test systems and by factors of 3 to 10 for all other test systems, except for the fluxgate sensor~\eqref{eq:fluxgate}. The fluxgate sensor is, strictly speaking, a 6D SDE, but we treated it as a 3D SDE with colored noise instead. Fig.~\ref{fig:AMD-ft} suggests that FMint-SDE trained on SDEs with white noise yields little error reduction for SDEs with colored noise. To extend FMint-SDEs for colored noise, we should include both white- and colored-noise SDEs into the training set. We leave this investigation for future work. }

{\bf FMint-SDE vs baselines.}
{ FMint-SDE outperforms the baseline models, black-box surrogate and single-SDE specialized, across all systems, except for the perturbed nonlinear oscillator~\eqref{eq:Dank}, where FMint-SDE is slightly worse than the single-SDE specialized model.}

{\bf Textual prompts.} The effect of textual prompts is marginal compared to the much larger error reduction from finetuning. 
Nonetheless, we observe that fine-tuning with prompts improves error metrics for more complex systems, such as fluxgate sensor, OLD-Mueller, and stochastic Lorenz. However, for simpler cases, such as IOU and coupled double-well, it introduces a small degradation. 

{ We believe,  we have not utilized the full potential of the textual prompt.
Our current scheme is not designed to explicitly establish a connection between numerical demos and textual information; rather, the textual component only serves as auxiliary input. Our further analysis on the role of textual prompts  is reported in Section~\ref{sec:impact_text}. We anticipate that a more refined design could make textual prompts consistently beneficial. 
We leave the exploration of this direction for future work.}

 \begin{figure}[h]
    \centering
    \includegraphics[width=0.75\linewidth]{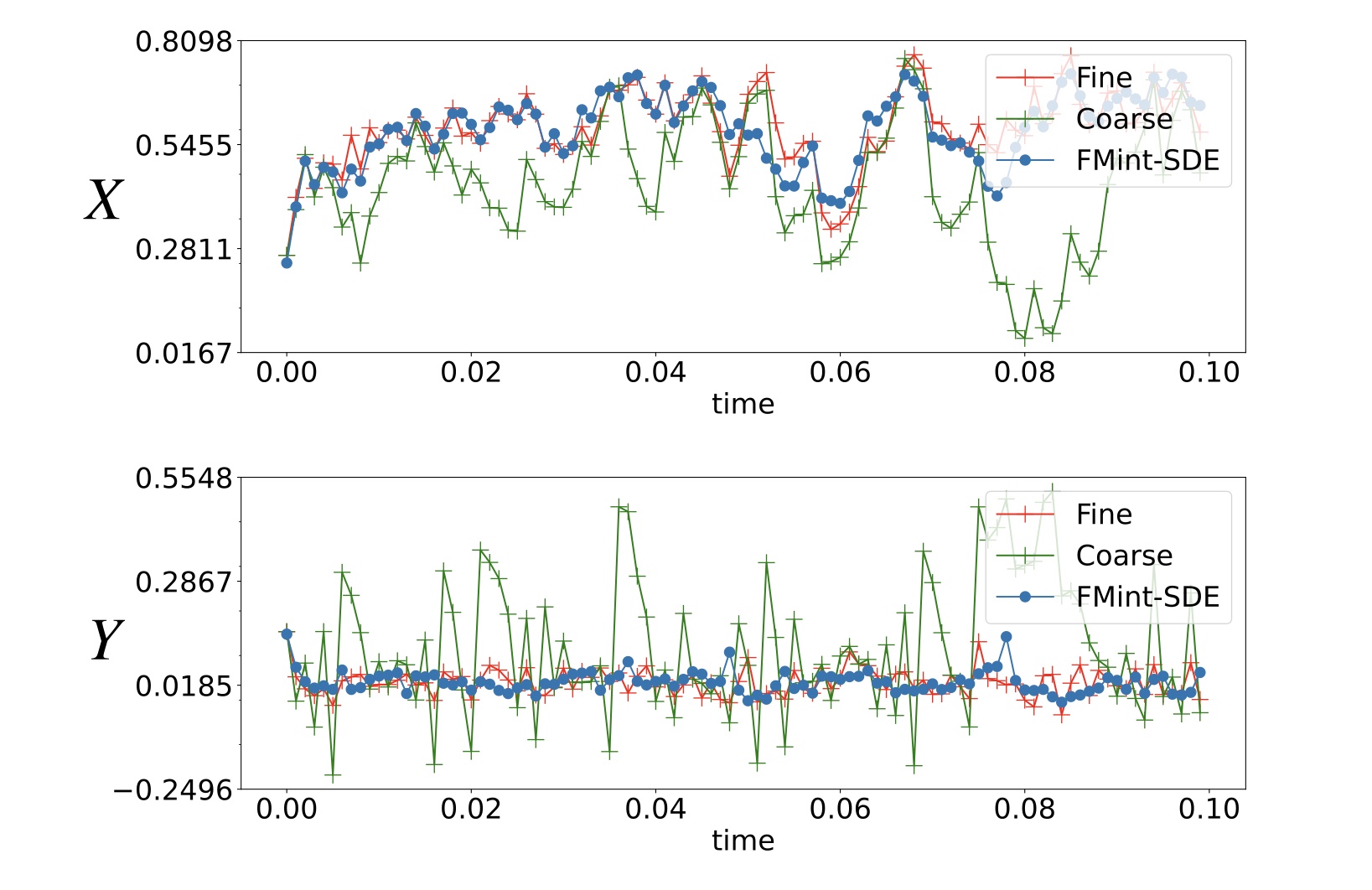}
    \caption{\footnotesize Roll-out scheme in Section \ref{sec:rollout} applied to overdamped Langevin with Mueller's potential for time $t \in [0,0.1]$.}
    \label{fig:rollout}
\end{figure}

\subsection{Performance on the roll-out}
Fig.~\ref{fig:rollout} displays the results of the roll-out scheme applied to the OLD-Mueller. Fig.~\ref{fig:Mueller_visual} shows the FMint-SDE correction for $t \in [0, 0.05]$. Applying the roll-out scheme described in Section~\ref{sec:rollout}, we obtained the FMint-SDE correction for $t \in [0, 0.1]$.
The trajectory of the coarse solution for $t \in [0.05, 0.1]$ deviates from the fine solution more than it does for $t \in [0, 0.05]$. By contrast, FMint-SDE under the roll-out scheme maintains its performance.

\begin{figure}[h]
    \centering
    \includegraphics[width=1.0\linewidth]{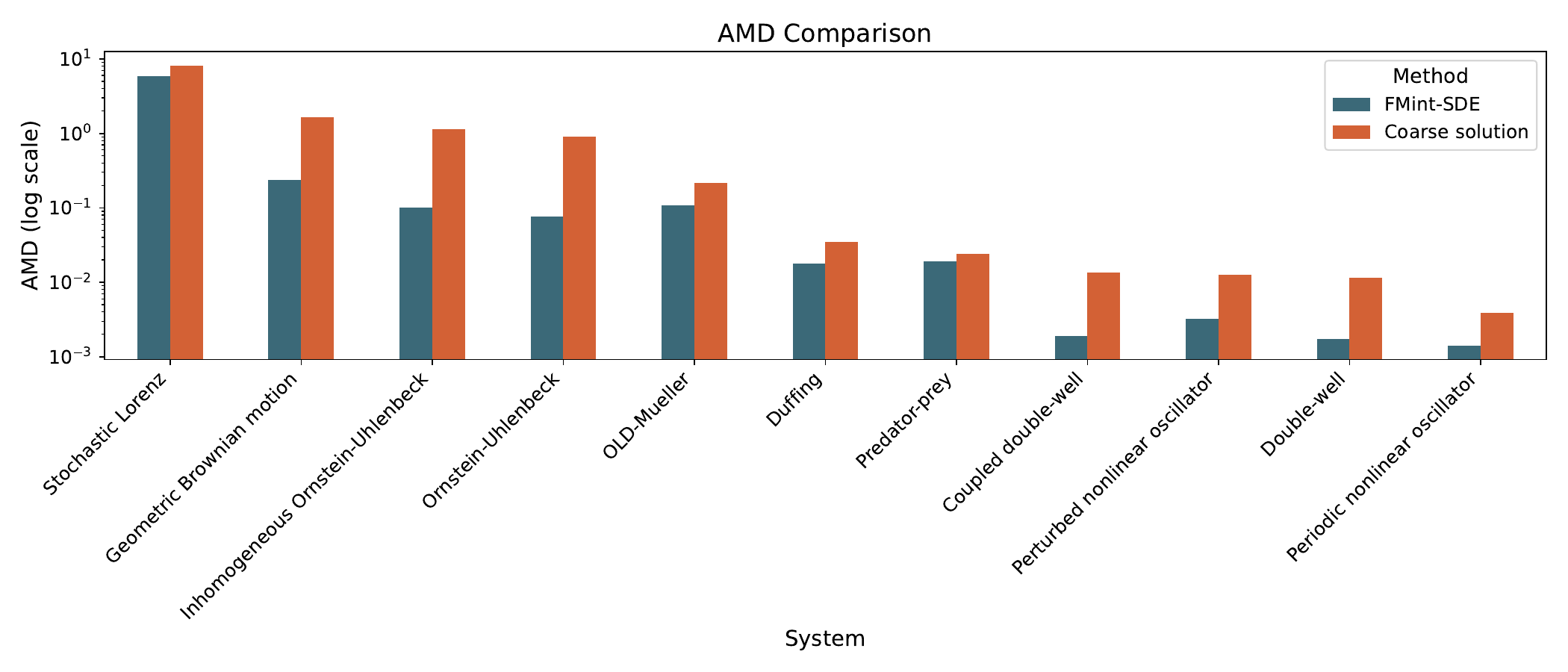}
    \caption{\footnotesize AMD error for all systems under the roll-out scheme for a total of 500 time steps correction.}
    \label{fig:rollout_AMD}
\end{figure}

{ Fig.~\ref{fig:rollout_AMD} reports AMD error for all test systems under the roll-out scheme. All trajectories are simulated for 500 time steps, which requires 10 applications of the roll-out scheme. 
FMint-SDE achieves significant improvement over the coarse solver across most systems, often reducing errors by multiple folds and, in several cases, by nearly an order of magnitude.
Unsurprisingly, the improvement for the stochastic Lorenz and predator-prey systems is marginal due to the chaotic nature of its drift field, resulting in the divergence of close trajectories for the corresponding ODE. 
}

\subsection{Error dependence on the fine-tuning training data size}
\begin{figure}[h]
    \centering
    \includegraphics[width=\linewidth]{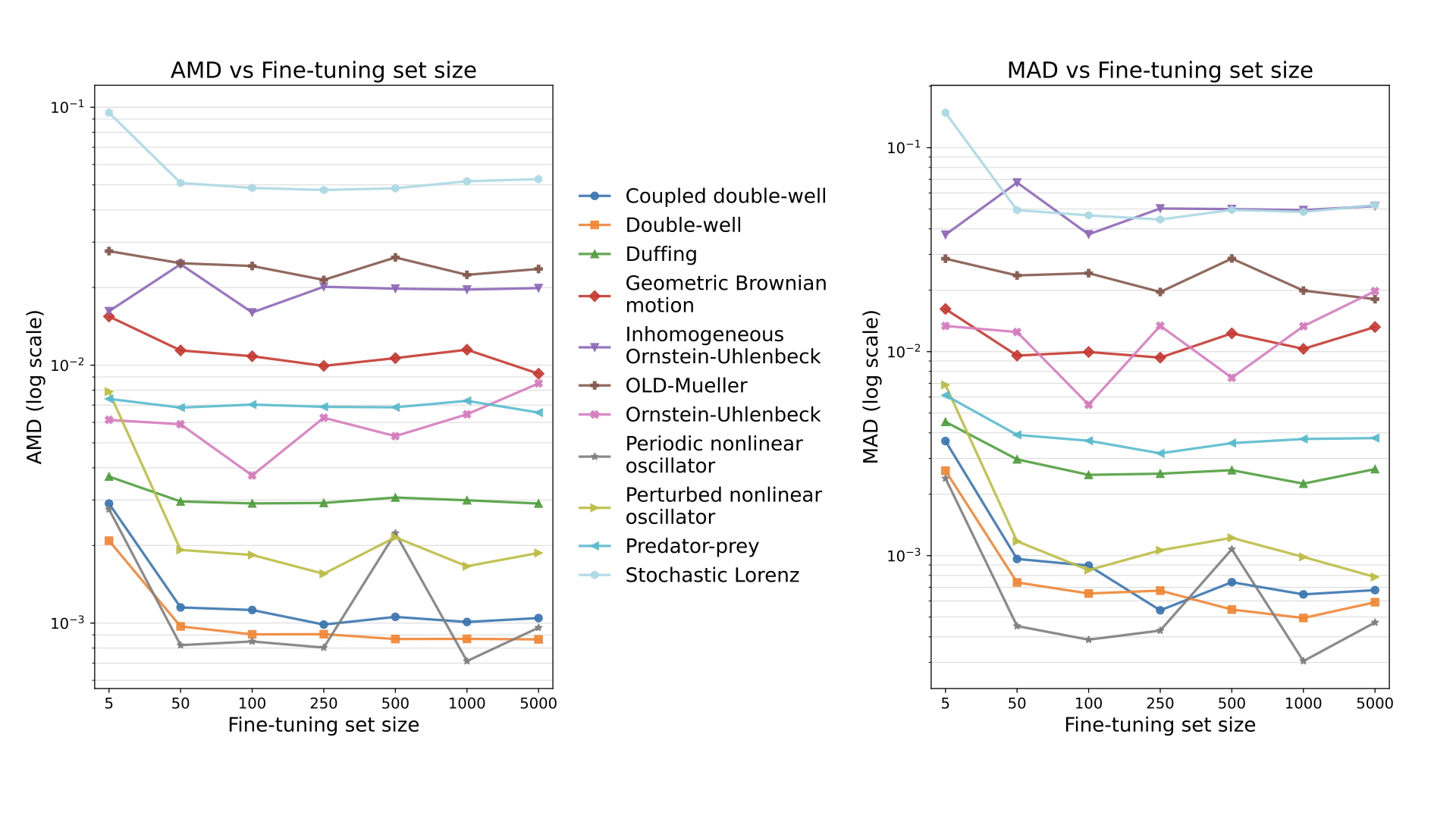}
    \caption{\footnotesize AMD (left) and MAD (right) v.s. fine-tuning set size for all systems.}
    \label{fig:error_vs_samples}
    \vspace{-2em}
\end{figure}

Here, we examine the effect of the size of the fine-tuning training data on FMint-SDE's performance. 
We test the fine-tuning training sets consisting of $N \in \{5, 50, 100, 250, 500, 1000, 5000\}$ trajectories. 
We fine-tune the model for 1000 or 2000 iterations, depending on the system, while keeping the number of fine-tuning iterations constant for each system, regardless of $N$.
Fig.~\ref{fig:error_vs_samples} shows the  AMD~\eqref{eq:AMD} and MAD~\eqref{eq:MAD} errors in the log scale with respect to the fine-tuning size $N$. Figure SM20  displays error plots of MAE and RMSE versus fine-tuning set size $N$. 

The increase of the fine-tuning size from $N = 5$ to $N = 50$ significantly reduces error across most systems, { whereas further increases lead to error saturation or oscillations.}  
{ For the double-well, the coupled double-well, duffing oscillators, geometric Brownian motion, predator–prey, stochastic Lorenz, and the OLD-Mueller, no sign of overfitting is observed.
For OU and IOU, the performance clearly degrades at larger $N$, suggesting potential overfitting. Overall, these results suggest that $N = 50$ trajectories are sufficient for fine-tuning.}
Tables SM3-SM6 contain a complete report, of errors versus fine-tuning set sizes.

\subsection{Performance on SDEs with different behaviors under different parameter coefficients}

Different parameter ranges for the same dynamical system may lead to qualitatively different behaviors due to bifurcations or stochastic resonance.
Simulations with large time steps may result in significant deviations from the true trajectories and may overlook critical system behaviors. 
{ To evaluate robustness, we test FMint-SDE on systems whose dynamics vary significantly under different parameter settings. These variations alter stability and trajectory behavior, allowing us to assess performance consistency. We report results on parameterized versions of the Duffing oscillator \eqref{eq:Duffing},  the stochastic Lorenz system, and the predator–prey system \eqref{eq:predator-prey}.}

\subsubsection*{Duffing oscillator}
We tested FMint-SDE on trajectories of periodically forced noisy Duffing oscillator \eqref{eq:Duffing} \cite{Zhang_2022} 
with three different behavior types: 
chaotic, overdamped with noise-induced transitions between two wells,
 and stochastic resonance with stable oscillations.
 { Details are included in Section SM7.}

\subsubsection*{Stochastic Lorenz}
For the stochastic Lorenz, we examine FMint-SDE's performance on trajectories simulated with $\sigma = 10$, $\beta = \frac{8}{3}$, and $\rho \in \{0.5, ~1, ~13.926, ~20, ~24.06,$
$ ~24.5, ~24.76, ~100\}$. The noise coefficients are set to $\eta_1 = \eta_2 = \eta_3 = 0.8$. In addition, we include one more case with $\rho = 28$ and $\eta_1 = \eta_2 = \eta_3 = 2.0$.
As $\rho$ increases from 0 to 100, the Lorenz ODE system experiences a series of bifurcations and exhibits various dynamical behaviors. This system attracted attention of many researchers. The topological structure of the Lorenz attractor was studied in~\cite{lorenz_Rand1978,Guckenheimer1979}. The preturbulence regime was identified and described in~\cite{Kaplan1979,Yorke1979MetastableChaos}.
{ See Section SM7 for more details.}

\subsubsection*{Predator-Prey}

Ref.~\cite{predator_prey_LiLiuWei} introduced and examined the predator-prey system with role reversal and added geometric noise ~\eqref{eq:predator-prey}. 
The variables $x$, $y_1$, and $y_2$ describe the prey, juvenile predator, and adult predator population sizes, respectively. We borrow parameter settings from~\cite{predator_prey_LiLiuWei} that correspond to biologically meaningful types of behavior.
Additionally, we consider a modification of Eq. \eqref{eq:predator-prey} in which stochasticity is introduced into the predator reproduction rate, i.e, the parameter $k$ is replaced with $k + \sigma_4dB_3$:
\begin{equation}
\label{eq:variant_predator_prey}
\begin{aligned}
    dx & = [x(r - ax + sy_1 - by_2)]dt + \sigma_1 x dB_1(t),\\
    dy_1 & = [kxy_2 - y_1(gx + D + v_1)]dt + \sigma_2 y_1 dB_2(t) + \sigma_4xy_2dB_3,\\
    dy_2 & = (D y_1 - v_2 y_2) dt + \sigma_3 y_2 dB_4(t),
\end{aligned}
\end{equation}
where $dB_j$, $j=1,2,3,4$, are increments of independent Brownian motions, and  $\sigma_4 = 0.005$.
Trajectories of \eqref{eq:variant_predator_prey} look more chaotic than those of \eqref{eq:predator-prey} with the same parameter values.
{ Details are provided in Section SM7.}

\begin{figure}[h]
    \centering
    \includegraphics[width=\linewidth]{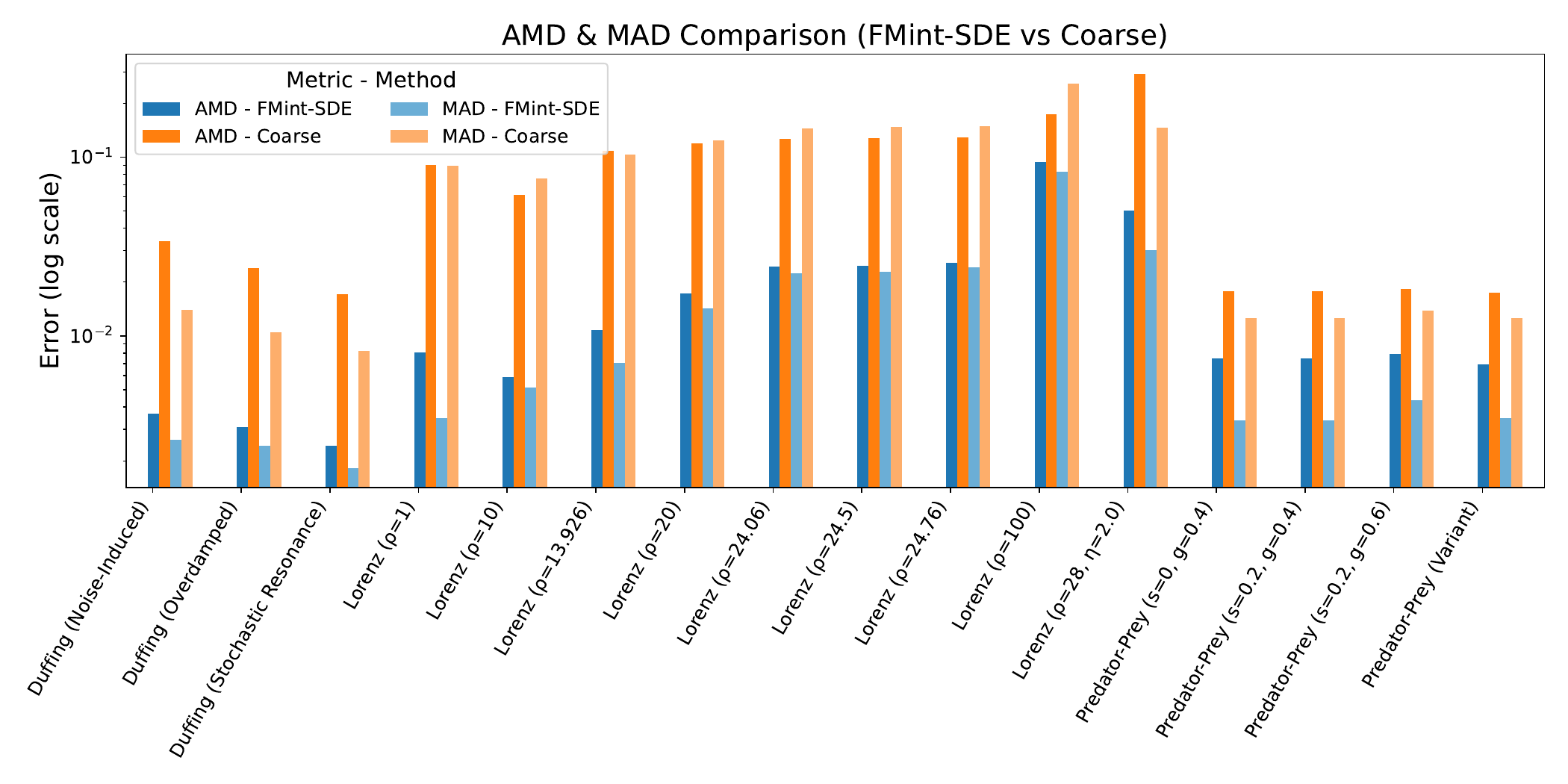}
    \caption{\footnotesize Performance of fine-tuned FMint-SDE on unseen SDE systems exhibiting various behaviors in log scale.}
    \label{fig:AMD-MAD-diff-coeff}
\end{figure}

Fig.~\ref{fig:AMD-MAD-diff-coeff} compares the AMD and MAD errors of fine-tuned FMint-SDE and the coarse integrator across various behavior types of the described above Duffing, predator-prey, and Lorenz systems. A complete report on all four evaluation metrics is provided in Table SM8.
On the Duffing oscillator, FMint-SDE achieves minimal residual errors with an improvement of accuracy of approximately one order of magnitude. 
For the predator-prey model, FMint-SDE achieves roughly half to one-third the error of the coarse solution across all metrics.

On the other hand, the coarse solution of the Lorenz system exhibits an apparent monotonic increase in error as $\rho$ increases from $1$ to $100$, or with a higher noise level.
Across all parameter settings, FMint-SDE consistently reduces errors by a factor of four to eight relative to the coarse solution across MAE, RMSE, AMD, and MAD.
FMint-SDE shows a significant improvement in the chaotic regime. 
Specifically, at $\rho=24.5$, $24.76$, $\rho = 28$ with  $\eta=2.0)$, and $\rho = 100$, FMint-SDE improves the relative MAE by 80 to 85 percent.
This indicates that FMint-SDE is robust to increasing system complexity and highlights its ability to improve the accuracy even in highly sensitive cases.

\subsection{ Effect of textual prompt}
\label{sec:impact_text}

{
To further analyze the role of multi-modality, we conduct an ablation study to assess the effect of textual prompts and to examine how textual information can be incorporated into scientific computing.

Specifically, we generate a diverse dataset comprising all systems with 250 different parameter settings, 10 different initial conditions, and 40 noise realizations. Two FMint-SDE models, one with and one without textual information, are pre-trained on this dataset for 50 epochs using the same architecture and training configuration. To simplify training, we use only the system name as the caption. The models are then evaluated on the test sets without any further fine-tuning.

Table~\ref{tab:amd_caption_vs_nocap_by_demos} reports the AMD error for different numbers of demos $K$ ranging from 0 to 4, with and without captions. We observe that textual prompts are most helpful in the low-data regime, especially at $K = 0$. The AMD error decreases sharply as $ K$ increases from $ 0$ to $ 1$ across nearly all systems. For $K \ge 1$, the error generally stabilizes for both models with and without captions. At best, we observe only marginal improvements as $K$ increases from 1 to 4. In many systems, the lowest errors are already achieved at $K=1$ or $K=2$, indicating diminishing returns from additional demonstrations. 
We conclude that the model consistently benefits from textual prompts in the absence of demos; however, the gain from a single demo exceeds that of a textual prompt. 
}

\begin{table}[h!]
\centering
\scriptsize
\setlength{\tabcolsep}{4pt}
\renewcommand{\arraystretch}{1.1}
\begin{tabular}{p{2.5cm}|cc|cc|cc|cc|cc}
\hline
\textbf{System} 
& \multicolumn{2}{c|}{\textbf{K=0 AMD}} 
& \multicolumn{2}{c|}{\textbf{K=1 AMD}} 
& \multicolumn{2}{c|}{\textbf{K=2 AMD}} 
& \multicolumn{2}{c|}{\textbf{K=3 AMD}} 
& \multicolumn{2}{c}{\textbf{K=4 AMD}} \\
\cline{2-11}
& \textbf{Coarse} & \textbf{w/} 
& \textbf{w/o} & \textbf{w/}
& \textbf{w/o} & \textbf{w/}
& \textbf{w/o} & \textbf{w/}
& \textbf{w/o} & \textbf{w/} \\
\hline

Coupled double-well 
& 8.9e-3 & 8.1e-3 
& 4.0e-3 & 5.0e-3 
& 4.0e-3 & 5.3e-3 
& 4.0e-3 & 6.5e-3 
& 4.3e-3 & 7.2e-3 \\

Double-well 
& 7.4e-3 & 8.3e-3 
& 3.8e-3 & 6.8e-3 
& 3.9e-3 & 8.0e-3 
& 4.2e-3 & 9.3e-3 
& 5.0e-3 & 1.0e-2 \\

Duffing 
& 2.2e-2 & 9.8e-3 
& 8.3e-3 & 8.2e-3 
& 7.9e-3 & 8.6e-3 
& 7.5e-3 & 8.6e-3 
& 7.6e-3 & 8.3e-3 \\

Fluxgate sensor 
& 5.3e-2 & 4.3e-2 
& 5.1e-2 & 4.1e-2 
& 5.1e-2 & 4.2e-2 
& 5.1e-2 & 4.2e-2 
& 5.1e-2 & 4.3e-2 \\

Geometric Brownian Motion 
& 5.1e-1 & 3.2e-1 
& 2.3e-2 & 1.5e-2 
& 1.8e-2 & 1.3e-2 
& 1.9e-2 & 1.2e-2 
& 2.1e-2 & 1.1e-2 \\

Inhomogeneous Ornstein-Uhlenbeck 
& 1.1  & 8.5e-2 
& 2.7e-2 & 2.9e-2 
& 2.8e-2 & 2.8e-2 
& 3.1e-2 & 2.8e-2 
& 3.4e-2 & 3.1e-2 \\

OLD-Mueller 
& 7.2e-2 & 3.0e-2 
& 3.0e-2 & 2.4e-2 
& 2.9e-2 & 2.4e-2 
& 3.0e-2 & 2.5e-2 
& 3.0e-2 & 2.4e-2 \\

Ornstein-Uhlenbeck 
& 5.4e-1 & 9.9e-2 
& 8.2e-3 & 9.7e-3 
& 7.8e-3 & 9.8e-3 
& 8.6e-3 & 9.0e-3 
& 8.9e-3 & 9.1e-3 \\

Periodic nonlinear oscillator 
& 3.1e-3 & 5.8e-3 
& 2.7e-3 & 3.0e-3 
& 2.6e-3 & 3.0e-3 
& 2.7e-3 & 3.1e-3 
& 2.8e-3 & 3.3e-3 \\

Perturbed nonlinear oscillator 
& 8.9e-3 & 9.8e-3 
& 6.8e-3 & 7.3e-3 
& 6.7e-3 & 7.7e-3 
& 7.0e-3 & 8.2e-3 
& 7.2e-3 & 8.5e-3 \\

Predator-prey 
& 1.7e-2 & 1.2e-2 
& 1.3e-2 & 1.3e-2 
& 1.3e-2 & 1.4e-2 
& 1.3e-2 & 1.4e-2 
& 1.2e-2 & 1.3e-2 \\

Stochastic Lorenz 
& 1.9e-1 & 9.8e-2 
& 6.4e-2 & 6.4e-2 
& 5.8e-2 & 6.2e-2 
& 6.1e-2 & 6.3e-2 
& 6.4e-2 & 6.5e-2 \\

\hline
\end{tabular}
\caption{\footnotesize Comparison of the strong error (AMD) obtained with and without a textual prompt for demonstration counts 
$K$ from 0 to 4. Reported values are rounded to one decimal place.}
\vspace{-2em}
\label{tab:amd_caption_vs_nocap_by_demos}
\end{table}


\section{Conclusion}
\label{sec:conclusion}
In this work, we proposed FMint-SDE, a multi-modal foundation model using a decoder-only transformer architecture that speeds up large-scale simulations of
stochastic differential equations.
It takes supplemental textual prompts and numerical demos of coarse solutions with large time steps, noise realizations, and corresponding errors. 
Using these inputs, FMint-SDE reproduces fine solutions with high accuracy and efficiency through error correction.
The model is trained using an in-context learning framework and the language-like next-item prediction learning scheme.

FMint-SDE is pre-trained on SDEs from four diverse families of dynamical systems and further fine-tuned on eight additional SDE families as downstream tasks.
Extensive experiments show that for the pre-training systems, zero-shot inference, without any additional fine-tuning, achieves high accuracy on new trajectories with parameter settings and initial conditions different from those used during training.
{Its performance consistently outperforms two heuristically designed baseline models across both accuracy and efficiency metrics.}
Runtime analyses further demonstrate that FMint-SDE attains the accuracy of fine solutions while maintaining the runtime efficiency of coarse solvers. 
For downstream tasks, slight fine-tuning with a small amount of data yields high accuracy, highlighting the strong transfer learning capability of FMint-SDE.
Although the model accepts inputs of fixed length, we show that FMint-SDE can be readily extended to long simulations through the proposed roll-out scheme.

Further improvements to the proposed methodology can proceed along two directions. First, the current model is confined to three-dimensional state spaces. Despite this limitation, FMint-SDE remains useful for oscillatory dynamics, whose intrinsic dimensionality is typically modest. In principle, FMint-SDE can scale to higher-dimensional settings, which would be especially valuable for molecular dynamics applications, where the state dimensionality can easily reach the hundreds. 
Second, although our model has roughly 16 million parameters, it is still small for a foundation model. At this scale, we observe only a slight benefit from textual prompting, likely due to limited model capacity and the scope of the training families. 
We expect the advantages of textual information to become more pronounced as the model size increases.
We leave these research topics for future work.

\section*{Acknowledgments}
MC and JY were partially supported by the Office of Naval Research Grant No. N000142412547  and the Air Force Office of Scientific Research MURI grant FA9550-20-1-0397. HY was partially supported by the US National Science Foundation under awards IIS-2520978, GEO/RISE-5239902, the Office of Naval Research Award N00014-23-1-2007, DOE (ASCR) Award DE-SC0026052, and the DARPA D24AP00325-00.

\section*{Code availability}
Our codes are published on GitHub \cite{yuan2025fmintSDE}. 



\section*{Supplementary Information}

  \setcounter{table}{0}
  \renewcommand{\thetable}{S\arabic{table}}%
  \setcounter{figure}{0}
  \renewcommand{\thefigure}{S\arabic{figure}}%
\renewcommand{\thepage}{S\arabic{page}}
 \setcounter{page}{1}
 
\renewcommand{\theequation}{S\arabic{equation}}
  \setcounter{equation}{0}
\renewcommand{\thesection}{S\arabic{section}}
\setcounter{section}{0}

\section{Input data structure}
Table~\ref{tab:token} shows the input data structure for a 2D SDE demo,  including one demo and one query. Each column is treated as a single token, comprising a time stamp, noise values, and coarse solution or error-term values. 
The first two columns form a demo pair and are available to the model when no textual prompts are provided. Although query error terms are included in the input during training, masked self-attention prevents the model from accessing their values. During inference, query error terms are neither available nor provided to the model.
 \begin{landscape}
\input{Tables/input_demo}
\end{landscape}

\section{ Baseline models implementation details}

{
Here, we provide the implementation details for the heuristically designed baseline models.
For the single-SDE specialized models, we use a standard architecture for tasks of comparable scale. 
This architecture is significantly simpler than the one in FMint-SDE. For the black-box surrogate, the architecture is the same as for FMint-SDE.

\textbf{Single-SDE specialized model architecture.}
The model incorporates a multi-layer perceptron (MLP) with a custom activation function, defined as follows:
\begin{itemize}
    \item \textbf{Input layer:} The state dimension of the system.
    \item \textbf{Hidden layer:} 1024 neurons with a custom rational activation function.
    \item \textbf{Output layer:} The error correction term for the state update.
\end{itemize}

The rational activation function is defined by:
\[
f(x) = \frac{a_3x^3 + a_2x^2 + a_1x + a_0}{b_2x^2 + b_1x + b_0},
\]
where the parameters are learnable and initialized as: \( a_0 = 0.0218, a_1 = 0.5000, a_2 = 1.5957, a_3 = 1.1915, b_0  = 1.0000, b_1 = 0.0000, \) and \( b_2 = 2.3830 \).

\textbf{Training Details.}
We trained each specialized model with the following training details:

\begin{itemize}
    \item Learning Rate: 0.001
    \item Optimizer: Adam
    \item Learning Rate Decay: cosine annealing schedule
    \item Batch Size: 256
    \item Number of Epochs: 500
    \item Loss Function: Mean Squared Error (MSE)
\end{itemize}
}

\section{Pretraining SDE systems}
\label{App:systems_pretrain}

\subsubsection*{Geometric Brownian motion (1D)} The system is governed by SDE 
\begin{equation}
\label{eq:GMB}
    dX_t = \mu X_t \, dt + \sigma X_t \, dW_t
\end{equation}
where $\mu \in \mathbb{R}$ is the drift term, usually referring to growth rate or trend; $\sigma > 0$ is the diffusion coefficient, and often refers to the volatility (random fluctuation intensity).
For pre-training, we used $\mu \sim \text{Unif}~[0.01, 0.15], ~\sigma \sim \text{Unif}~[0.01, 0.2]$. 
We simulate trajectories with $~\Delta t = 0.0005, ~k = 100$ and initial conditions $X_0 \sim \text{Unif}~[50, 100]$.
An example trajectory is included in Figure \ref{fig:traj_GMB_mueller} (left). 
We set the number of iterations to 1000 to fine-tune the model on this system.

\subsubsection*{Overdamped Langevin dynamics with Mueller's potential (2D)}
\begin{figure}[h]
    \centering
    \includegraphics[width=\linewidth]{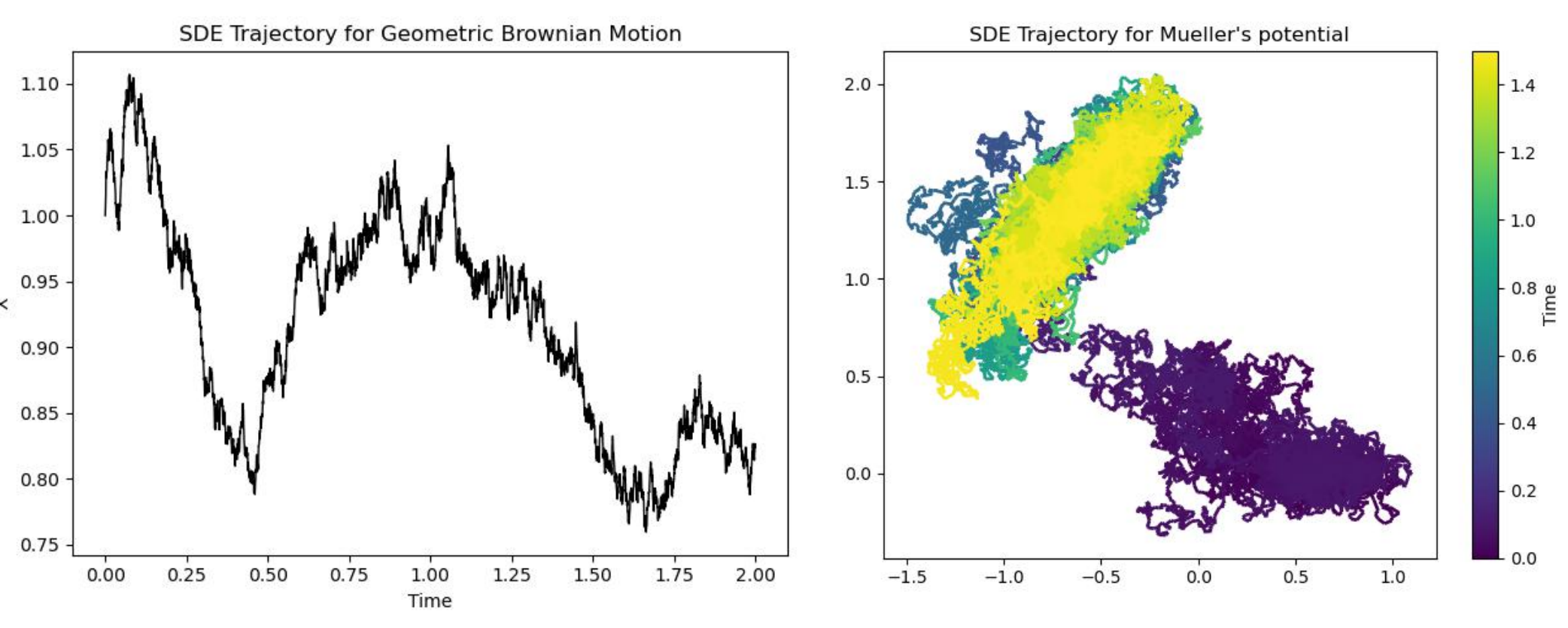}
     \caption{(left) Geometric Brownian motion~\eqref{eq:GMB} with $\theta = 0.25, ~\mu = 1, ~\sigma = 0.3$ (right) overdamped Langevin equation~\eqref{eq:overdamped} with Mueller's potential~\eqref{mueller} with $\beta = 0.05$.}
    \label{fig:traj_GMB_mueller}
\end{figure}

Overdamped Langevin dynamics is often used to describe molecular dynamics with a governing SDE
\begin{equation}
\label{eq:overdamped}
    dX_t = -\nabla V(X_t) dt + \sqrt{2\beta^{-1}} dW_t,
\end{equation}
where $V(x)$ is a smooth and coercive potential with a finite number of isolated minima and $\beta^{-1}$ 
is a small parameter often interpreted in chemical physics applications as the inverse temperature.  
Mueller’s potential \cite{LiLinRen2019} describes the system with a potential energy function
\begin{equation}
\label{mueller}
    V(x_1, x_2) = \sum\limits_{i=1}^4 D_i \exp\left\{a_i(x_1 - X_i)^2 + b_i(x_1 - X_i)(x_2 - Y_i) + c_i(x_2 - Y_i)^2 \right\}
\end{equation}
where
\begin{align*}
    [a_1, a_2, a_3, a_4] &= [-1, -1, -6.5, 0.7] \\
    [b_1, b_2, b_3, b_4] &= [0, 0, 11, 0.6] \\
    [c_1, c_2, c_3, c_4] &= [-10, -10, -6.5, 0.7] \\
    [D_1, D_2, D_3, D_4] &= [-200, -100, -170, 15] \\
    [X_1, X_2, X_3, X_4] &= [1, 0, -0.5, -1] \\
    [Y_1, Y_2, Y_3, Y_4] &= [0, 0.5, 1.5, 1].
\end{align*}
Since the only variation for this system is the temperature, we added four scaling hyperparameters and two shifting hyperparameters such that the potential energy for the simulation used is 
\begin{equation}
    V(x_1, x_2) = \sum\limits_{i=1}^4 \tilde{D}_i \exp\left\{\tilde{a}_i(x_1 - \tilde{X}_i)^2 + \tilde{b}_i(x_1 - \tilde{X}_i)(x_2 - \tilde{Y}_i) + \tilde{c}_i(x_2 - \tilde{Y}_i)^2 \right\}
\end{equation}
with
\begin{equation*}
\begin{aligned}
   &\tilde{a}_i = a^{\text{scale}} a_i, ~\tilde{b}_i = b^{\text{scale}} b_i, ~ \tilde{c}_i = c^{\text{scale}} c_i, \\
   & \tilde{D}_i = D^{\text{scale}} D_i, ~\tilde{X}_i = X^{\text{shift}} + X_i, ~\tilde{Y}_i = Y^{\text{shift}} + Y_i
   \end{aligned}
\end{equation*}
In this work, we used 
\begin{align*}
&a^{\text{scale}} \sim \text{Unif}~[0.8,1.2],\quad b^{\text{scale}} \sim \text{Unif}~[0.8,1.2]\\
&c^{\text{scale}} \sim \text{Unif}~[0.8,1.2],\quad D^{\text{scale}} \sim \text{Unif}~[0.7,1.3],\\
&X^{\text{shift}} \sim \text{Unif}~[-0.1,0.1],\quad Y^{\text{shift}} \sim \text{Unif}~[-0.1,0.1].\\
&\beta \sim \text{Unif}~[0.05, 2].
\end{align*}
We simulate trajectories with $~\Delta t = 10^{-5}, ~k = 100$ and initial conditions $X_0 \sim \text{Unif}~ [-0.5,0.5],X_1 \sim \text{Unif}~[-0.5,1.5]$
An example trajectory is included in Figure \ref{fig:traj_GMB_mueller} (right).
We set the number of iterations to 1000 to fine-tune the model on this system. 

\subsubsection*{Stochastically forced nonlinear oscillator with a time-periodic modulation in the drift term and state-dependent (multiplicative) noise (2D, periodic nonlinear oscillator)}
This oscillator system was first introduced in \cite{Cycles_Tori_Dankowicz}. The 2D system is governed by non-autonomous SDEs
\begin{equation}
\label{eq:periodic_osci}
\begin{aligned}
    dx_1 &= \frac{2\pi}{\omega}\!\left( -\Omega x_2 + x_1 \!\left( 1 + \sqrt{x_1^2 + x_2^2}\, (\cos 2\pi t - 1)\right) \right) dt 
            + \sigma \sqrt{\frac{2\pi}{\omega}}\, x_1 x_2\, dW_t, \\
    dx_2 &= \frac{2\pi}{\omega}\!\left( \Omega x_1 + x_2 \!\left( 1 + \sqrt{x_1^2 + x_2^2}\, (\cos 2\pi t - 1)\right) \right) dt 
            + \sigma \sqrt{\frac{2\pi}{\omega}}\, x_2^2\, dW_t.
\end{aligned}
\end{equation}
The hyper-parameters used in this work are $~\omega \sim \text{Unif}~[\pi, 2\pi], ~\Omega \sim \text{Unif}~[0.1, \pi], ~\sigma \sim \text{Unif}~[0.01, 1]$. 
We simulate trajectories with $~\Delta t =10^{-5}, ~k = 10$ and initial condition $X_0, X_1 \sim \text{Unif}~[-1,1]$.
We set the number of iterations to 1000 to fine-tune the model on this system. 

\subsubsection*{Stochastic Lorenz (3D)} 
\begin{figure}[h]
    \centering
    \includegraphics[width=0.45\linewidth]{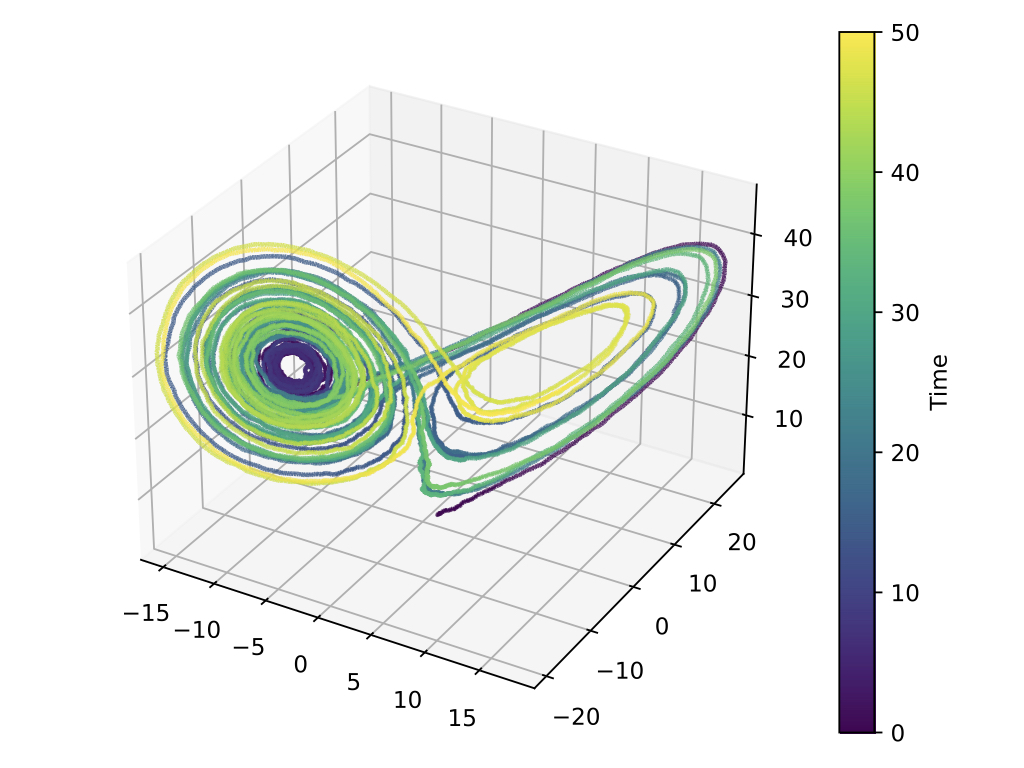}
    \caption{(left) Stochastically forced nonlinear oscillator with a time-periodic modulation in the drift term and state-dependent (multiplicative) noise ~\eqref{eq:periodic_osci} with $\Omega = 1.0, ~\omega = \pi, ~\sigma = 0.8$ (right) Stochastic Lorenz~\eqref{eq:stochastic_lorenz} with $\rho = 28,~\beta = \frac{8}{3}, ~\sigma = 10.$.}
    \label{fig:traj_periodic_lorenz}
\end{figure}

The system is governed by SDE \cite{lorenz_Rand1978, lorenz_Sparrow1982, lorenz_Cameron_2019}
\begin{equation}
\label{eq:stochastic_lorenz}
\begin{aligned}
    dX_t &= \sigma (Y_t - X_t) dt + \eta_1 dW_{1t}, \\
    dY_t &= (X_t (\rho - Z_t) - Y_t) dt + \eta_2 dW_{2t}, \\
    dZ_t &= (X_t Y_t - \beta Z_t) dt + \eta_3 dW_{3t},
\end{aligned}
\end{equation}
For pretraining, we consider $\sigma \sim \text{Unif}~[5,15], ~\rho \sim \text{Unif}~[20,40], ~\beta \sim \text{Unif}~[1,3], ~\eta_i \sim \text{Unif}~[0.1, 2], ~i = 1,2,3. $
We simulate trajectories with $ ~\Delta t = 10^{-4}, ~k = 100$ and initial condition $X_0, X_1 X_2 \sim \text{Unif}~[-1,1]$.
An example trajectory is included in Figure \ref{fig:traj_periodic_lorenz}.
We set the number of iterations to 1000 to fine-tune the model on this system.

\section{Testing SDE systems}
\label{App:systems_test}

\subsubsection*{Ornstein-Uhlenbeck (1D)}
The Ornstein-Uhlenbeck SDE
\begin{equation}
\label{eq:OU}
    dX_t = \theta (\mu - X_t) dt + \sigma dW_t, \quad \theta >0, \quad \sigma >0,
\end{equation}
 has applications in financial mathematics and the physical sciences. 

In this work, we used parameters randomly sampled from $\theta \sim \text{Unif}~[0.1, 0.5], ~\mu \sim \text{Unif}~[1,5], ~\sigma \sim \text{Unif}~[0.1, 0.5].$
We simulate trajectories with $ ~\Delta t = 0.001, ~k = 100$ and initial condition $X_0 \sim \text{Unif}~[50,100].$
We set the number of iterations to 1000 to fine-tune the model on this system. 

\subsubsection*{Inhomogeneous Ornstein-Uhlenbeck (1D)}
\begin{figure}[h]
    \centering
    \includegraphics[width=0.45\linewidth]{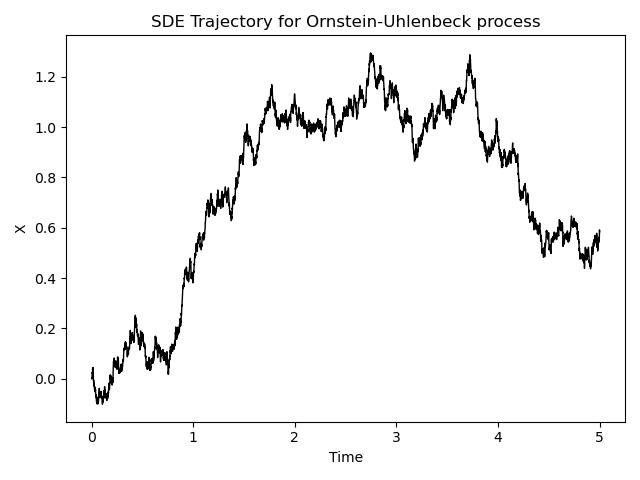}
    \includegraphics[width=0.45\linewidth]{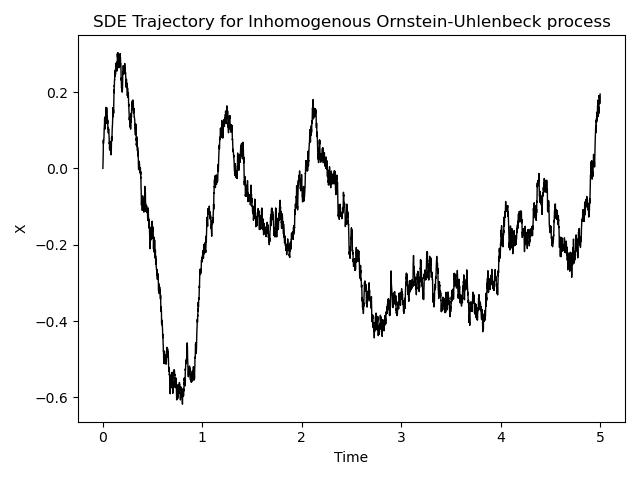}
    \caption{(left) Ornstein-Uhlenbeck process~\eqref{eq:OU} with $\theta = 0.25, ~\mu = 1.0, ~\sigma = 0.3$ (right) Inhomogeneous Ornstein-Uhlenbeck process~\eqref{eq:IOU} with $\theta = 1.0, ~a = 1.0, ~\omega = 2\pi, ~\sigma = 0.3$.}
    \label{fig:traj_OUs}
\end{figure}

An inhomogeneous Ornstein-Uhlenbeck (OU) process is a mean-reverting diffusion process. We consider the process written as 
\begin{equation}
\label{eq:IOU}
    dX_t = [a\cos(\omega t) - \theta X_t] dt + \sigma dW_t
\end{equation}
Parameters used in this work are $a \sim \text{Unif}~[0.5, 2.0], ~\omega \sim \text{Unif}~[\pi, 4\pi], ~\theta \sim \text{Unif}~[0.5, 2.0], \sigma \sim \text{Unif}~[0.1, 0.5].$ 
We simulate trajectories with $ ~\Delta t = 0.001, ~k = 100$ and initial condition $X_0 \sim \text{Unif}~[50,100].$
Example trajectories for both Ornstein-Uhlenbeck processes are included in Figure \ref{fig:traj_OUs}.
We set the number of iterations to 1000 to finetune the model on this system. 

\subsubsection*{Overdamped Langevin equation with double-well potential (2D)}
The system follows the overdamped Langevin dynamics~\eqref{eq:overdamped} with potential energy 
\begin{equation}
\label{eq:double_well}
    V(x) = (x_1^2 - 1)^2 + \alpha x_2^2
\end{equation}
Variation of this system stems from $\alpha$ and inverse temperature $\beta$. In this work, we used $\alpha \sim \text{Unif}~[0.1, 0.5], ~\beta \sim \text{Unif}~[5, 20].$
We simulate trajectories with $ ~\Delta t = 10^{-5}, ~k = 100$ and initial condition $X_0, X_1 \sim \text{Unif}~[-1,1]$.
We set the number of iterations to 1000 to fine-tune the model on this system. 

\begin{figure}[h]
    \centering
    \includegraphics[width=0.45\linewidth]{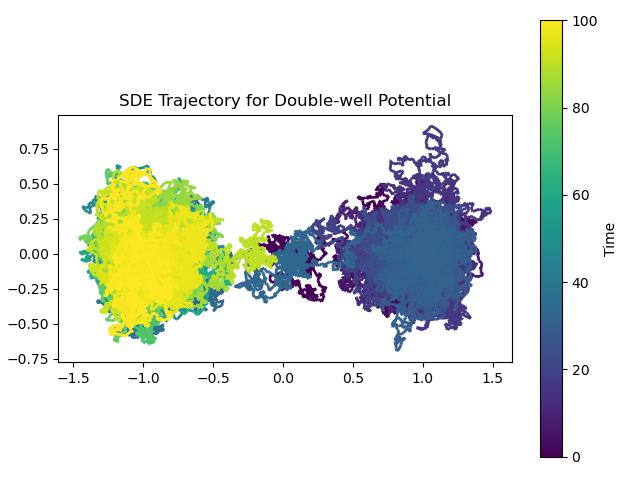}
    \includegraphics[width=0.45\linewidth]{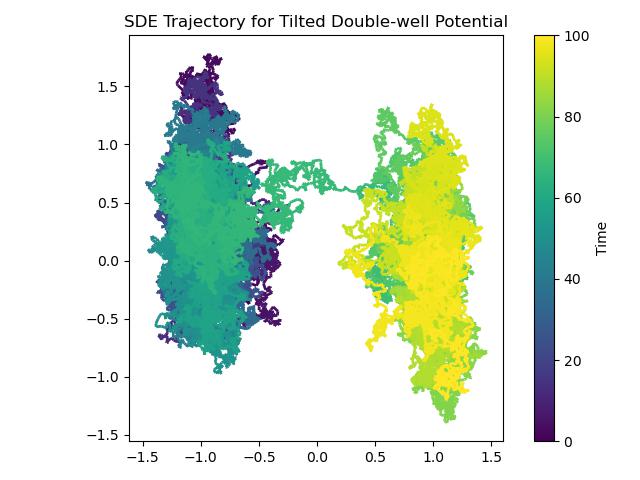}
    \caption{(left) Overdamped Langevin equations with double well potential~\eqref{eq:double_well} with $\alpha = 3.0,  ~\beta = 5$ (right) Coupled double well potential~\eqref{eq:coupled_DW} with $\alpha = 0.25,  ~\beta = 5$.}
    \label{fig:traj_double_wells}
\end{figure}

\textbf{Overdamped Langevin equation with coupled double-well potential (2D).}
The coupled double-well potential breaks the independence of $x_1$ and $x_2$ and is governed by the SDE
\begin{equation}
\label{eq:coupled_DW}
    V(x_1,x_2) = (x_1^2 - 1)^2 + \frac{1}{2}x_2^2 + \alpha x_1x_2
\end{equation}
We used the same parameter range as for the double-well potential: $\alpha \sim \text{Unif}~[0.1, 0.5], ~\beta \sim \text{Unif}~[5, 20].$
We simulate trajectories with $ ~\Delta t = 10^{-5}, ~k = 100$ and initial condition $X_0, X_1 \sim \text{Unif}~[-1,1]$.
Example trajectories for both double-well and coupled double-well potentials are included in Figure \ref{fig:traj_double_wells}.

\subsubsection*{Duffing oscillator (2D)}
The bistable Duffing oscillator dynamics are governed by the Langevin SDE
\begin{equation}
\label{Duffing}
    d\begin{bmatrix} X_t \\ V_t \end{bmatrix} = \begin{bmatrix} V_t \\ -\delta V_t - \alpha X_t - \beta X_t^3 + \gamma \cos(\omega t) \end{bmatrix}dt + \sqrt{\epsilon}\begin{bmatrix} 0 \\ dW_t \end{bmatrix}.
\end{equation}

The system has two stable equilibria $a = (-1, 0)$ and $b = (1, 0)$ and an unstable equilibrium at the origin.
In this work, we consider parameters $\delta \sim \text{Unif}~[0.05, 0.5], ~\alpha \sim \text{Unif}~[-1, 1], ~\beta \sim \text{Unif}~[1, 10], ~\gamma \sim \text{Unif}~[0.1,1], ~\omega \sim \text{Unif}~[0.5,6], ~\epsilon \sim \text{Unif}~[0.01, 0.1].$
We simulate trajectories with $\Delta t = 10^{-4}, ~k = 100$ and initial condition $X_0, X_1 \sim \text{Unif}~[-1,1]$.
An example trajectory is included in Figure \ref{fig:traj_duffing_perturbed} (left).
We set the number of iterations to 1000 to fine-tune the model on this system. 

\subsubsection*{Stochastically perturbed limit cycles (2D)}
\begin{figure}[h]
    \centering
    \includegraphics[width=0.45\linewidth]{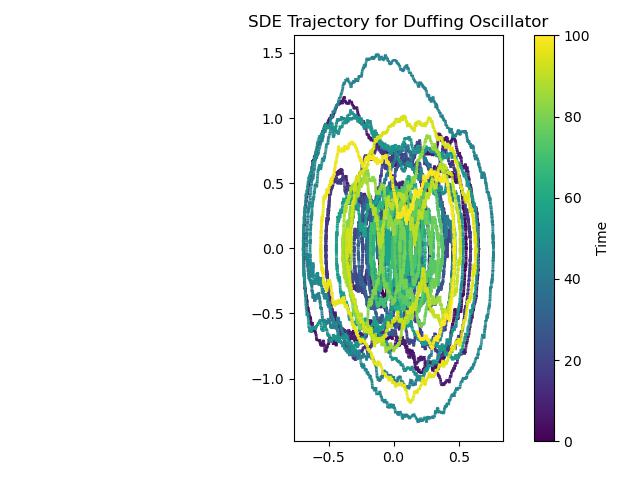}
    \includegraphics[width=0.45\linewidth]{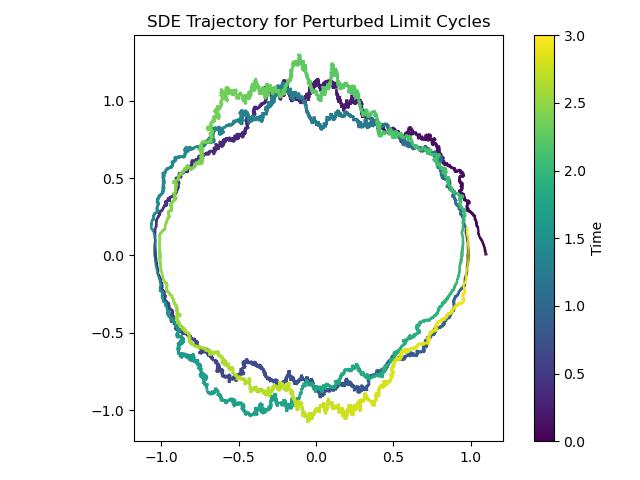}
    \caption{(left) Duffing oscillator~\eqref{Duffing} with $\delta = 0.2, ~\alpha = 1.0, ~\beta = 5.0, ~\gamma = 1.0, ~\omega = 1.6\pi, ~\epsilon = 0.05$ (right) Stochastically perturbed limit cycles~\eqref{eq:perturbed_osci} with $T = 2\pi,  ~\sigma = 0.25$.}
    \label{fig:traj_duffing_perturbed}
\end{figure}

This numerical example was taken from \cite{Cycles_Tori_Dankowicz}. The authors considered a system
\begin{equation}
\label{eq:perturbed_osci}
    dx = T f(x) dt + \sigma \sqrt{T} F(x)dW_t
\end{equation}
where
$T$ is the period of the deterministic limit cycle in the noiseless case,
$f(x)$ is the drift field,
$F(x)$ is the noise vector; the diffusion matrix is ${\rm diag} (F(x))$,
$\sigma$ is the noise intensity. 
Here we consider the case with
\begin{equation*}
    f(x) = \left(\begin{matrix}
x_1 - x_2 - x_1(x_1^2 + x_2^2) \\
x_1 + x_2 - x_2(x_1^2 + x^2_2)
\end{matrix} \right),  \quad F(x) = \left(\begin{matrix}
x_1x_2 \\
x_2^2
\end{matrix} \right)
\end{equation*}
In particular, we used $T \sim \text{Unif}~[0.01, 2\pi], ~\sigma \sim \text{Unif}~[0.01, 1].$
We simulate trajectories with $\Delta t = 10^{-5}, ~ k = 100$ and initial condition $X_0, X_1 \sim \text{Unif}~[-1,1]$.
An example trajectory is included in Figure \ref{fig:traj_duffing_perturbed} (right).
We set the number of iterations to 2000 to fine-tune the model on this system. 

\subsubsection*{Predator-prey model (3D)}
The predator and prey model is a famous biological model. In particular, we consider the same system as \cite{predator_prey_LiLiuWei}
\begin{equation}
\label{eq:predator-prey}
    \begin{aligned}
    dx & = [x(r - ax + sy_1 - by_2)]dt + \sigma_1 x dB_1(t)\\
    dy_1 & = [kxy_2 - y_1(gx + D + v_1)]dt + \sigma_2 y_1 dB_2(t)\\
    dy_2 & = (D y_1 - v_2 y_2) dt + \sigma_3 y_2 dB_3(t).
\end{aligned}
\end{equation}
The physical meaning of each variable and parameter is listed as follows:
\begin{itemize}
    \item $x$ denotes the density of the prey
    \item $y_1$ and $y_2$ denote the densities of juvenile and
adult predators at time t, respectively.
\item $r$ represents the intrinsic growth rate
of the prey
\item $a$ denotes the intraspecific competition rate of the prey
\item $b$ and $g$ represent the predation rates of the adult predator and the prey, respectively
\item $k = \delta_1 b > 0$ and $s = \delta_2 g \geq 0$ denote the reproduction rates among which $\delta_i, ~(i = 1,2)$ are
the conversion eﬃciency ($\delta_i < 1$ due to the biological significance)
\item $D$ represents the rate at which juvenile predators become
adult predators
\item $v_i, ~(i = 1,2)$ represent the death rates of juvenile and adult predators, respectively.
\end{itemize}
An example trajectory is included in Figure \ref{fig:traj_predator_fluxgate} (left).
We set the number of iterations to 2000 to fine-tune the model on this system.

\subsubsection*{Fluxgate sensor}
\begin{figure}[h]
    \centering
    \includegraphics[width=0.45\linewidth]{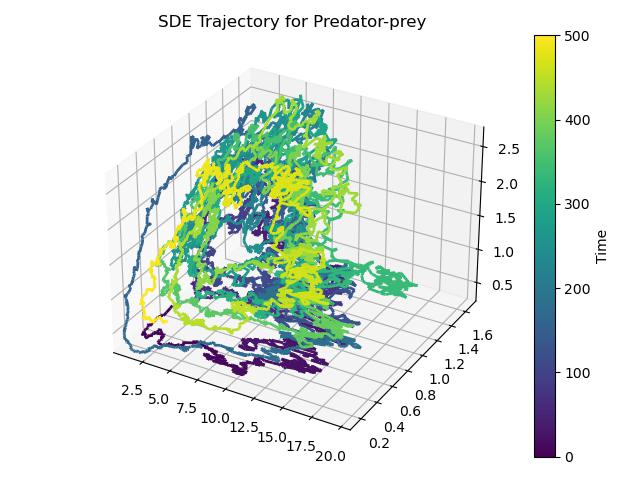}
    \includegraphics[width=0.45\linewidth]{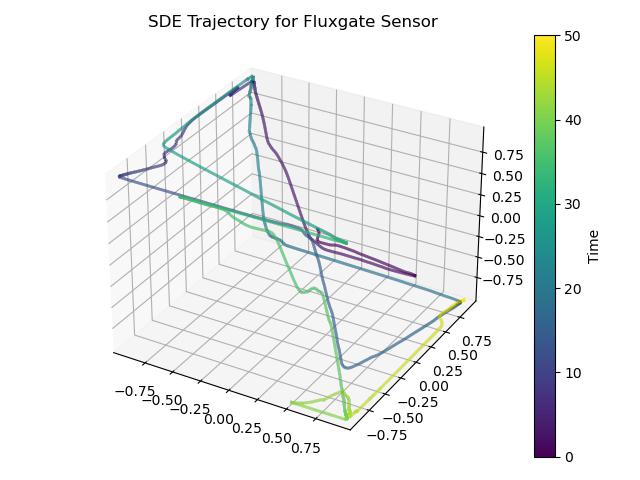}
    \caption{(left) Predator-prey model~\eqref{eq:predator-prey} with $r = 0.4, ~a = 0.02, ~s = 0.35, ~b = 0.4, ~k = .4, ~g = 0.5, ~D = 0.4, ~v_1 = 0.25, ~v_2 = 0.25, ~\sigma_1 = 0.15, ~\sigma_2 = 0.12, ~\sigma_3 = 0.1$ (right) Fluxgate sensor~\eqref{eq:fluxgate} with $c = 3.0,  ~\lambda = -0.435, ~\omega = 1.0, \epsilon = 0.5$.}
    \label{fig:traj_predator_fluxgate}
\end{figure}

This system describes a single magnetic core that acts
as an overdamped oscillator with a double-well
potential with a quadratic growth as $|x|\rightarrow\infty$:
\begin{equation}
\label{eq:fluxgate}
\begin{aligned}
    \dot{x}_j &= -x_j +\tanh(c(x_j + \lambda x_{j+1} +y_j)) \\
    dy_j &= -\omega y_j dt + \omega\sqrt{\epsilon}dW.
\end{aligned}
\end{equation}
The composite system may consist of an arbitrary number of magnetic cores. Here, we consider three coupled cores with colored noise
\begin{equation}
\begin{cases}
    dx_1 & = -x_1 + \tanh[c(x_1 + \lambda x_2 + y_1)], \quad dy_1 = -\omega y_1 dt + \omega \sqrt{\epsilon} dW\\
    dx_2 & = -x_2 + \tanh[c(x_2 + \lambda x_3 + y_2)], \quad dy_2 = -\omega y_2 dt + \omega \sqrt{\epsilon} dW\\
    dx_3 & = -x_3 + \tanh[c(x_3 + \lambda x_1 + y_3)], \quad dy_3 = -\omega y_3 dt + \omega \sqrt{\epsilon} dW\\
\end{cases}
\end{equation}
In this work, we used $c \sim \text{Unif}~[3,5],~\lambda \sim \text{Unif}~[0.1, 1.0], ~\epsilon \sim \text{Unif}~[0.1, 0.5], ~\omega = 3, ~\Delta t = 10^{-3}, ~k = 100$.
An example trajectory is included in Figure \ref{fig:traj_predator_fluxgate} (right).
We set the number of iterations to 2000 to fine-tune the model on this system. 

\section{Textual data generation}
\label{App:textual}
For more well-known systems with enough publicly available data, we generate the supplemental textual data with the assistance of GPT-4. We use the following prompt for generation. 
\texttt{\footnotesize Please provide [number required] different explanations of the [SDE system], [mathematical formula].\\
Explanations should cover the following aspects: the contexts or domains where this system can be applied; \\
the meanings and roles of its parameters; and the types of behaviors or dynamics the system exhibits under \\
different parameter ranges.}

For systems that are less commonly known or have specific forms introduced in particular reference papers, we manually created the textual prompts by extracting the necessary information from the corresponding sources. For these systems, in addition to the system name and mathematical formulation, we also include the application context, each parameter's meaning, and the underlying dynamics' potential behaviors.

The systems with manually created textual prompts include the periodic nonlinear oscillator, perturbed nonlinear oscillator, fluxgate sensor, and predator–prey system. The textual prompts for the remaining systems were generated with the assistance of GPT-5.

We generated ten textual prompts for each system. For those produced with GPT-5, slight manual adjustments were made after generation to ensure correctness and fluency. During training and inference, a textual prompt is randomly selected from the available set corresponding to the given SDE system.

Here we show several examples of the textual data for GPT-5-assisted and manually crafted textual prompts for the Duffing oscillator and stochastically perturbed limit cycles and tori (perturbed nonlinear oscillator)
\begin{itemize}
    \item Prompts for Duffing oscillator:
    \begin{itemize}
        \item The non-linear restoring force, $-\alpha X_t - \beta X_t^3$, can be derived from a potential energy function $V(X) = \frac{1}{2}\alpha X^2 + \frac{1}{4}\beta X^4$.
        \item If $\alpha > 0$ and $\beta > 0$, the potential is a single-well (harmonic-like) potential, and the system is an oscillator. If $\alpha < 0$ and $\beta > 0$, the potential is a double-well potential, and the system is a bistable oscillator, similar to the previous examples but with inertial effects.
    \end{itemize}
    \item Prompts for stochastically perturbed limit cycles and tori (perturbed nonlinear oscillator):
    \begin{itemize}
        \item The system is defined by the SDE $dx = T f(x) dt + \sigma \sqrt{T} F(x)dW_t$, where $x$ is the state vector.
        \item The $f(x)$ term represents the deterministic drift field, governing the system's inherent oscillatory dynamics, as exemplified by 
        \begin{equation*}
        f(x) = \left(\begin{matrix} x_1 - x_2 - x_1(x_1^2 + x_2^2) \\ x_1 + x_2 - x_2(x_1^2 + x^2_2) \end{matrix} \right).
        \end{equation*}
        \item $\sigma$ denotes the noise intensity, and $\sqrt{T}$ scales the noise, linking its magnitude to the deterministic cycle's period.
    \end{itemize}
\end{itemize}

\section{Additional numerical experimental results of FMint-SDE}
\label{App:results}
Table \ref{tab:finetune-OOD} includes a complete numerical report of the performance of FMint-SDE {and baseline models} on all out-of-distribution SDE systems. 
For these systems, the models are fine-tuned for 1000 to 2000 steps on training data of size 50. The errors are then obtained by averaging over 500 trajectories.

\input{Tables/finetune}

\begin{figure}
    \centering
    \includegraphics[width=\linewidth]{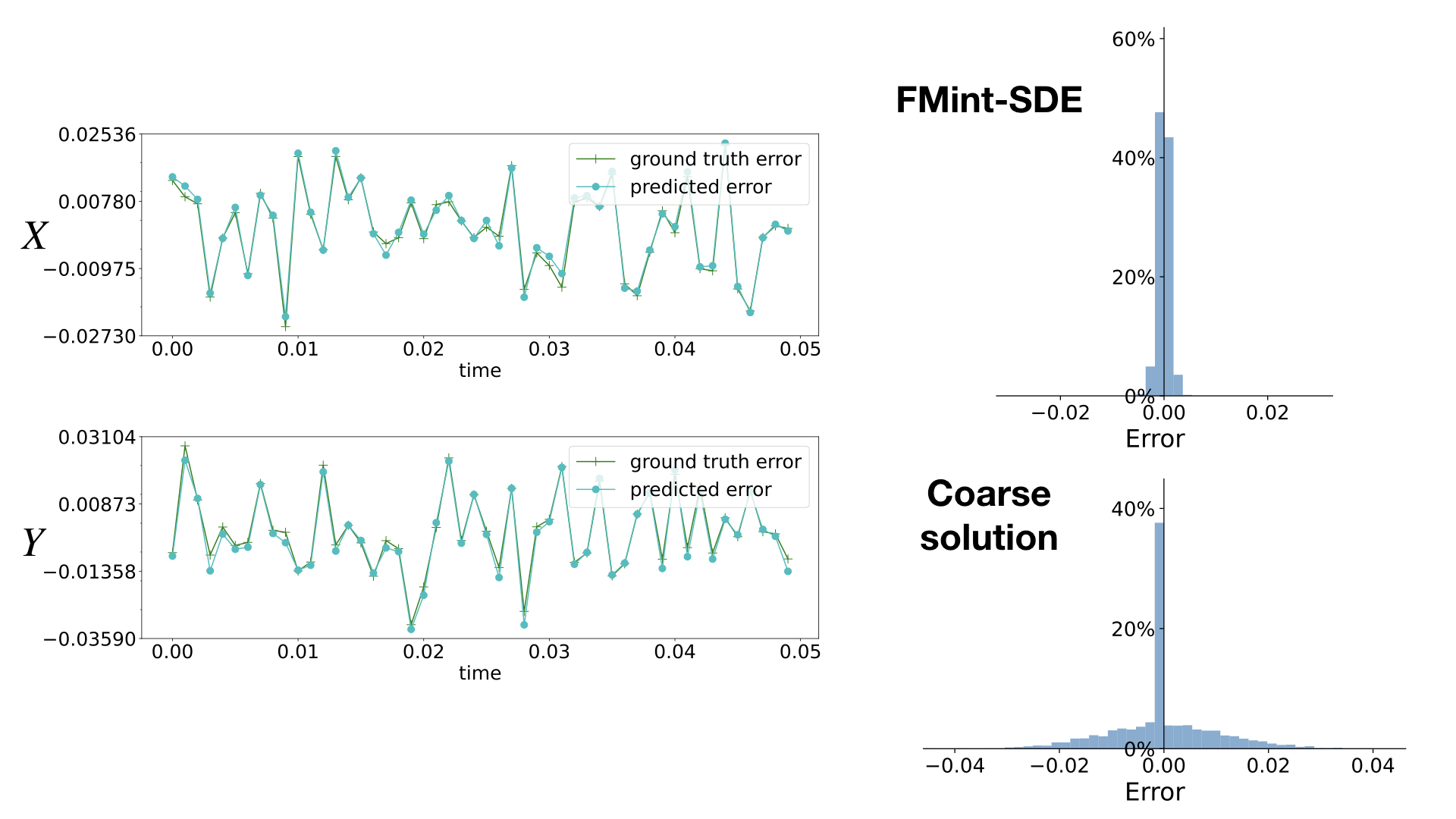}
    \caption{Overdamped Langevin with coupled double well potential~\eqref{eq:coupled_DW}}
    \label{fig:appdx_couple_dw}
\end{figure}

\begin{figure}
    \centering
    \includegraphics[width=\linewidth]{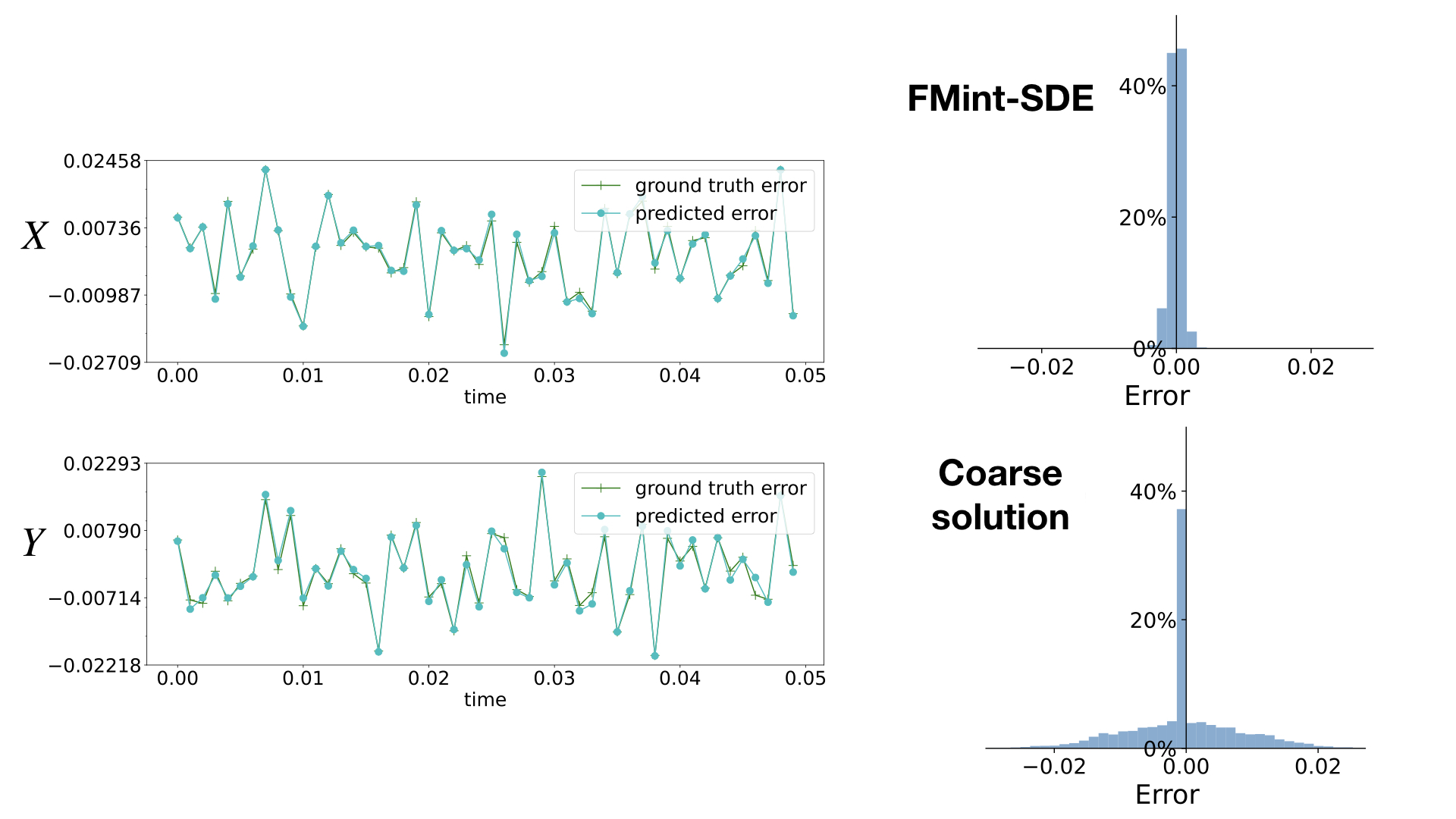}
    \caption{Overdamped Langevin with double-well potential~\eqref{eq:double_well}}
    \label{fig:appdx_dw}
\end{figure}

\begin{figure}
    \centering
    \includegraphics[width=\linewidth]{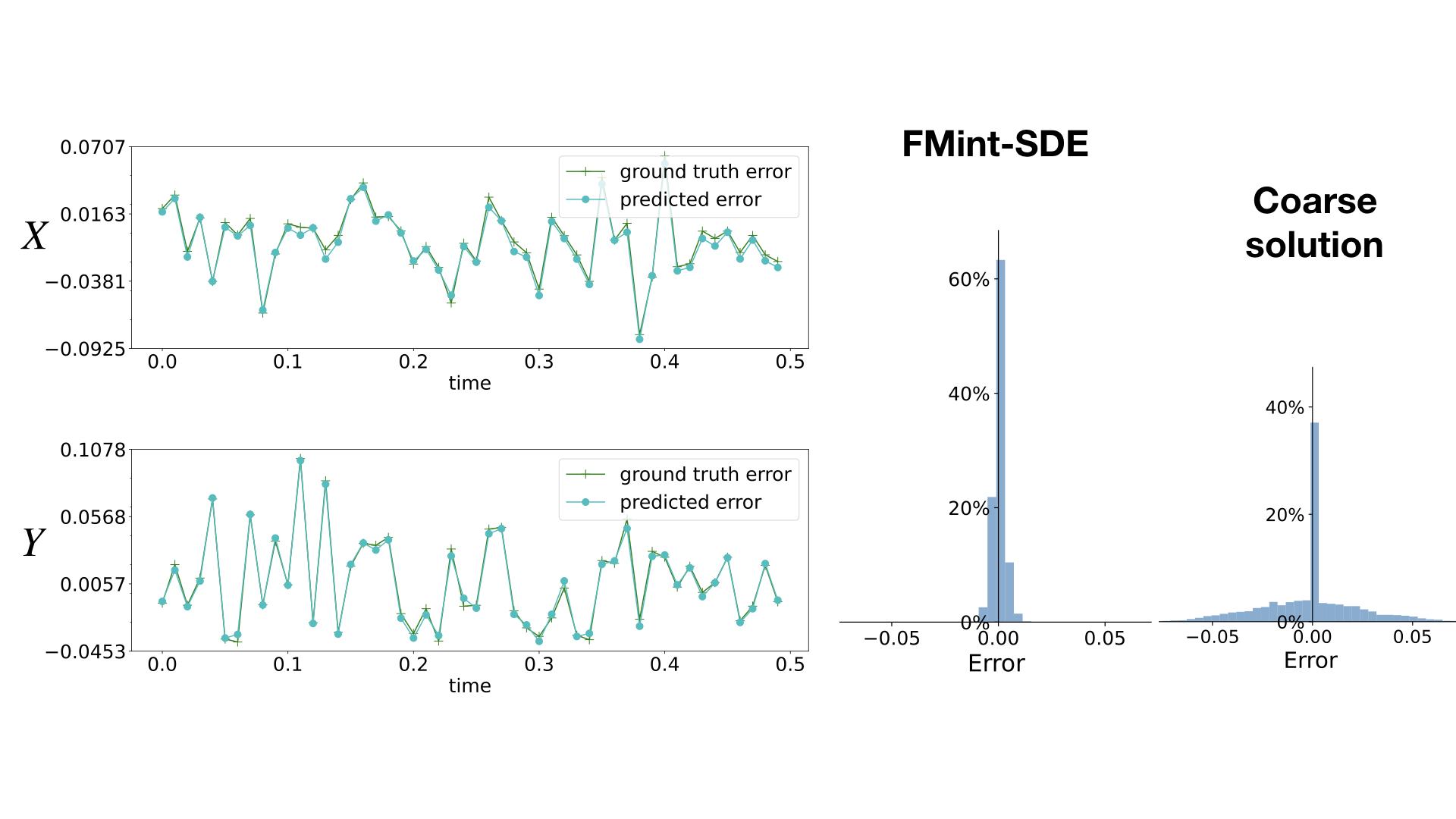}
    \caption{The Duffing oscillator~\eqref{Duffing}}
    \label{fig:appdx_duffing}
\end{figure}

\begin{figure}
    \centering
    \includegraphics[width=\linewidth]{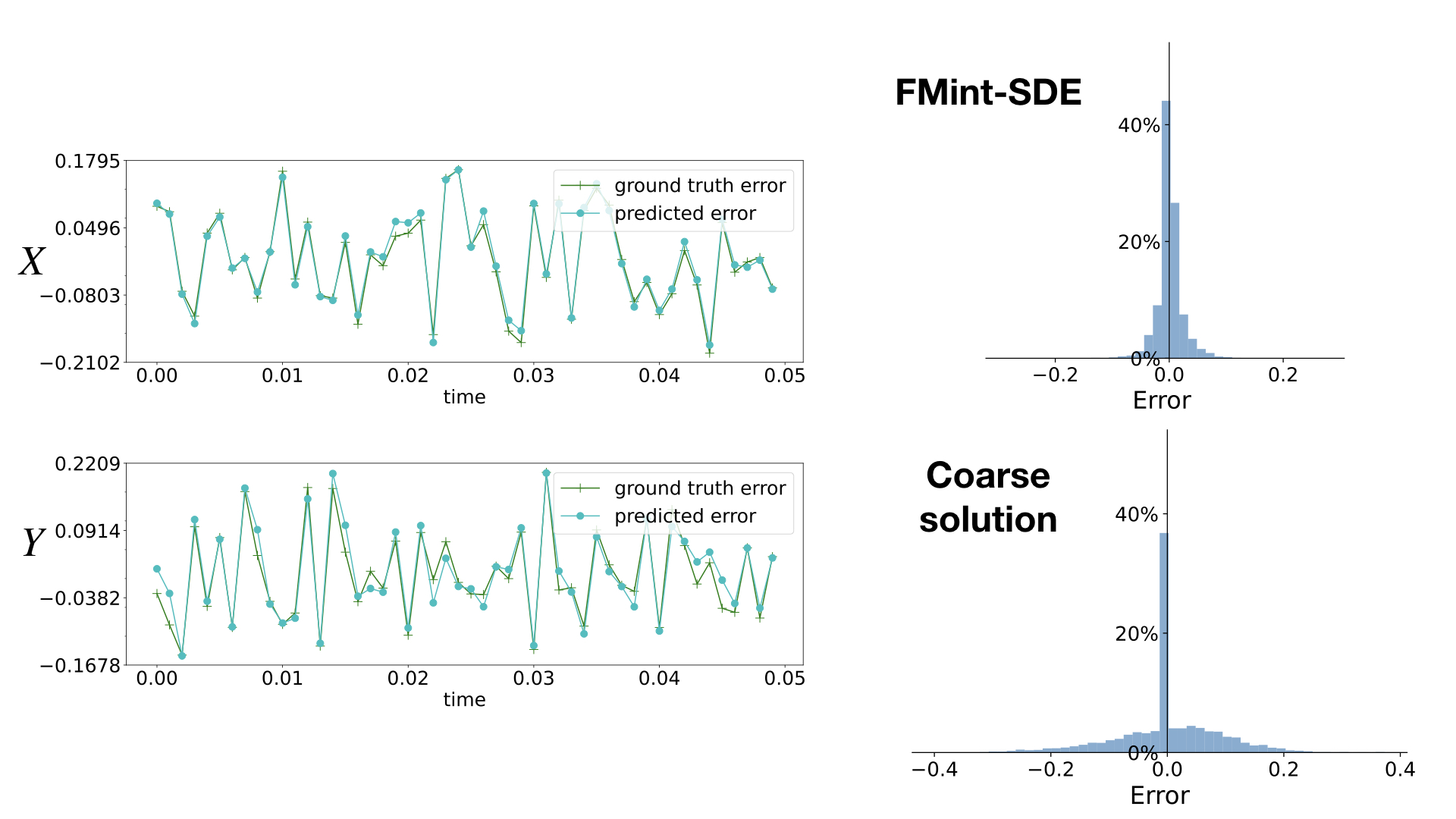}
    \caption{Overdamped Langevin with Mueller's potential~\eqref{mueller}}
    \label{fig:appdx_mueller}
\end{figure}

\begin{figure}
    \centering
    \includegraphics[width=\linewidth]{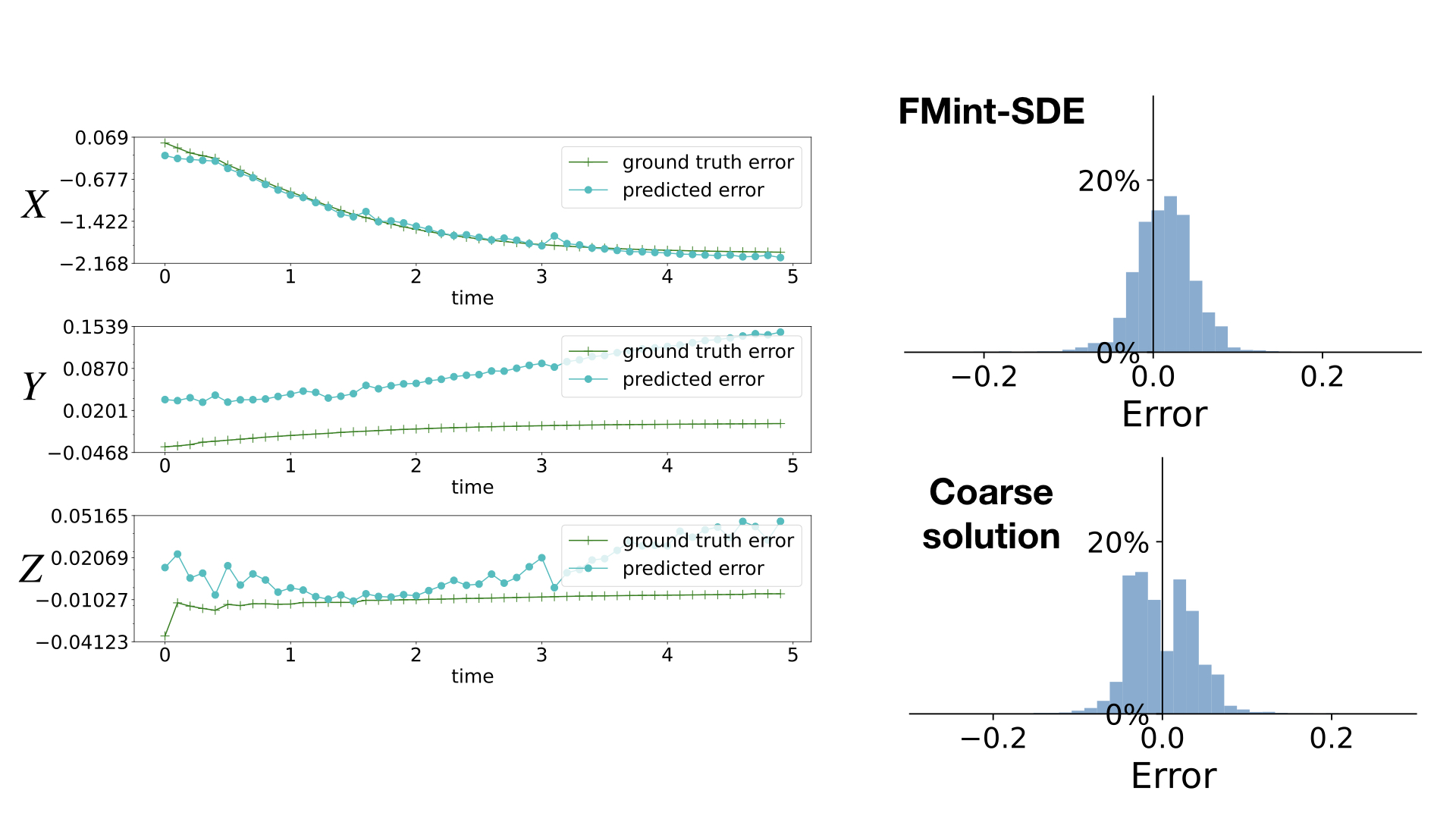}
    \caption{The fluxgate sensor~\eqref{eq:fluxgate}}
    \label{fig:appdx_fluxgate}
\end{figure}

\begin{figure}
    \centering
    \includegraphics[width=\linewidth]{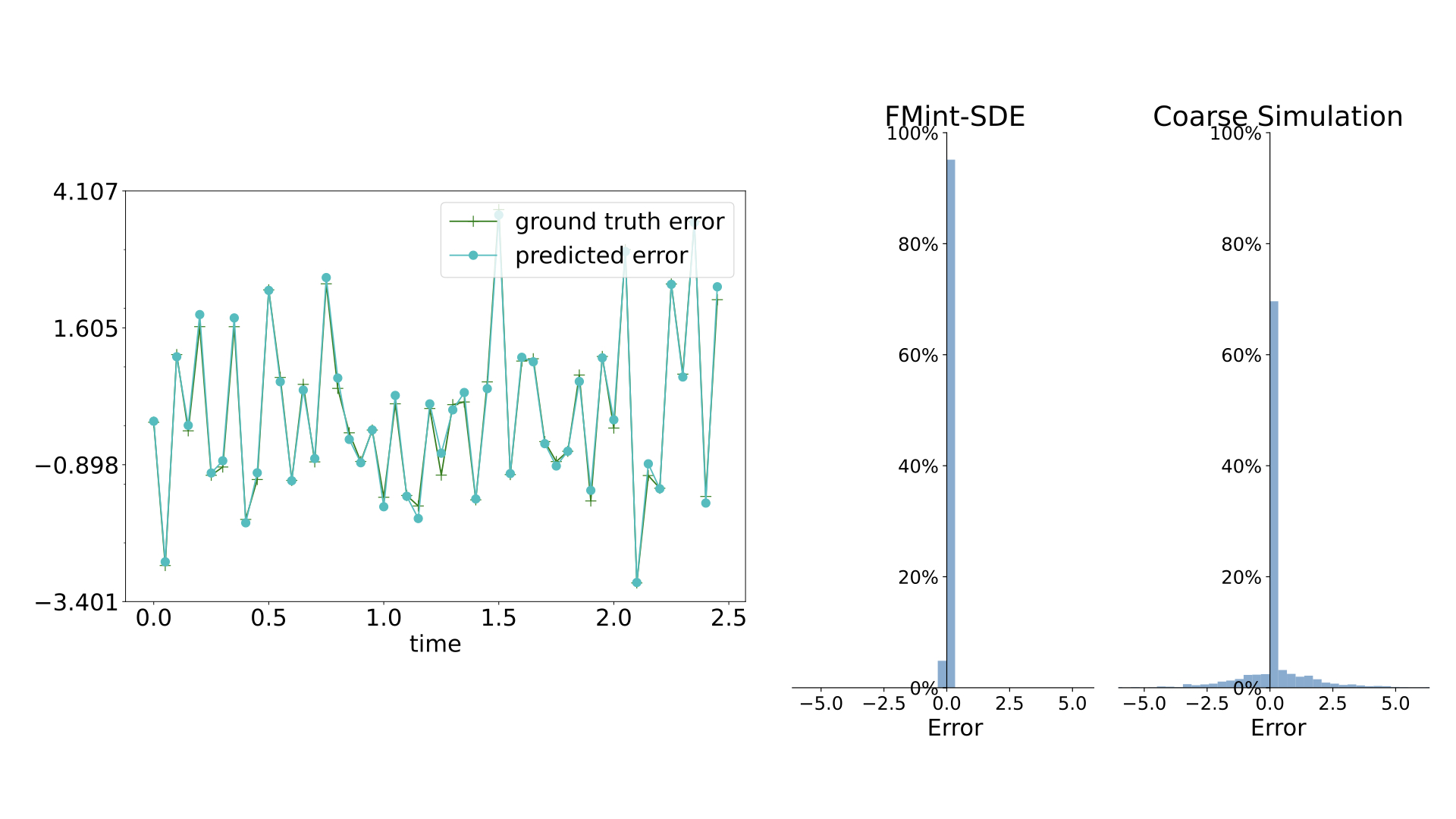}
    \caption{The geometric Brownian motion~\eqref{eq:GMB}}
    \label{fig:appdx_GBM}
\end{figure}

\begin{figure}
    \centering
    \includegraphics[width=\linewidth]{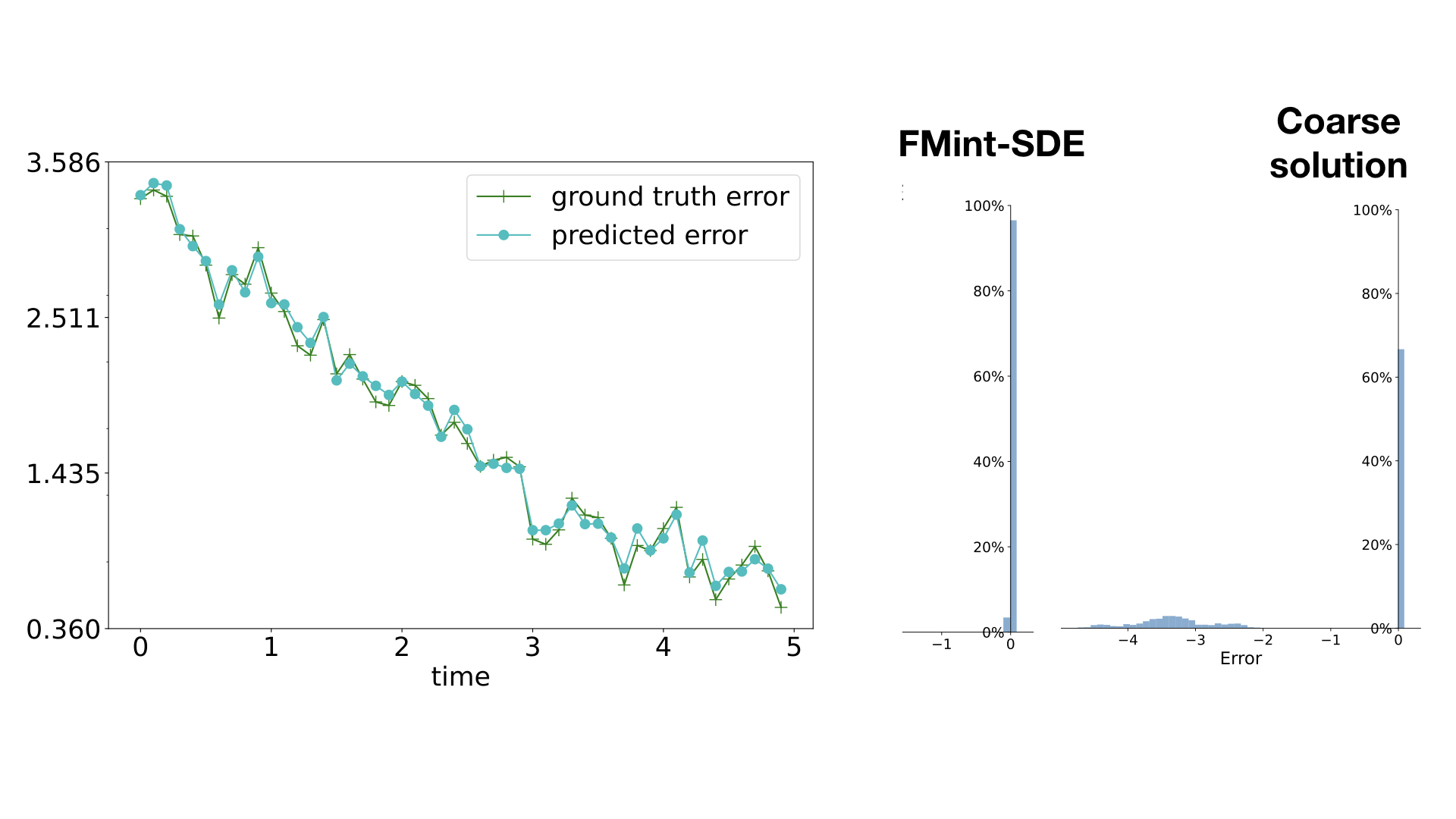}
    \caption{Inhomogeneous Ornstein-Uhlenbeck~\eqref{eq:IOU}}
    \label{fig:appdx_IOU}
\end{figure}

\begin{figure}
    \centering
    \includegraphics[width=\linewidth]{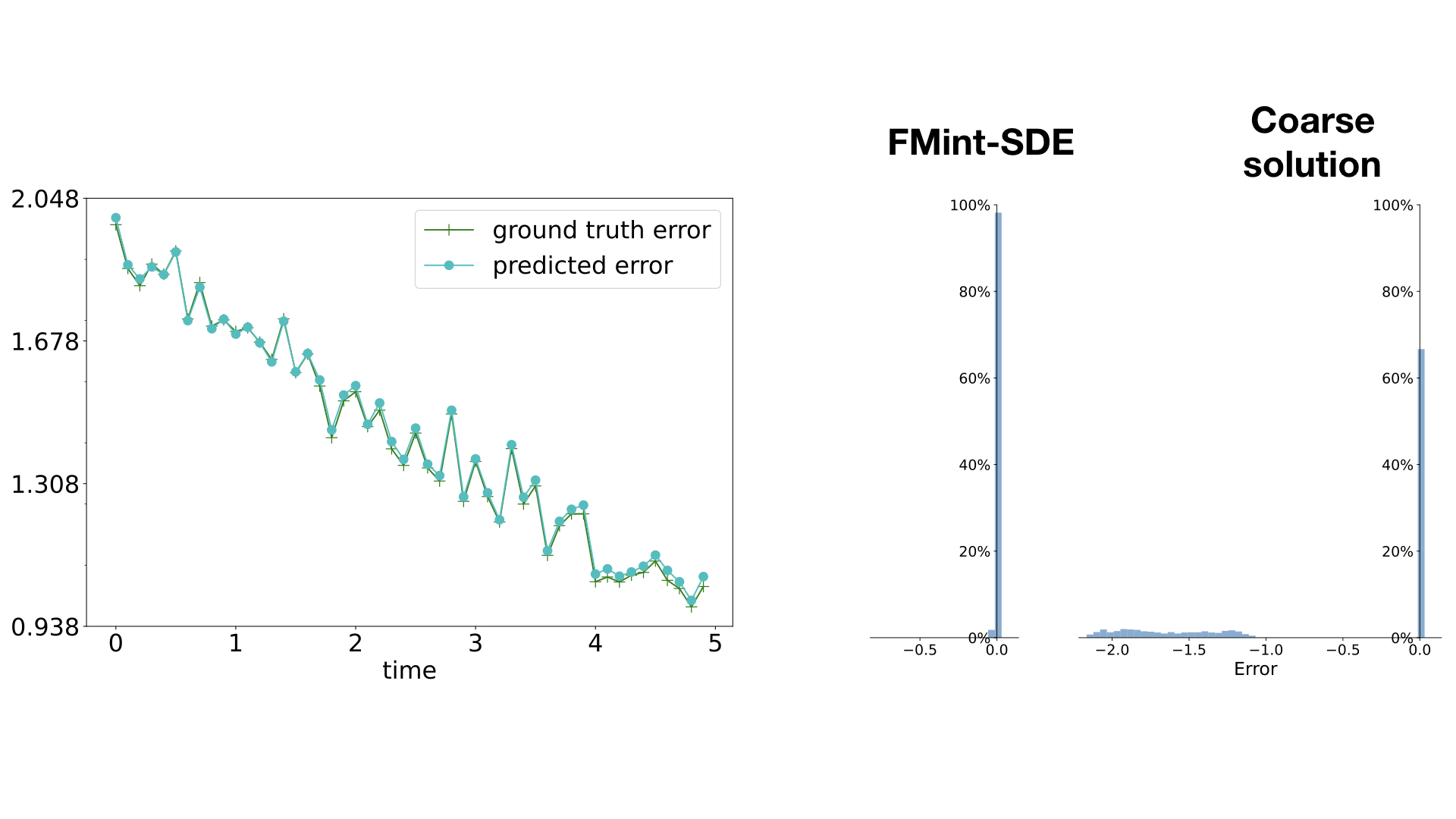}
    \caption{Ornstein-Uhlenbeck~\eqref{eq:OU}}
    \label{fig:appdx_OU}
\end{figure}

\begin{figure}
    \centering
    \includegraphics[width=\linewidth]{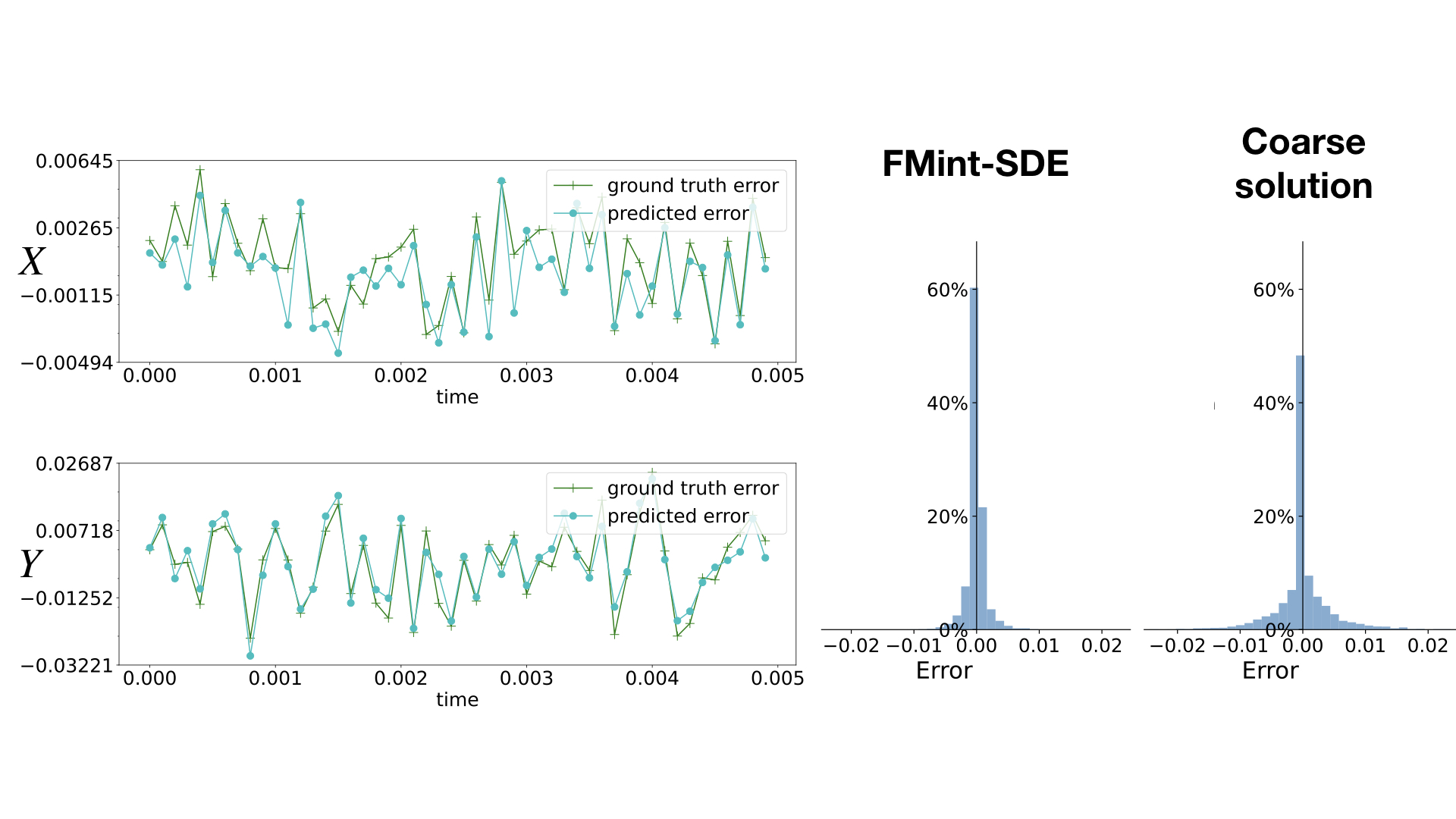}
    \caption{The periodic nonlinear oscillator~\eqref{eq:periodic_osci}}
    \label{fig:appdx_periodic}
\end{figure}

\begin{figure}
    \centering
    \includegraphics[width=\linewidth]{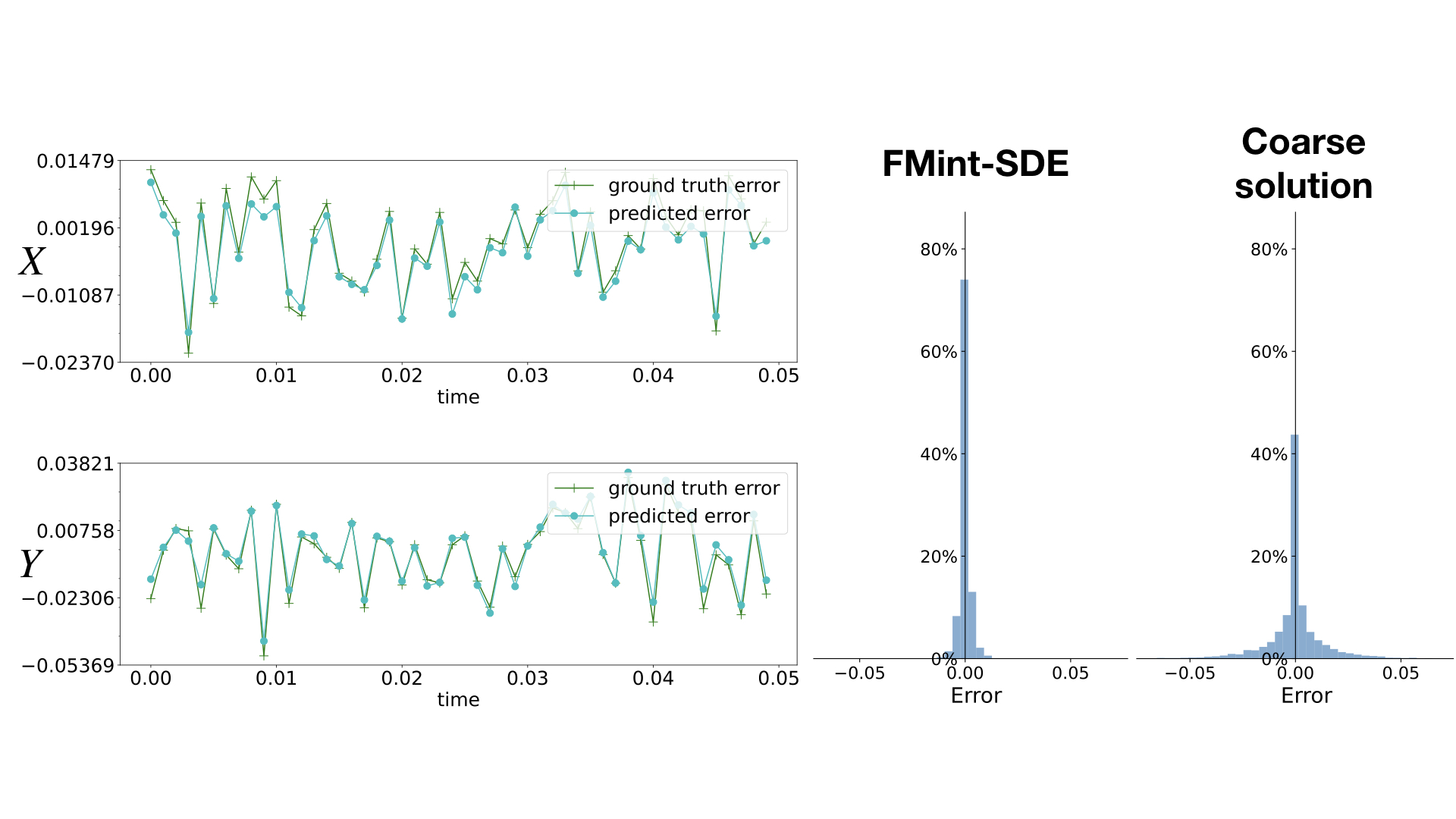}
    \caption{The perturbed nonlinear oscillator~\eqref{eq:perturbed_osci}}
    \label{fig:appdx_perturbed}
\end{figure}

\begin{figure}
    \centering
    \includegraphics[width=\linewidth]{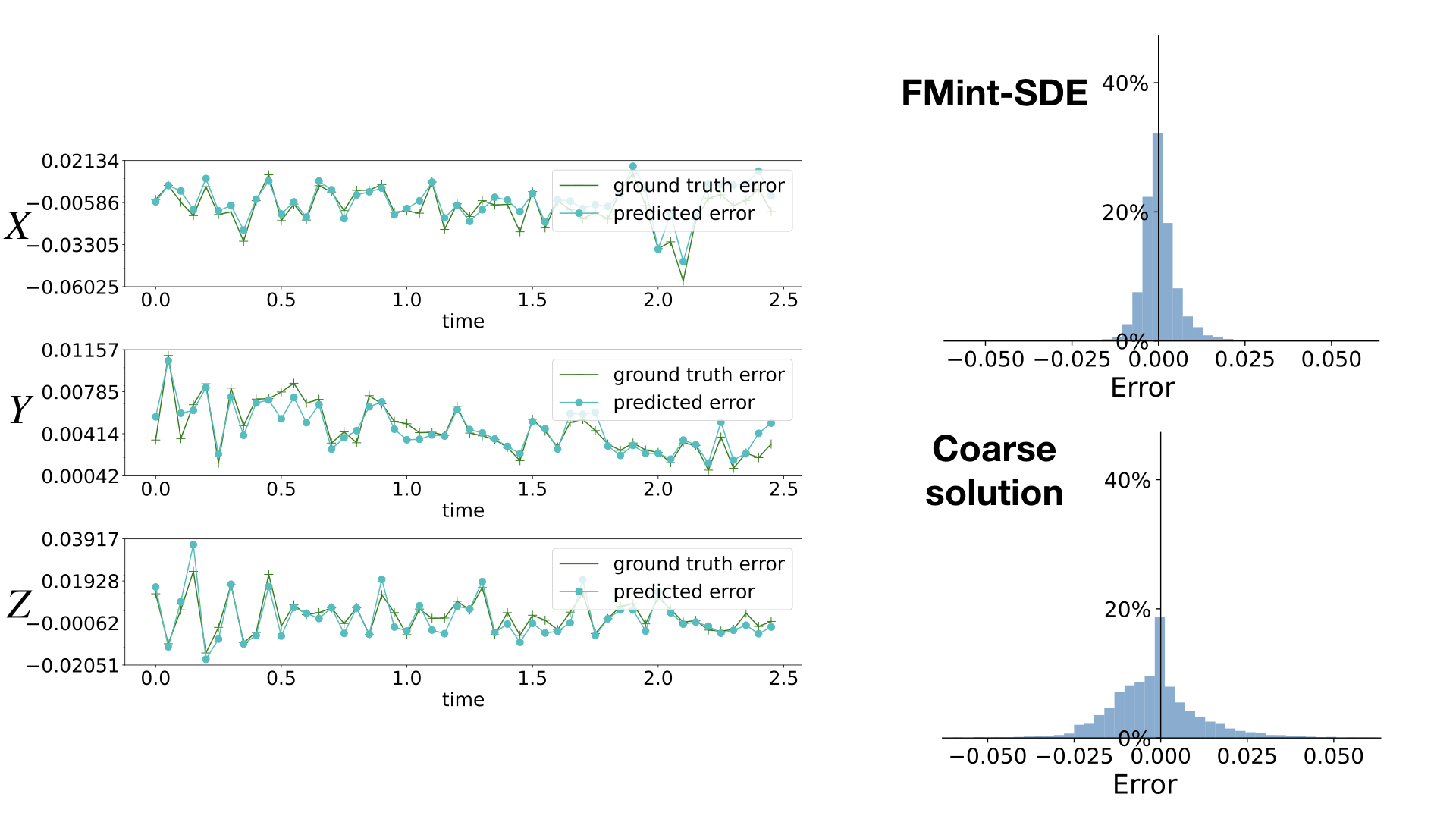}
    \caption{Predator-prey~\eqref{eq:predator-prey}}
    \label{fig:appdx_predator}
\end{figure}

\begin{figure}
    \centering
    \includegraphics[width=\linewidth]{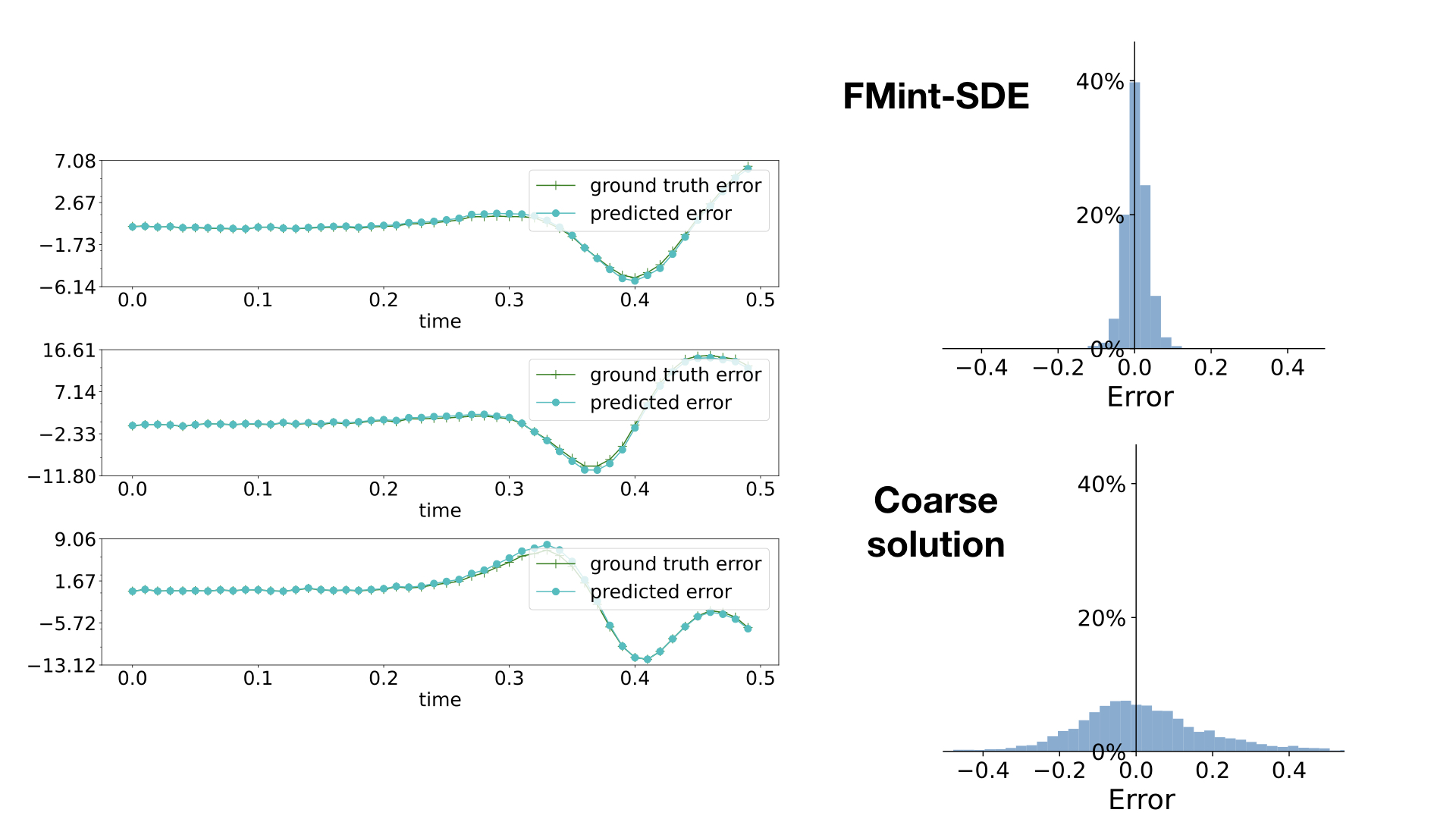}
    \caption{Stochastic Lorenz~\eqref{eq:stochastic_lorenz}}
    \label{fig:appdx_lorenz}
\end{figure}

Visualization of the error correction performance of FMint-SDE for all systems is shown in Figure \ref{fig:appdx_couple_dw} - \ref{fig:appdx_lorenz}.
The predicted error by FMint-SDE is represented by connected blue dots, and the ground truth errors between the fine-grained and coarse-grained trajectories are represented by green. 

{ \textit{Notes on fluxgate sensor:} Although the AMD error for the fluxgate sensor (Fig. 5, main text) shows no significant improvement over the coarse solution, Figure \ref{fig:appdx_fluxgate} indicates a nicer error distribution compared to coarse simulations. We further examine two cases with identical parameters but different initial conditions. In Figure \ref{fig:fluxgate_example}(left), FMint-SDE correctly recovers the fine trajectory in the first dimension, while the coarse solution diverges in the opposite direction. In contrast, Figure \ref{fig:fluxgate_example}(right) shows a case where FMint-SDE fails to improve upon the coarse result.
The fluxgate sensor can be formulated either as a 6D SDE with degenerate white noise or as a 3D SDE with colored noise; we adopt the latter. Since FMint-SDE was trained only on white-noise systems, this setting tests its robustness to colored noise, which has quite different statistical properties from white noise due to the time correlation of its increments.
}

\begin{figure}
    \centering
    \includegraphics[width=\linewidth]{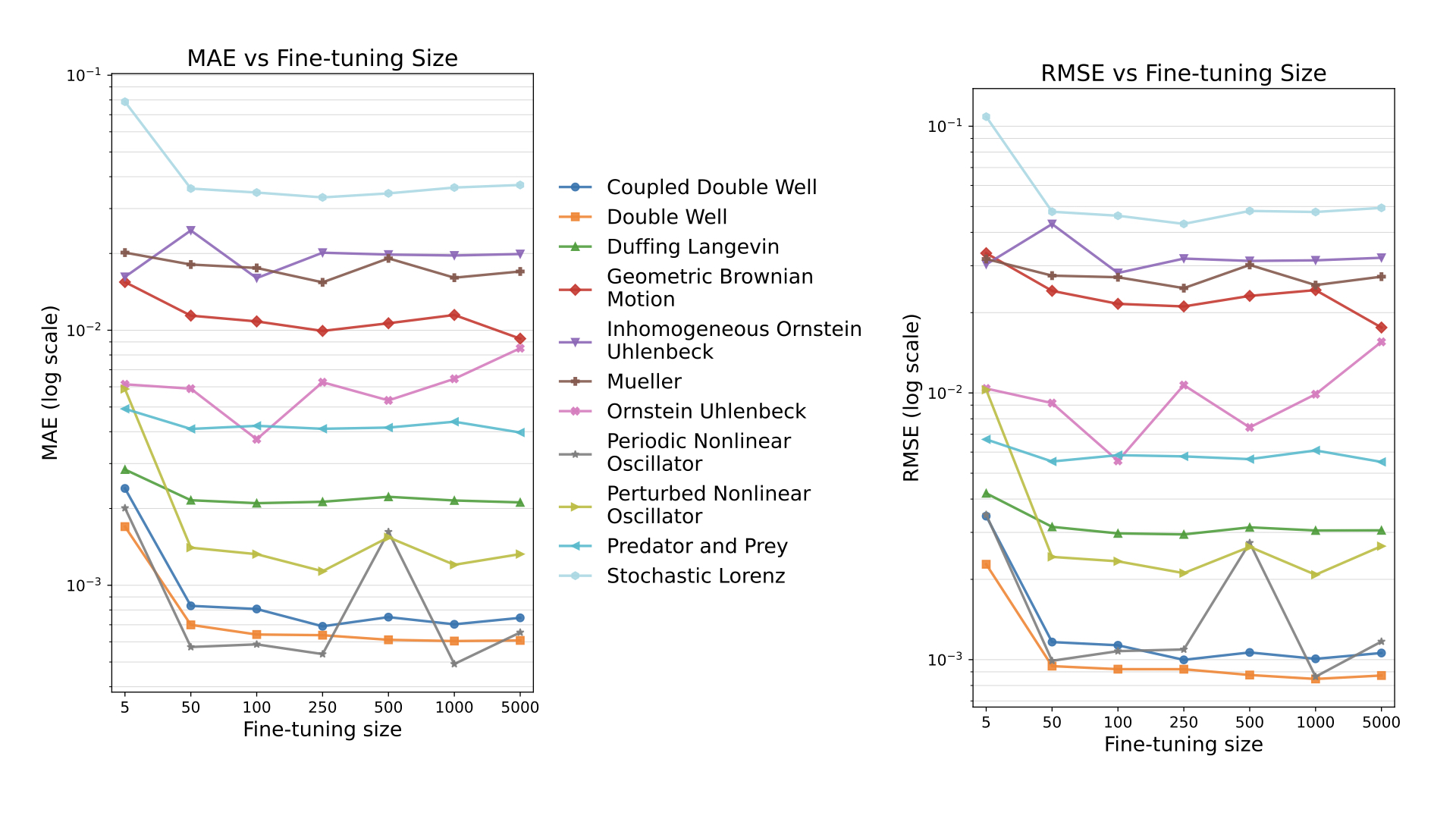}
    \caption{MAE (left) and RMSE (right) vs. fine-tuning size for all systems}
    \label{fig:mae_rmse_vs_samples}
\end{figure}

\begin{figure}[h]
    \centering
    \includegraphics[width=\linewidth]{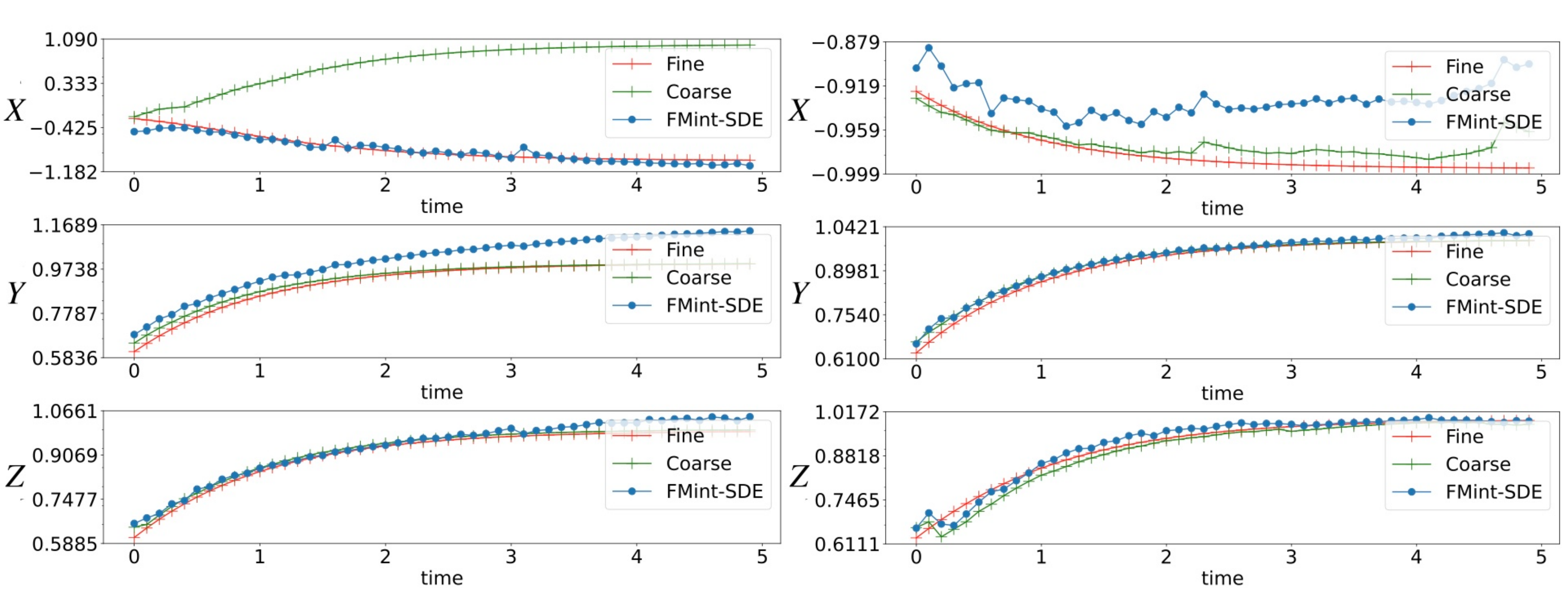}
    \caption{Performance of FMint-SDE on two particular example of fluxgate sensor}
    \label{fig:fluxgate_example}
\end{figure}

Figure \ref{fig:mae_rmse_vs_samples} and Tables \ref{tab:AMD_samples} - \ref{tab:RMSE_samples} include the complete report on the performance of FMint-SDE vs. finetuning size for all systems across all four evaluation metrics.


\begin{landscape}
\input{Tables/strong_samples}
\end{landscape}
\begin{landscape}
\input{Tables/weak_samples}
\end{landscape}
\begin{landscape}
\input{Tables/mae_samples}
\end{landscape}
\begin{landscape}
\input{Tables/rmse_samples}
\end{landscape}

\subsection{Details on the performance on the Roll-out}

{Table~\ref{tab:rollout} reports the performance of FMint-SDE compared to the coarse solution under the roll-out scheme. The total length of the trajectories are 500 and hence the roll-out is carried out for 10 times. }

\begin{table}[h!]
\centering
\small
\setlength{\tabcolsep}{4pt}
\begin{tabular}{p{2.9cm}cccccccc}
\hline
System & MAE & MAE$_{\text{coarse}}$ & RMSE & RMSE$_{\text{coarse}}$ & AMD & AMD$_{\text{coarse}}$ & MAD & MAD$_{\text{coarse}}$ \\
\hline
Coupled double-well & 1.32e-03 & 9.49e-03 & 1.69e-03 & 1.19e-02 & 1.88e-03 & 1.34e-02 & 9.48e-04 & 1.61e-03 \\
Double-well & 1.21e-03 & 8.06e-03 & 1.56e-03 & 1.01e-02 & 1.72e-03 & 1.14e-02 & 8.45e-04 & 1.43e-03 \\
Duffing & 1.26e-02 & 2.46e-02 & 1.96e-02 & 3.10e-02 & 1.78e-02 & 3.49e-02 & 1.24e-02 & 9.14e-03 \\
Geometric Brownian Motion & 2.37e-01 & 1.65  & 3.34e-01 & 2.21  & 2.37e-01 & 1.65  & 9.59e-02 & 2.66e-01 \\
Inhomogeneous Ornstein-Uhlenbeck & 1.00e-01 & 1.13  & 1.35e-01 & 1.58  & 1.00e-01 & 1.13  & 5.42e-02 & 1.12  \\
OLD-Mueller & 8.53e-02 & 1.67e-01 & 1.93e-01 & 2.55e-01 & 1.08e-01 & 2.17e-01 & 8.36e-02 & 1.46e-01 \\
Ornstein-Uhlenbeck & 7.68e-02 & 9.09e-01 & 1.19e-01 & 1.01  & 7.68e-02 & 9.09e-01 & 4.43e-02 & 9.09e-01 \\
Periodic nonlinear oscillator & 9.58e-04 & 2.58e-03 & 1.55e-03 & 4.29e-03 & 1.42e-03 & 3.90e-03 & 4.51e-04 & 4.62e-04 \\
Perturbed nonlinear oscillator & 2.19e-03 & 8.26e-03 & 4.47e-03 & 1.46e-02 & 3.24e-03 & 1.25e-02 & 1.79e-03 & 3.34e-03 \\
Predator-prey & 1.04e-02 & 1.07e-02 & 1.79e-02 & 2.07e-02 & 1.92e-02 & 2.39e-02 & 9.18e-03 & 8.58e-03 \\
Stochastic Lorenz & 3.78  & 5.06  & 6.21  & 7.20  & 5.86  & 8.21  & 4.40  & 5.41  \\
\hline
\end{tabular}
\caption{Roll-out performance of FMint-SDE across systems. Total length is 500 for each system, with 10 roll-out scheme carried out.}
\label{tab:rollout}
\end{table}


\section{Details on coefficients used for exhibiting various behaviors}
\label{App:coeffs}

\subsubsection*{Forced noisy Duffing oscillator}

{Concretely, at $\delta = 1, ~\alpha = -1, ~\beta = 1, ~\gamma = 1, ~\omega = 1.4, ~\epsilon = 0.2$, the system exhibits nonlinear oscillations with chaotic behavior. At $\delta = 1.5, ~\alpha = -1, ~\beta = 1, ~\gamma = 0.1, ~\omega = 1, ~\epsilon = 0.1$, the stronger damping $\delta$ and weaker forcing $\gamma$ lead to oscillations around one of the potential wells with little chance of switching to the other well.
Lastly, at $\delta = 0.5, ~\alpha = -1, ~\beta = 1, ~\gamma = 0.5, ~\omega = 1, ~\epsilon = 0.05$, the light damping and the low noise level give us stable oscillations with occasional transitions between wells.}

\begin{figure}[ht]
    \centering
        \includegraphics[width=\textwidth]{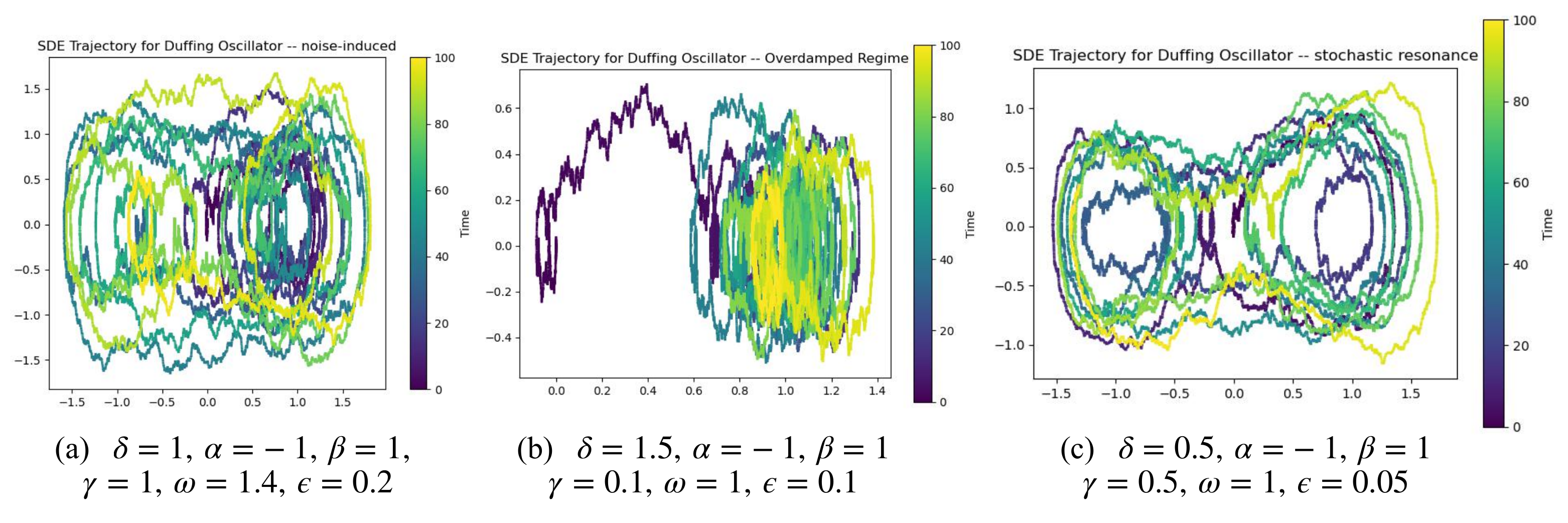}
    \caption{Simulation of Duffing oscillator. The trajectories are sampled with time step $\Delta t = $1e-3 for $T = 100$, starting at initial condition $[0, ~0]$.}
    \label{fig:appdx_duffing_traj}
\end{figure}

Figure~\ref{fig:appdx_duffing_traj} includes three sample trajectories of the forced noisy Duffing oscillator that exhibit different behaviors.

\subsubsection*{Predator-Prey}

\begin{figure}[ht]
    \centering
        \includegraphics[width=\textwidth]{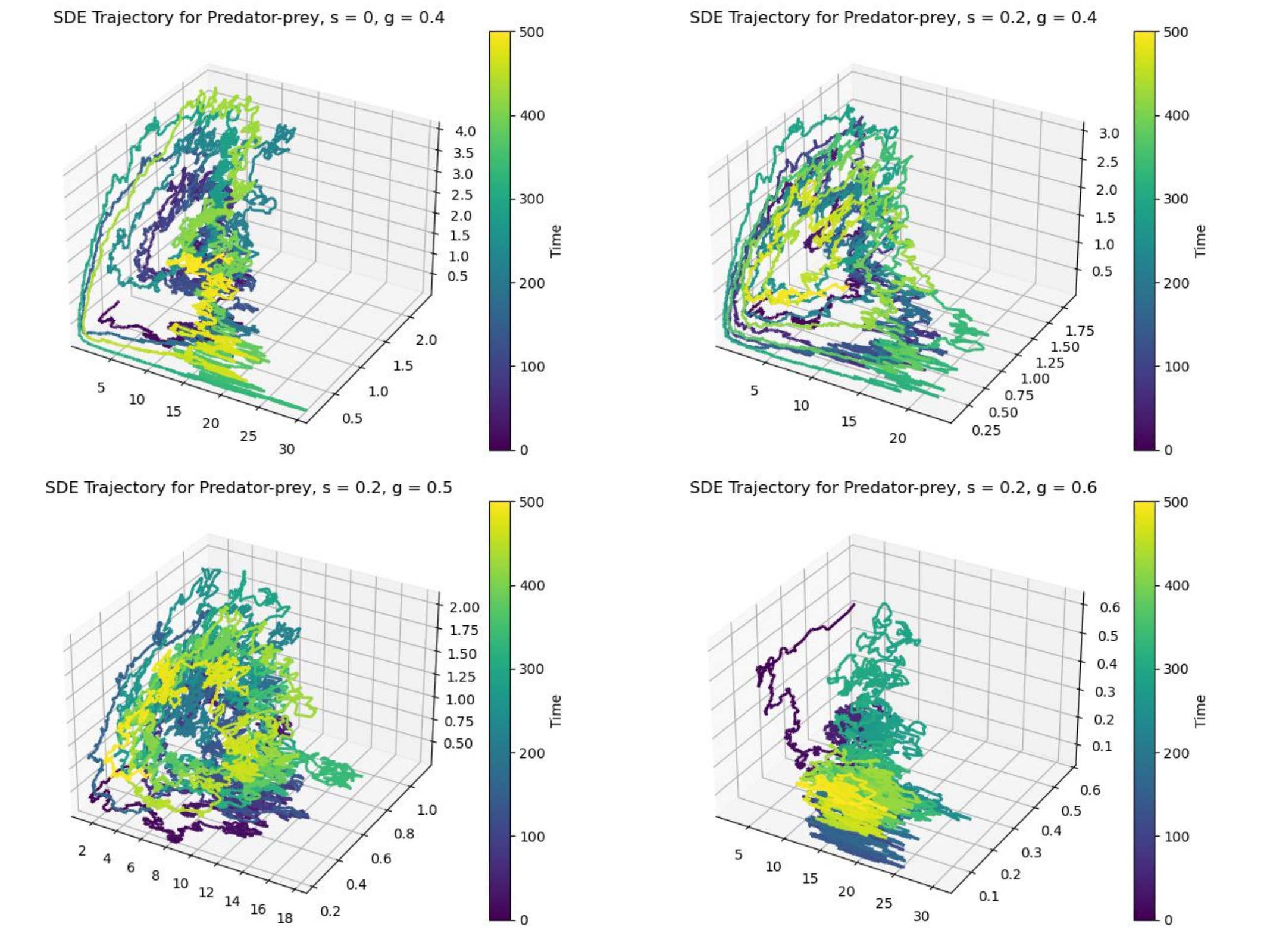}
    \caption{Simulation of predator and prey model. The trajectories are sampled with $r = 0.4, ~a = 0.02, ~b = 0.4, ~k = 0.4, ~D = 0.4, ~v_1 = 0.25, ~v_2 = 0.25, \sigma_1 = 0.15, ~\sigma_2 = 0.12, ~\sigma_3 = 0.1$ and time step $\Delta t = 0.05$ for $T = 500$, starting at initial condition $[1, 0.6, 0.4]$ for $[x, y_1, y_2]$.}
    \label{fig:appdx_predator_traj}
\end{figure}

{We examined the following combinations of parameters $s$ and $g$, while keeping other parameters constant: 
(1) $s = 0,~ g = 0.4$, representing a classic predator--prey coexistence with oscillations; 
(2) $s = 0.2,~ g = 0.4$, showing larger cycles and possible bistability; and 
(3) $s = 0.2,~ g = 0.6$, which tends toward a prey-dominated equilibrium or predator collapse, with noise accelerating the extinction risk.}
Figure~\ref{fig:appdx_predator_traj} includes the sample trajectories for four scenarios we considered for the predator-prey model. 

\subsubsection*{Stochastic Lorenz}

{For $0 < \rho < 1$, the origin is the unique globally attracting equilibrium. At $\rho = 1$, a supercritical pitchfork bifurcation occurs, creating two symmetric equilibria. When $1 < \rho < 13.926$, the origin becomes a saddle while the two symmetric equilibria are asymptotically stable. At $\rho = 13.926$, homoclinic orbits connecting back to the origin appear.
For $13.926 < \rho < 24.06$, the two symmetric equilibria are surrounded by saddle cycles, and preturbulence emerges. As $\rho$ approaches 24.74, the Lorenz attractor appears, coexisting with the stable equilibria. Finally, at $\rho = 24.74$, a subcritical Hopf bifurcation causes the cycles to shrink into the equilibria.}

\begin{figure}[h]
    \centering
        \includegraphics[width=\textwidth]{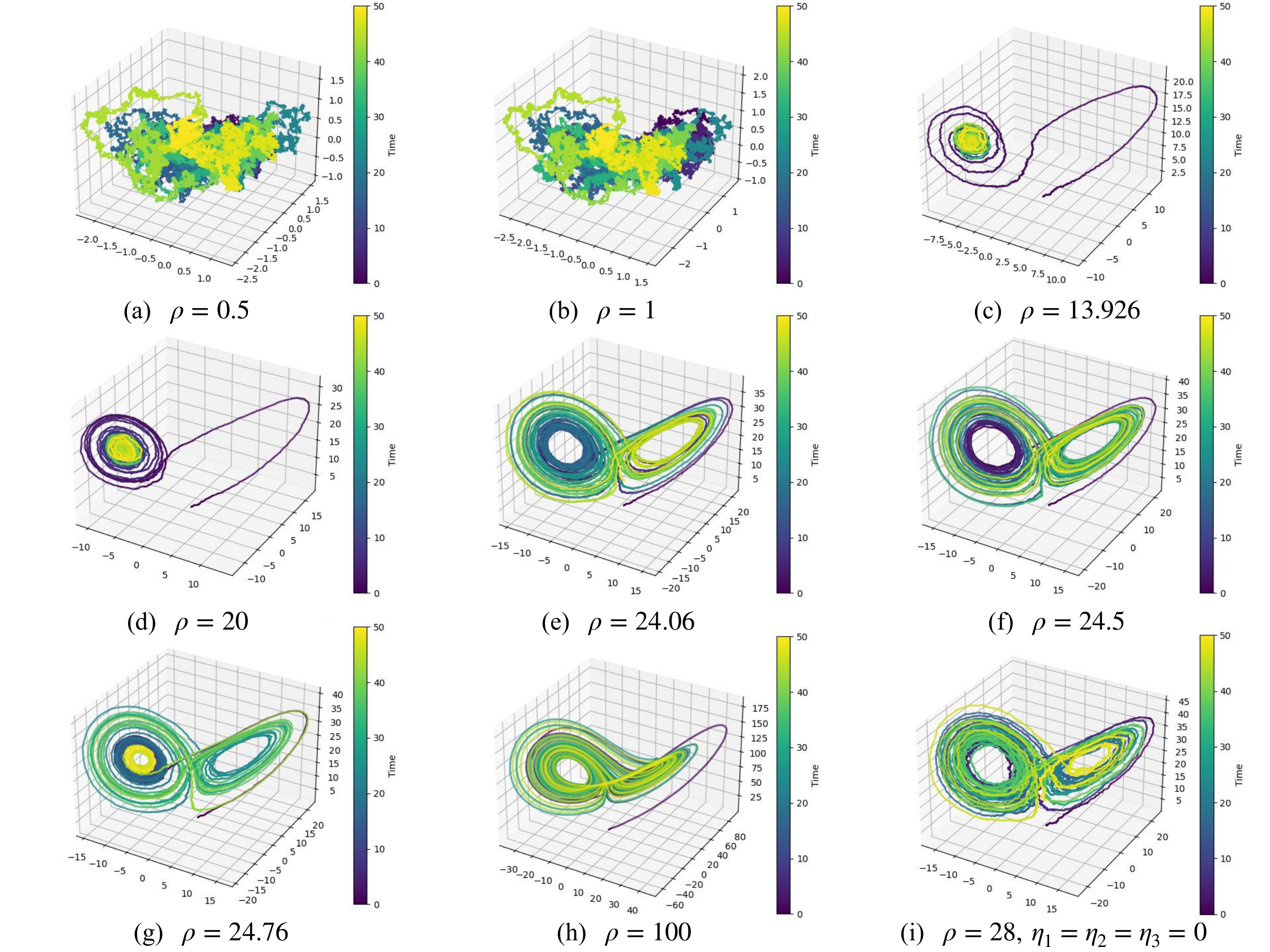}
    \caption{Simulation of stochastic Lorenz with $\sigma = 10$, $\beta = \frac{8}{3}$ and different values of $\rho$. The trajectories are sampled with time step $\Delta t = $1e-4 for $T = 50$, starting at initial condition $[1.0, 1.0, 1.0]$. The stochasticity level is kept at $\eta_1 = \eta_2 = \eta_3 = 0.8$ unless specified otherwise.}
    \label{fig:appdx_lorenz_traj}
\end{figure}

Figure~\ref{fig:appdx_lorenz_traj} includes trajectories of stochastic Lorenz with $\sigma = 10$, $\beta = \frac{8}{3}$ and \\
$\rho \in \{0.5, ~1.0, ~13.926, ~20, ~24.06, ~24.5, ~24.76, ~100\}$. 
The randomness factors are set to be $\eta_1 = \eta_2 = \eta_3 = 0.8$ unless specified otherwise.

Table~\ref{tab:diff_coeff} presents the performance of FMint-SDE compared to coarse solutions across four evaluation metrics. 

\input{Tables/diff_coeff}

\clearpage

\bibliographystyle{abbrv}

\end{document}

%% file: Tables/in-distribution-ft.tex
\begin{table}[h!]
\centering
\footnotesize
\begin{tabular}{p{1.9cm}|p{2.9cm}cccc}
\hline
\textbf{System} & \textbf{Method} & \textbf{MAE} & \textbf{RMSE} & \textbf{AMD} & \textbf{MAD} \\
\hline
\multirow{7}{*}{\shortstack[l]{Geometric \\ Brownian \\ Motion}}
 & Few-shot w/o prompt & \textbf{6.38e-3} & \textbf{1.11e-2} & \textbf{6.38e-3} & 8.28e-3 \\
 & Few-shot w/ prompt & 7.28e-3 & 1.22e-2 & 7.28e-3 & \textbf{8.10e-3} \\
 & Fine-tune w/o prompt & 1.11e-2 & 2.14e-2 & 1.11e-2 & 1.78e-2 \\
 & Fine-tune w/ prompt  & 1.14e-2 & 2.41e-2 & 1.14e-2 & 9.57e-3 \\
[1pt]
\cdashline{2-6}
\\[-8pt]
 & Coarse solution     & 4.90e-1 & 1.14  & 4.90e-1 & 3.40e-1 \\
 & Black-box surrogate & 6.23e-2 & 1.22e-1 & 6.23e-2 & 1.34e-1 \\
 & Single-SDE specialized & 9.02e-1 & 1.46  & 9.02e-1 & 8.67e-1 \\
\hline

\multirow{7}{*}{\shortstack[l]{OLD- \\ Mueller}} 
 & Few-shot w/o prompt & 1.23e-2 & 1.92e-2 & 1.76e-2 & 1.25e-2 \\
 & Few-shot w/ prompt & \textbf{1.05e-2} & \textbf{1.56e-2} & \textbf{1.49e-2} & \textbf{9.39e-3} \\
 & Fine-tune w/o prompt & 1.35e-2 & 2.23e-2 & 1.92e-2 & 1.59e-2 \\
 & Fine-tune w/ prompt  & 1.81e-2 & 2.75e-2 & 2.49e-2 & 2.36e-2 \\
 [1pt]
\cdashline{2-6}
\\[-8pt]
 & Coarse solution     & 5.56e-2 & 8.60e-2 & 7.83e-2 & 7.45e-2 \\
 & Black-box surrogate & 4.84e-2 & 6.97e-2 & 6.21e-2 & 8.39e-2 \\
 & Single-SDE specialized & 2.29e-1 & 3.58e-1 & 2.84e-1 & 2.72e-1 \\
\hline

\multirow{7}{*}{\shortstack[l]{Periodic \\ Nonlinear \\ Oscillator}}
 & Few-shot w/o prompt & 2.26e-3 & 3.40e-3 & 3.05e-3 & 1.91e-3 \\
 & Few-shot w/ prompt & 2.50e-3 & 4.01e-3 & 3.32e-3 & 2.57e-3 \\
 & Fine-tune w/o prompt & 8.40e-4 & 1.42e-3 & 1.20e-3 & 6.60e-4 \\
 & Fine-tune w/ prompt  & \textbf{5.73e-4} & \textbf{9.90e-4} & \textbf{8.21e-4} & \textbf{4.52e-4} \\
 [1pt]
\cdashline{2-6}
\\[-8pt]
 & Coarse solution     & 2.41e-3 & 4.45e-3 & 3.41e-3 & 8.78e-4 \\
 & Black-box surrogate & 7.86e-3 & 1.08e-2 & 9.24e-3 & 1.21e-2 \\
 & Single-SDE specialized & 6.51e-3 & 1.00e-2 & 9.52e-3 & 6.73e-3 \\
\hline

\multirow{7}{*}{\shortstack[l]{Stochastic \\ Lorenz}} 
 & Few-shot w/o prompt & 2.31e-2 & 3.16e-2 & 3.58e-2 & \textbf{1.68e-2} \\
 & Few-shot w/ prompt & \textbf{2.02e-2} & \textbf{2.83e-2} & \textbf{3.16e-2} & 2.15e-2 \\
 & Fine-tune w/o prompt & 2.31e-2 & 3.07e-2 & 3.53e-2 & 2.36e-2 \\
 & Fine-tune w/ prompt  & 3.59e-2 & 4.78e-2 & 5.09e-2 & 4.94e-2 \\
 [1pt]
\cdashline{2-6}
\\[-8pt]
 & Coarse solution     & 1.32e-1 & 1.74e-1 & 2.02e-1 & 1.94e-1 \\
 & Black-box surrogate & 8.66e-2 & 1.16e-1 & 1.34e-1 & 8.60e-2 \\
 & Single-SDE specialized & 9.45e-1 & 1.26  & 1.56  & 7.54e-1 \\
\hline
\end{tabular}
\caption{\footnotesize Comparison of error metrics (MAE, RMSE, AMD, MAD) on the in-distribution SDEs. The smallest error in each column is highlighted in bold.}
\label{tab:in_dist_error_all}
\end{table}

%% file: Tables/input_demo.tex
\begin{table}[h]
\caption{Input data structure for a 2D SDE demo.}
\centering
\scriptsize
\begin{tabular}{ccccccc|ccccccccc}
\toprule
& \multicolumn{3}{c}{Coarse solution} & \multicolumn{3}{c}{Error term} & \multicolumn{3}{c}{Query coarse solution} & \multicolumn{3}{c}{Query error term} \\
\cmidrule(r){2-4}
\cmidrule(r){5-7}
\cmidrule(r){8-10}
\cmidrule(r){11-13}

time & 0 & $\ldots$ & $t_n$ & 0 &$\ldots$  & $t_n$ & 0 &$\ldots$  & $t_q$ & 0 &$\ldots$  & $t_q$ \\
1st-dim noise & $\Delta W^1_0$ & $\ldots$ & $\Delta W^1_n$ & $\Delta W^1_0$ &$\ldots$  & $\Delta W^1_n$ & $\Delta \tilde{W}^1_0$ &$\ldots$  & $\Delta \tilde{W}^1_q$ & $\Delta \tilde{W}^1_0$ &$\ldots$  & $\Delta \tilde{W}^1_q$ \\
2nd-dim noise & $\Delta W^2_0$ & $\ldots$ & $\Delta W^2_n$ & $\Delta W^2_0$ &$\ldots$  & $\Delta W^2_n$ & $\Delta \tilde{W}^2_0$ &$\ldots$  & $\Delta \tilde{W}^2_q$ & $\Delta \tilde{W}^2_0$ &$\ldots$  & $\Delta \tilde{W}^2_q$ \\
\midrule
value  & $\hat{X}_1(0)$  & $\ldots$ & $\hat{X}_1(t_{n})$ & $\text{err}_{\hat{X}_1}(0)$  & $\ldots$ & $\text{err}_{\hat{X}_1}(t_{n})$ & $\hat{X}_1^q(0)$  & $\ldots$ & $\hat{X}_1^q(t_{q})$ & $\text{err}_{\hat{X}_1^q}(0)$  & $\ldots$ & $\text{err}_{\hat{X}_1^q}(t_{q})$ \\
 & $\hat{X}_2(0)$&  $\ldots$ & $\hat{X}_2(t_{n})$ & $\text{err}_{\hat{X}_2}(0)$ & $\ldots$ & $\text{err}_{\hat{X}_2}(t_{n})$ & $\hat{X}_2^q(0)$&  $\ldots$ & $\hat{X}_2^q(t_{q})$ & $\text{err}_{\hat{X}_2^q}(0)$  & $\ldots$ & $\text{err}_{\hat{X}_2^q}(t_{q})$\\
\bottomrule
\end{tabular}
\label{tab:token}
\end{table} 


%% file: Tables/finetune.tex
\begin{table}[htbp]
\centering
\footnotesize
\begin{tabular}{l|lcccc}
\hline
\textbf{System} & \textbf{Method} & \textbf{MAE} & \textbf{RMSE} & \textbf{AMD} & \textbf{MAD} \\
\hline

\multirow{5}{*}{Coupled double-well}
  & Finetune w/o prompt & \textbf{7.016e-4} & \textbf{9.928e-4} & \textbf{9.992e-4} & \textbf{6.580e-4} \\
  & Finetune w/ prompt  & 8.304e-4 & 1.164e-3 & 1.150e-3 & 9.637e-4 \\
  & Coarse solution     & 6.306e-3 & 9.741e-3 & 9.011e-3 & 2.725e-3 \\
  & Black-box surrogate & 2.219e-2 & 3.279e-2 & 2.489e-2 & 1.841e-2 \\
  & Single-SDE specialized & 2.067e-3 & 3.614e-3 & 3.700e-3 & 3.666e-3 \\
\hline

\multirow{5}{*}{Double-well}
  & Finetune w/o prompt & \textbf{5.814e-4} & \textbf{8.323e-4} & \textbf{8.318e-4} & \textbf{5.706e-4} \\
  & Finetune w/ prompt  & 6.994e-4 & 9.456e-4 & 9.706e-4 & 7.395e-4 \\
  & Coarse solution     & 5.273e-3 & 8.092e-3 & 7.517e-3 & 2.141e-3 \\
  & Black-box surrogate & 1.844e-2 & 2.690e-2 & 2.085e-2 & 1.680e-2 \\
  & Single-SDE specialized & 2.877e-3 & 4.062e-3 & 4.052e-3 & 3.827e-3 \\
\hline

\multirow{5}{*}{Duffing}
  & Finetune w/o prompt & \textbf{1.897e-3} & \textbf{2.723e-3} & \textbf{2.694e-3} & \textbf{2.117e-3} \\
  & Finetune w/ prompt  & 2.155e-3 & 3.147e-3 & 2.962e-3 & 2.966e-3 \\
  & Coarse solution     & 1.565e-2 & 2.412e-2 & 2.222e-2 & 8.800e-3 \\
  & Black-box surrogate & 3.155e-2 & 4.290e-2 & 3.851e-2 & 3.842e-2 \\
  & Single-SDE specialized & 1.207e-1 & 1.746e-1 & 1.891e-1 & 1.790e-1 \\
\hline

\multirow{5}{*}{Fluxgate sensor}
  & Finetune w/o prompt & 3.327e-2 & 4.278e-2 & 4.726e-2 & 4.146e-2 \\
  & Finetune w/ prompt  & \textbf{2.812e-2} & \textbf{3.729e-2} & \textbf{4.108e-2} & \textbf{3.213e-2} \\
  & Coarse solution     & 5.067e-2 & 6.460e-2 & 6.314e-2 & 8.161e-2 \\
  & Black-box surrogate & 1.042e-1 & 1.319e-1 & 1.441e-1 & 1.327e-1 \\
  & Single-SDE specialized & 2.149e-1 & 4.262e-1 & 4.959e-1 & 3.305e-1 \\
\hline

\multirow{5}{*}{Inhomogeneous Ornstein-Uhlenbeck}
  & Finetune w/o prompt & \textbf{1.067e-2} & \textbf{1.731e-2} & \textbf{1.067e-2} & \textbf{2.105e-2} \\
  & Finetune w/ prompt  & 2.462e-2 & 4.298e-2 & 2.462e-2 & 6.744e-2 \\
  & Coarse solution     & 1.118 & 1.963 & 1.118 & 3.354 \\
  & Black-box surrogate & 1.348e-1 & 2.396e-1 & 1.348e-1 & 2.505e-1 \\
  & Single-SDE specialized & 7.562 & 7.804 & 7.562 & 7.562 \\
\hline

\multirow{5}{*}{Ornstein-Uhlenbeck}
  & Finetune w/o prompt & 6.603e-3 & \textbf{8.874e-3} & 6.603e-3 & \textbf{9.875e-3} \\
  & Finetune w/ prompt  & \textbf{5.899e-3} & 9.162e-3 & \textbf{5.899e-3} & 1.249e-2 \\
  & Coarse solution     & 5.481e-1 & 9.658e-1 & 5.481e-1 & 1.644 \\
  & Black-box surrogate & 8.278e-2 & 1.585e-1 & 8.278e-2 & 2.108e-1 \\
  & Single-SDE specialized & 6.071 & 6.944 & 6.071 & 6.071 \\
\hline

\multirow{5}{*}{Perturbed Nonlinear Oscillator}
  & Finetune w/o prompt & 1.518e-3 & 2.517e-3 & 2.090e-3 & 1.303e-3 \\
  & Finetune w/ prompt  & 1.404e-3 & 2.428e-3 & 1.922e-3 & 1.177e-3 \\
  & Coarse solution     & 7.656e-3 & 1.411e-2 & 1.086e-2 & 3.353e-3 \\
  & Black-box surrogate & 2.237e-2 & 3.507e-2 & 2.544e-2 & 2.716e-2 \\
  & Single-SDE specialized & \textbf{9.011e-4} & \textbf{1.316e-3} & \textbf{1.295e-3} & \textbf{1.153e-3} \\
\hline

\multirow{5}{*}{Predator-prey}
  & Finetune w/o prompt & \textbf{3.470e-3} & \textbf{4.867e-3} & \textbf{5.748e-3} & \textbf{3.018e-3} \\
  & Finetune w/ prompt  & 4.102e-3 & 5.533e-3 & 6.846e-3 & 3.910e-3 \\
  & Coarse solution     & 9.708e-3 & 1.366e-2 & 1.516e-2 & 1.017e-2 \\
  & Black-box surrogate & 2.523e-2 & 3.233e-2 & 3.442e-2 & 2.906e-2 \\
  & Single-SDE specialized & 5.647e-2 & 2.074e-1 & 1.346e-1 & 1.206e-1 \\
\hline

\end{tabular}
\caption{Comparison of error metrics (MAE, RMSE, AMD, MAD) across systems for Finetuned models (with/without prompt), Coarse solutions, Black-box surrogates, and Single-SDE specialized models. Best (lowest) in each column is highlighted, except for Fluxgate Sensor.}
\label{tab:finetune-OOD}
\end{table}

%% file: Tables/strong_samples.tex
\begin{table}[h]
\centering
\footnotesize
\begin{tabular}{l|cccccccc}
\hline
\textbf{System} & \textbf{5} & \textbf{50} & \textbf{100} & \textbf{250} & \textbf{500} & \textbf{1000} & \textbf{5000} & \textbf{Coarse} \\
\hline
Coupled Double Well & 2.907e-3 & 1.150e-3 & 1.124e-3 & 9.868e-4 & 1.057e-3 & 1.010e-3 & 1.046e-3 & 8.910e-3 \\
Double Well & 2.087e-3 & 9.706e-4 & 9.044e-4 & 9.058e-4 & 8.673e-4 & 8.685e-4 & 8.648e-4 & 7.413e-3 \\
Duffing Langevin & 3.700e-3 & 2.962e-3 & 2.911e-3 & 2.923e-3 & 3.068e-3 & 2.994e-3 & 2.908e-3 & 2.201e-2 \\
Geometric Brownian Motion & 1.545e-2 & 1.141e-2 & 1.083e-2 & 9.938e-3 & 1.065e-2 & 1.148e-2 & 9.267e-3 & 5.114e-1 \\
Inhomogeneous Ornstein Uhlenbeck & 1.619e-2 & 2.462e-2 & 1.601e-2 & 2.014e-2 & 1.980e-2 & 1.964e-2 & 1.990e-2 & 1.122 \\
Mueller & 2.769e-2 & 2.485e-2 & 2.424e-2 & 2.141e-2 & 2.617e-2 & 2.240e-2 & 2.361e-2 & 7.229e-2 \\
Ornstein Uhlenbeck & 6.138e-3 & 5.899e-3 & 3.736e-3 & 6.255e-3 & 5.307e-3 & 6.449e-3 & 8.500e-3 & 5.399e-1 \\
Periodic Nonlinear Oscillator & 2.762e-3 & 8.211e-4 & 8.487e-4 & 8.036e-4 & 2.232e-3 & 7.123e-4 & 9.600e-4 & 3.130e-3 \\
Perturbed Nonlinear Oscillator & 7.892e-3 & 1.922e-3 & 1.837e-3 & 1.553e-3 & 2.151e-3 & 1.661e-3 & 1.871e-3 & 8.914e-3 \\
Predator and Prey & 7.402e-3 & 6.846e-3 & 7.035e-3 & 6.895e-3 & 6.866e-3 & 7.275e-3 & 6.551e-3 & 1.668e-2 \\
Stochastic Lorenz & 9.528e-2 & 5.089e-2 & 4.866e-2 & 4.779e-2 & 4.850e-2 & 5.166e-2 & 5.263e-2 & 1.928e-1 \\
\hline
\end{tabular}
\caption{AMD vs fine-tuning size for all systems (scientific notation).}
\label{tab:AMD_samples}
\end{table}

%% file: Tables/weak_samples.tex
\begin{table}[htbp]
\centering
\footnotesize
\begin{tabular}{l|cccccccc}
\hline
\textbf{System} & \textbf{5} & \textbf{50} & \textbf{100} & \textbf{250} & \textbf{500} & \textbf{1000} & \textbf{5000} & \textbf{Coarse} \\
\hline
Coupled Double Well & 3.647e-3 & 9.637e-4 & 8.942e-4 & 5.398e-4 & 7.416e-4 & 6.460e-4 & 6.781e-4 & 2.137e-3 \\
Double Well & 2.610e-3 & 7.395e-4 & 6.525e-4 & 6.742e-4 & 5.444e-4 & 4.950e-4 & 5.912e-4 & 2.133e-3 \\
Duffing Langevin & 4.529e-3 & 2.966e-3 & 2.489e-3 & 2.523e-3 & 2.624e-3 & 2.255e-3 & 2.657e-3 & 7.996e-3 \\
Geometric Brownian Motion & 1.619e-2 & 9.565e-3 & 9.963e-3 & 9.351e-3 & 1.230e-2 & 1.032e-2 & 1.321e-2 & 3.152e-1 \\
Inhomogeneous Ornstein Uhlenbeck & 3.744e-2 & 6.744e-2 & 3.759e-2 & 5.034e-2 & 5.001e-2 & 4.943e-2 & 5.176e-2 & 3.366 \\
Mueller & 2.857e-2 & 2.364e-2 & 2.425e-2 & 1.962e-2 & 2.861e-2 & 1.993e-2 & 1.810e-2 & 6.405e-2 \\
Ornstein Uhlenbeck & 1.337e-2 & 1.249e-2 & 5.495e-3 & 1.339e-2 & 7.450e-3 & 1.334e-2 & 1.981e-2 & 1.620 \\
Periodic Nonlinear Oscillator & 2.393e-3 & 4.519e-4 & 3.874e-4 & 4.296e-4 & 1.076e-3 & 3.043e-4 & 4.706e-4 & 1.061e-3 \\
Perturbed Nonlinear Oscillator & 6.871e-3 & 1.177e-3 & 8.496e-4 & 1.062e-3 & 1.223e-3 & 9.857e-4 & 7.860e-4 & 3.173e-3 \\
Predator and Prey & 6.112e-3 & 3.910e-3 & 3.660e-3 & 3.175e-3 & 3.562e-3 & 3.732e-3 & 3.772e-3 & 1.365e-2 \\
Stochastic Lorenz & 1.484e-1 & 4.940e-2 & 4.660e-2 & 4.444e-2 & 4.955e-2 & 4.841e-2 & 5.239e-2 & 1.805e-1 \\
\hline
\end{tabular}
\caption{MAD vs fine-tuning size for all systems (scientific notation).}
\label{tab:MAD_samples}
\end{table}

%% file: Tables/mae_samples.tex
\begin{table}[htbp]
\centering
\footnotesize
\begin{tabular}{l|cccccccc}
\hline
\textbf{System} & \textbf{5} & \textbf{50} & \textbf{100} & \textbf{250} & \textbf{500} & \textbf{1000} & \textbf{5000} & \textbf{Coarse} \\
\hline
Coupled Double Well & 2.397e-3 & 8.304e-4 & 8.071e-4 & 6.903e-4 & 7.500e-4 & 7.029e-4 & 7.452e-4 & 6.290e-3 \\
Double Well & 1.698e-3 & 6.994e-4 & 6.410e-4 & 6.371e-4 & 6.111e-4 & 6.043e-4 & 6.078e-4 & 5.320e-3 \\
Duffing Langevin & 2.843e-3 & 2.155e-3 & 2.098e-3 & 2.125e-3 & 2.223e-3 & 2.150e-3 & 2.113e-3 & 1.569e-2 \\
Geometric Brownian Motion & 1.545e-2 & 1.141e-2 & 1.083e-2 & 9.938e-3 & 1.065e-2 & 1.148e-2 & 9.267e-3 & 5.114e-1 \\
Inhomogeneous Ornstein Uhlenbeck & 1.619e-2 & 2.462e-2 & 1.601e-2 & 2.014e-2 & 1.980e-2 & 1.964e-2 & 1.990e-2 & 1.122 \\
Mueller & 2.016e-2 & 1.810e-2 & 1.755e-2 & 1.541e-2 & 1.914e-2 & 1.607e-2 & 1.700e-2 & 5.146e-2 \\
Ornstein Uhlenbeck & 6.138e-3 & 5.899e-3 & 3.736e-3 & 6.255e-3 & 5.307e-3 & 6.449e-3 & 8.500e-3 & 5.399e-1 \\
Periodic Nonlinear Oscillator & 2.008e-3 & 5.726e-4 & 5.865e-4 & 5.371e-4 & 1.620e-3 & 4.912e-4 & 6.530e-4 & 2.170e-3 \\
Perturbed Nonlinear Oscillator & 5.883e-3 & 1.404e-3 & 1.324e-3 & 1.135e-3 & 1.542e-3 & 1.203e-3 & 1.325e-3 & 6.314e-3 \\
Predator and Prey & 4.918e-3 & 4.102e-3 & 4.219e-3 & 4.107e-3 & 4.153e-3 & 4.378e-3 & 3.972e-3 & 1.105e-2 \\
Stochastic Lorenz & 7.872e-2 & 3.594e-2 & 3.467e-2 & 3.318e-2 & 3.441e-2 & 3.627e-2 & 3.712e-2 & 1.245e-1 \\
\hline
\end{tabular}
\caption{MAE vs fine-tuning size for all systems (scientific notation).}
\label{tab:MAE_samples}
\end{table}

%% file: Tables/rmse_samples.tex
\begin{table}[htbp]
\centering
\footnotesize
\begin{tabular}{l|cccccccc}
\hline
\textbf{System} & \textbf{5} & \textbf{50} & \textbf{100} & \textbf{250} & \textbf{500} & \textbf{1000} & \textbf{5000} & \textbf{Coarse} \\
\hline
Coupled Double Well & 3.452e-3 & 1.164e-3 & 1.132e-3 & 9.982e-4 & 1.064e-3 & 1.007e-3 & 1.059e-3 & 9.648e-3 \\
Double Well & 2.278e-3 & 9.456e-4 & 9.210e-4 & 9.206e-4 & 8.757e-4 & 8.464e-4 & 8.719e-4 & 8.074e-3 \\
Duffing Langevin & 4.205e-3 & 3.147e-3 & 2.974e-3 & 2.948e-3 & 3.134e-3 & 3.050e-3 & 3.054e-3 & 2.412e-2 \\
Geometric Brownian Motion & 3.341e-2 & 2.414e-2 & 2.158e-2 & 2.110e-2 & 2.311e-2 & 2.430e-2 & 1.759e-2 & 1.147 \\
Inhomogeneous Ornstein Uhlenbeck & 3.034e-2 & 4.298e-2 & 2.817e-2 & 3.190e-2 & 3.126e-2 & 3.142e-2 & 3.212e-2 & 1.987 \\
Mueller & 3.179e-2 & 2.753e-2 & 2.715e-2 & 2.471e-2 & 3.020e-2 & 2.537e-2 & 2.729e-2 & 7.990e-2 \\
Ornstein Uhlenbeck & 1.039e-2 & 9.162e-3 & 5.561e-3 & 1.070e-2 & 7.426e-3 & 9.888e-3 & 1.554e-2 & 9.490e-1 \\
Periodic Nonlinear Oscillator & 3.495e-3 & 9.903e-4 & 1.076e-3 & 1.092e-3 & 2.741e-3 & 8.631e-4 & 1.169e-3 & 4.269e-3 \\
Perturbed Nonlinear Oscillator & 1.028e-2 & 2.428e-3 & 2.337e-3 & 2.110e-3 & 2.648e-3 & 2.081e-3 & 2.664e-3 & 1.270e-2 \\
Predator and Prey & 6.696e-3 & 5.533e-3 & 5.843e-3 & 5.782e-3 & 5.646e-3 & 6.087e-3 & 5.512e-3 & 1.483e-2 \\
Stochastic Lorenz & 1.086e-1 & 4.780e-2 & 4.618e-2 & 4.303e-2 & 4.813e-2 & 4.769e-2 & 4.942e-2 & 1.660e-1 \\
\hline
\end{tabular}
\caption{RMSE vs fine-tuning size for all systems (scientific notation).}
\label{tab:RMSE_samples}
\end{table}

%% file: Tables/diff_coeff.tex
\begin{table}[htbp]
\centering
\scriptsize
\begin{tabular}{l|lcccc}
\hline
\textbf{System} & \textbf{Method} & \textbf{MAE} & \textbf{RMSE} & \textbf{AMD} & \textbf{MAD} \\
\hline
\multirow{2}{*}{Duffing Langevin (Noise-Induced)} 
 & Finetune         & 2.570e-3 & 3.772e-3 & 3.682e-3 & 2.629e-3 \\
 & Coarse solution  & 2.379e-2 & 3.663e-2 & 3.381e-2 & 1.392e-2 \\
\hline
\multirow{2}{*}{Duffing Langevin (Overdamped)} 
 & Finetune         & 2.212e-3 & 3.127e-3 & 3.090e-3 & 2.432e-3 \\
 & Coarse solution  & 1.690e-2 & 2.605e-2 & 2.402e-2 & 1.053e-2 \\
\hline
\multirow{2}{*}{Duffing Langevin (Stochastic Resonance)} 
 & Finetune         & 1.726e-3 & 2.408e-3 & 2.418e-3 & 1.824e-3 \\
 & Coarse solution  & 1.205e-2 & 1.856e-2 & 1.712e-2 & 8.217e-3 \\
\hline
\multirow{2}{*}{Predator and Prey ($s=0, g=0.4$)} 
 & Finetune         & 4.439e-3 & 6.250e-3 & 7.451e-3 & 3.355e-3 \\
 & Coarse solution  & 1.123e-2 & 1.562e-2 & 1.786e-2 & 1.264e-2 \\
\hline
\multirow{2}{*}{Predator and Prey ($s=0.2, g=0.4$)} 
 & Finetune         & 4.442e-3 & 6.251e-3 & 7.452e-3 & 3.373e-3 \\
 & Coarse solution  & 1.123e-2 & 1.562e-2 & 1.786e-2 & 1.264e-2 \\
\hline
\multirow{2}{*}{Predator and Prey ($s=0.2, g=0.6$)} 
 & Finetune         & 4.777e-3 & 6.747e-3 & 7.919e-3 & 4.354e-3 \\
 & Coarse solution  & 1.158e-2 & 1.611e-2 & 1.827e-2 & 1.387e-2 \\
\hline
\multirow{2}{*}{Predator and Prey (Variant)} 
 & Finetune         & 4.183e-3 & 5.816e-3 & 6.943e-3 & 3.469e-3 \\
 & Coarse solution  & 1.130e-2 & 1.566e-2 & 1.745e-2 & 1.259e-2 \\
\hline
\multirow{2}{*}{Lorenz ($\rho=1$)} 
 & Finetune         & 4.979e-3 & 6.410e-3 & 8.094e-3 & 3.483e-3 \\
 & Coarse solution  & 5.948e-2 & 7.870e-2 & 9.042e-2 & 8.948e-2 \\
\hline
\multirow{2}{*}{Lorenz ($\rho=10$)} 
 & Finetune         & 4.065e-3 & 5.426e-3 & 5.878e-3 & 5.128e-3 \\
 & Coarse solution  & 4.608e-2 & 6.071e-2 & 6.149e-2 & 7.606e-2 \\
\hline
\multirow{2}{*}{Lorenz ($\rho=13.926$)} 
 & Finetune         & 6.967e-3 & 9.046e-3 & 1.083e-2 & 7.088e-3 \\
 & Coarse solution  & 7.357e-2 & 9.225e-2 & 1.083e-1 & 1.037e-1 \\
\hline
\multirow{2}{*}{Lorenz ($\rho=20$)} 
 & Finetune         & 1.153e-2 & 1.506e-2 & 1.723e-2 & 1.419e-2 \\
 & Coarse solution  & 8.385e-2 & 1.055e-1 & 1.196e-1 & 1.238e-1 \\
\hline
\multirow{2}{*}{Lorenz ($\rho=24.5$)} 
 & Finetune         & 1.722e-2 & 2.409e-2 & 2.462e-2 & 2.282e-2 \\
 & Coarse solution  & 9.178e-2 & 1.171e-1 & 1.280e-1 & 1.470e-1 \\
\hline
\multirow{2}{*}{Lorenz ($\rho=24.06$)} 
 & Finetune         & 1.705e-2 & 2.349e-2 & 2.433e-2 & 2.245e-2 \\
 & Coarse solution  & 9.099e-2 & 1.159e-1 & 1.272e-1 & 1.447e-1 \\
\hline
\multirow{2}{*}{Lorenz ($\rho=24.76$)} 
 & Finetune         & 1.813e-2 & 2.606e-2 & 2.573e-2 & 2.429e-2 \\
 & Coarse solution  & 9.224e-2 & 1.178e-1 & 1.285e-1 & 1.485e-1 \\
\hline
\multirow{2}{*}{Lorenz ($\rho=100$)} 
 & Finetune         & 6.383e-2 & 8.211e-2 & 9.392e-2 & 8.263e-2 \\
 & Coarse solution  & 1.268e-1 & 1.812e-1 & 1.733e-1 & 2.583e-1 \\
\hline
\multirow{2}{*}{Lorenz ($\rho=28, \eta=2.0$)} 
 & Finetune         & 3.146e-2 & 4.172e-2 & 5.020e-2 & 3.034e-2 \\
 & Coarse solution  & 1.781e-1 & 2.224e-1 & 2.927e-1 & 1.457e-1 \\
\hline
\end{tabular}
\caption{Comparison of error metrics (MAE, RMSE, AMD, MAD) for systems with different behaviors, showing finetuned models vs. Coarse solutions. All values are reported in scientific notation.}
\label{tab:diff_coeff}
\vspace{-2em}
\end{table}